\documentclass[pra,aps,twocolumn,eqsecnum,showpacs,letterpaper,10pt]{revtex4}
\usepackage{amsmath}
\usepackage{amssymb}
\usepackage{amsfonts}
\usepackage{bm} 
\usepackage{hyperref}
\usepackage{color}
\usepackage[ngerman]{babel}
\usepackage[T1]{fontenc}
\usepackage[latin1]{inputenc}

\begin{document}
\title{Der Weg von der mikroskopischen Vielteilchen-Theorie\newline zur makroskopischen Hydrodynamik}

\author{Rudolf Haussmann}
\affiliation{Fachbereich Physik, Universität Konstanz, D-78457 Konstanz, Deutschland}

\email[E-Mail:\enskip]{rudolf.haussmann@uni-konstanz.de}
\homepage[\newline Internet:\enskip]{https://www.kormoran-tech.com}

\date{24. Februar 2016; \enskip veröffentlicht in \ Journal of Physics Condensed Matter \textbf{28}, 113001 (2016)}

\begin{abstract}
Ausgehend von der mikroskopischen Beschreibung einer Flüssigkeit durch eine beliebige lokale 
Theorie für wechselwirkende Vielteilchensysteme wird mittels Methoden der statistischen Physik 
für Nichtgleichgewichtssysteme gezeigt, wie allgemein hydrodynamische Gleichungen hergeleitet 
werden können. Für die Dichten von den Erhaltungsgrößen werden Bewegungsgleichungen gefunden 
mit drei Arten von Termen: reversiblen, dissipativen und fluktuierenden. Zunächst sind diese 
Gleichungen vollständig exakt und enthalten räumlich und zeitlich nicht lokale Terme, die 
Gedächtniseffekte beschreiben. Mit wenigen Näherungen werden die nicht lokalen Eigenschaften 
und die Gedächtniseffekte entfernt, und man erhält die bekannten hydrodynamischen Gleichungen 
einer normalen Flüssigkeit mit gaußischen stochastischen Kräften. Anschließend wird untersucht, 
wie die Zeitumkehrinvarianz der mikroskopischen Theorie gebrochen wird und wie der zweite 
Hauptsatz der Thermodynamik zustande kommt. Weiterhin wird gezeigt, dass die hydrodynamischen 
Gleichungen mit Fluktuationen äquivalent zu stochastischen Langevin-Gleichungen und zugehörigen 
Fokker-Planck-Gleichungen sind. Zum Schluss wird das Fluktuations-Theorem untersucht und 
mit einem Zusatzterm erweitert.
\end{abstract}


\maketitle

\tableofcontents

\section{Einleitung}
\label{section::1}
Eine Flüssigkeit ist ein dicht gepacktes System aus vielen Atomen oder Molekülen, die über 
Kräfte elektromagnetischen Ursprungs miteinander wechselwirken. Die Teilchen bewegen sich 
nahezu in einem thermischen Gleichgewicht, das zumindest lokal auf kleinen Längenskalen 
ausgebildet ist. Die Temperatur ist hinreichend hoch, so dass das Vielteilchensystem nicht 
fest ist, sondern die fließenden Bewegungen einer Flüssigkeit ausführen kann \cite{HM86}.

Auf mikroskopischer Ebene wird das Vielteilchensystem durch eine klassische oder eine 
Quanten-Theorie beschrieben, welche die Bewegung von Atomen oder Molekülen modelliert, 
die über ein beliebiges Feld miteinander wechselwirken \cite{Ma00,LL09}. Eine detaillierte 
Ausarbeitung und Lösung einer solchen mikroskopischen Theorie ist in der Praxis nahezu 
unmöglich wegen der starken Wechselwirkung in der Flüssigkeit und der extrem großen 
Anzahl von Freiheitsgraden durch die extrem vielen Teilchen. Es sind nur Näherungen 
im Rahmen einer Störungstheorie mit Feynman-Diagrammen möglich.

Auf makroskopischer Ebene verhält sich eine Flüssigkeit deutlich einfacher. Auf kurzen Längen- 
und Zeitskalen befindet sich die Flüssigkeit lokal in einem thermischen Gleichgewicht. Auf 
großen Längen- und Zeitskalen ist sie im Nichtgleichgewicht. Sie fließt und transportiert 
Masse, Impuls und Wärme wie ein kontinuierliches Medium. Die Anzahl der Freiheitsgrade, 
die hierbei eine Rolle spielen, ist deutlich geringer. 

Die makroskopische Bewegung eines Kontinuums wird durch eine Theorie beschrieben, die 
unter dem Namen \emph{Hydrodynamik} bekannt ist \cite{LL06}. Es werden phänomenologische 
Gleichungen aufgestellt welche die Erhaltungssätze der physikalischen Größen Masse, Impuls 
und Energie auf lokaler Ebene erfüllen. Diese phänomenologischen Gleichungen sind 
Kontinuitätsgleichungen für die Dichten und Stromdichten der physikalischen Größen. 
Man erhält ein geschlossenes System von solchen Gleichungen, indem man geeignete Ansätze 
für die Stromdichten macht und diese wiederum durch die Dichten und die Gradienten von 
den Dichten ausdrückt.

Die Stromdichten enthalten zum einen reversible Terme, welche aus der Kinematik der zugrunde 
liegenden mikroskopischen Theorie bestimmt werden. Zum zweiten enthalten sie dissipative 
Terme, welche durch die komplizierten Wechselwirkungen der vielen Teilchen erzeugt und durch 
einen linearen Ansatz mit Gradienten der intensiven thermdynamischen Variablen wie Temperatur, 
Geschwindigkeit und chemisches Potential modelliert werden. Zum dritten enthalten sie 
fluktuierende Terme, welche durch gaußische stochastische Kräfte dargestellt werden \cite{LL09}.

In dieser Arbeit wird ein wohl etablierter Weg beschrieben, wie man die hydrodynamischen 
Gleichungen konsequent aus der mikroskopischen Theorie eines wechselwirkenden 
Vielteilchensystems herleiten kann. Wir verbinden die eher herkömmlichen Herleitungen 
mit neueren Themen der statistischen Theorie von Systemen fern vom Gleichgewicht, 
und zwar dem GENERIC-Formalismus \cite{GO97A,GO97B,Ot05} und dem Fluktuations-Theorem 
\cite{Ev93,Ev94,Ev02a,Ja97A,Ja97B}. Ein größerer Teil dieser Arbeit widmet sich den 
\emph{Fluktuationen}. Die herkömmlichen Darstellungen behandeln die Fluktuationen 
in linearer Antwort in der Nähe des Gleichgewichts, wobei die Korrelationsfunktionen 
im thermischen Gleichgewicht berechnet werden \cite{KM63,Fo75}. In dieser Arbeit jedoch 
betrachten wir die Hydrodynamik und die Fluktuationen im Nichtgleichgewicht in einer 
vollständig \emph{nichtlinearen} Weise. Wir leiten nichtlineare stochastische 
Differentialgleichungen mit multiplikativem Rauschen her, so dass die Fluktuationen 
nichtlinear von den hydrodynamischen Variablen abhängen.

Ausgangspunkt in Kapitel \ref{section::2} ist die Liouville-von-Neumann-Gleichung für die 
Dichtematrix $\varrho(t)$ des Quantensystems. Mittels Projektionsoperatoren \cite{Gr82} 
werden die Freiheitsgrade des Vielteilchensystems unterteilt in relevante, welche die 
hydrodynamischen Eigenschaften auf makroskopischer Ebene beschreiben, und restliche 
irrelevante. Eine relevante Dichtematrix $\tilde{\varrho}(t)$ wird definiert durch Maximieren 
der Entropie unter der Nebenbedingung, dass die relevanten Variablen die exakten 
Erwartungswerte haben.

In Kapitel \ref{section::3} werden die restlichen irrelevanten Freiheitsgrade eliminiert. Als 
Ergebnis erhält man für die relevante Dichtematrix $\tilde{\varrho}(t)$ eine Mastergleichung mit 
Gedächtniseffekten und fluktuierenden Termen, welche unter dem Namen \emph{Robertson-Gleichung} 
bekannt ist \cite{Ro66}. Durch Bildung von Erwartungswerten mit den relevanten Variablen erhält 
man daraus Bewegungsgleichungen für die makroskopischen Freiheitsgrade, welche ohne irgendeine 
Näherung bereits die Form von verallgemeinerten hydrodynamischen Gleichungen haben. Wir folgen 
hier der Darstellung von Fick und Sauermann \cite{FS90}. Anschließend zeigen wir, dass die 
verallgemeinerten hydrodynamischen Gleichungen bereits in ihrer exakten Form ohne Näherung in 
die Schreibweise des GENERIC-Formalismus von Grmela und Öttinger \cite{GO97A,GO97B,Ot05} 
übergeführt werden können, einschließlich aller Nebenbedingungen für die Funktional-Ableitungen 
von Entropie, Energie, Impuls und Teilchenzahl. An dieser Stelle sind unsere Gleichungen noch 
etwas allgemeiner als jene von Grmela und Öttinger, weil unsere Gleichungen die Gedächtniseffekte 
einschließen, die letzteren jedoch nicht.

Für eine korrekte Behandlung der Fluktuationen muss die Dichtematrix einen reinen 
Zustand darstellen und die Form $\varrho(t)= \vert\Psi(t)\rangle \langle\Psi(t)\vert$ 
haben, und zwar für alle Zeiten $t$. Der Projektionsoperator-Formalismus liefert 
eine Formel für die fluktuierenden Kräfte, welche explizit von der Dichtematrix des 
Anfangszustandes $\varrho(t_0)$ zur Anfangszeit $t_0$ abhängt. Aus diesem Grunde wählen 
wir $\varrho(t_0)= \vert\Psi_0\rangle \langle\Psi_0\vert$ und erhalten somit nichttriviale 
fluktuierende Kräfte. Andererseits verwenden die konventionellen Theorien \cite{Ro66,FS90} 
für den Anfangszustand eine relevante Dichtematrix $\varrho(t_0) = \tilde{\varrho}(t_0)$. 
In Folge dessen sind die fluktuierenden Kräfte null für alle Zeiten, so dass die 
konventionellen Theorien die Fluktuationen gar nicht erst berücksichtigen.

In Kapitel \ref{section::4} werden die Näherungen und Symmetrie-Überlegungen durchgeführt, 
welche auf die hydrodynamischen Gleichungen mit Fluktuationen in ihrer bekannten Form führen. 
Zunächst werden alle Gedächtniseffekte weggelassen, und man erhält die hydrodynamischen 
Gleichungen in der Schreibweise des GENERIC-Formalismus, wie sie ursprünglichen von Grmela 
und Öttinger \cite{GO97A,GO97B,Ot05} aufgestellt wurden. Die reversiblen Terme lassen sich 
durch Poisson-Klammern der Dichten mit der Energie des Systems darstellen wie dies zuvor von 
Dzyaloshinskii und Volovik \cite{DV80} gemacht wurde. Weiterhin werden die dissipativen Terme 
mit einer 
lokalen Näherung vereinfacht, und die fluktuierenden Terme werden durch gaußische stochastische 
Kräfte modelliert. Die Stärke der Dissipationen und der Fluktuationen wird gemäß dem 
Fluktuations-Dissipations-Theorem durch eine Onsager-Matrix parametrisiert, welche sich aus 
Symmetriegründen mit nur drei Parameter darstellen läßt, der Scherviskosität $\eta$, der 
Volumenviskosität $\zeta$ und der Wärmeleitfähigkeit $\varkappa$. Zwar können die 
Transportkoeffizienten $\eta$, $\zeta$, and $\varkappa$ im Prinzip explizit und quantitativ 
berechnet werden. Wir betrachten sie jedoch als phänomenologische Parameter, die experimentell 
bestimmt werden müssen. Als Ergebnis erhalten wir schließlich die aus allgemeinen Lehrbüchern 
bekannten hydrodynamischen Gleichungen einer normalen Flüssigkeit mit Fluktuationen 
\cite{LL06,LL09}.

In Kapitel \ref{section::5} untersuchen wir die in der Thermodynamik und Hydrodynamik 
bekannte Brechung der Zeitumkehrinvarianz und den zweiten Hauptsatz der Thermodynamik. 
Zur Überraschung finden wir, dass die Summe der dissipativen und der fluktuierenden Terme 
die Zeitumkehrinvarianz nicht bricht, zumindest in der exakten Theorie von Kapitel 
\ref{section::3} bevor irgendwelche Näherungen gemacht werden. In Kapitel \ref{section::6} 
zeigen wir einen Zusammenhang der hydrodynamischen Gleichungen mit stochastischen 
Langevin-Gleichungen und einer zugehörigen Fokker-Planck-Gleichung. Wir untersuchen die 
Lösung dieser Gleichungen  im thermischen Gleichgewicht und finden eine großkanonische 
Boltzmann-Verteilung. Weiterhin finden wir, dass im thermischen Gleichgewicht die Entropie 
im Mittel konstant bleibt, wie es auch sein muss. 

In Kapitel \ref{section::7} untersuchen wir, in wie weit das Fluktuations-Theorem 
von Evans \emph{et al.}\ \cite{Ev93,Ev94,Ev02a} und die Jarzynski-Gleichung 
\cite{Ja97A,Ja97B} auf eine normale Flüssigkeit anwendbar sind. Zunächst zeigen wir, 
dass die Herleitung des Fluktuations-Theorems von Crooks \cite{Cr98,Cr99,Cr00} 
erfolgreich auf den GENERIC-Formalsimus übertragen werden kann. Wir sind jedoch mit 
dem Ergebnis nicht zufrieden, weil die Variable des Fluktuations-Theorems nicht die 
Entropie-Änderung der Flüssigkeit ist. Aus diesem Grunde leiten wir aus unserer 
Theorie ein \emph{modifiziertes} Fluktuations-Theorem und eine \emph{modifiziertes} 
Jarzynski-Gleichung für die Entropie-Änderung her mit einem Zusatzterm. Diesen 
Zusatzterm berechnen wir explizit und finden eine Ultraviolett-Divergenz für eine 
normale Flüssigkeit. Nach der Regularisierung mit einer minimalen Längenskala für 
die räumlichen Variationen der Fluktuationen wird der Zusatzterm zwar endlich, hängt 
jedoch stark von der Größe dieser minimalen Länge ab. Abschließende Bemerkungen 
folgen dann in Kapitel \ref{section::8}.

\section{Quantenstatistik für Vielteilchensysteme}
\label{section::2}
Wir gehen davon aus, dass eine Flüssigkeit oder ein Gas beschrieben wird durch eine 
mikroskopische Theorie für ein System von vielen Teilchen, die miteinander über ein 
Feld wechselwirken. Eine ziemlich allgemeines fundamentales Modell für das System geht 
aus von den Atomkernen und den Elektronen in der Materie, welche mit einander über ein 
elektromagnetisches Feld wechselwirken. Im allgemeinsten Fall sind die Teilchen Quarks 
und Leptonen, die miteinander wechselwirken über Eichfelder für die starke, schwache und 
elektromagnetische Wechselwirkung.

Die mikroskopische Theorie kann nichtrelativistisch oder relativistisch sein, klassisch 
oder quantenmechanisch. Welches mikroskopisches Modell für das Vielteilchen-System im 
Speziellen gewählt wird, ist am Ende für unsere Betrachtungen nicht so wichtig. Das 
Modell muss nur einige wenige sehr allgemeine Bedingungen erfüllen. Es müssen einige 
\emph{Erhaltungsgrößen} vorhanden sein, welche durch Integrale von \emph{lokalen Dichten} 
dargestellt werden können. Für eine gewöhnliche nichtrelativistische Flüssigkeit wären 
dies die Masse, der Impuls und die Energie, definiert durch
\begin{eqnarray}
 \hat{M}(t) &=& \int d^d r\ \hat{\rho}(\mathbf{r},t) \ ,
 \label{equation::B_010} \\
 \hat{\mathbf{P}}(t) &=& \int d^d r\ \hat{\mathbf{j}}(\mathbf{r},t) \ ,
 \label{equation::B_020} \\
 \hat{E}(t) &=& \int d^d r\ \hat{\varepsilon}(\mathbf{r},t) \ .
 \label{equation::B_030}
\end{eqnarray}
Die Dimension des Raumes bezeichnen wir mit $d$. Um die Formeln so allgemein wie 
möglich zu halten, lassen wir zunächst beliebige Werte für $d$ zu. Es muss jedoch $d>2$ 
gelten, um das System von einem \emph{hydrodynamischer Selbstmord} abzuhalten \cite{Br89}. 
Am Ende setzen wir $d=3$ für eine gewöhnliche Flüssigkeit ein. Die Funktionen in den 
Integralen sind die Massendichte $\hat{\rho}(\mathbf{r},t)$, die Impulsdichte 
$\hat{\mathbf{j}}(\mathbf{r},t)$ und die Energiedichte $\hat{\varepsilon}(\mathbf{r},t)$. 
Die Dächer über den Symbolen in den Gleichungen \eqref{equation::B_010}-\eqref{equation::B_030} 
zeigen an, dass die physikalischen Größen quantenmechanische Operatoren sind. Im Folgenden 
gehen wir immer davon aus, dass das mikroskopische Vielteilchen-System durch eine 
Quanten-Theorie beschrieben wird. Wir treffen diese Wahl, weil in diesem Fall die 
Formeln einfacher und kompakter sind. Alle nachfolgenden Betrachtungen und Berechnungen 
können alternativ auch mit einer klassischen Theorie für das mikroskopische System 
durchgeführt werden. 

Erhaltungsgrößen sind nach Definition konstant in der Zeit. Folglich werden sich die 
Dichten von diesen Erhaltungsgrößen langsam mit der Zeit verändern, 
wenn man Variationen auf großen Längenskalen betrachtet. Wir schließen daraus, dass sich 
die Dichten der Erhaltungsgrößen als relevante Variablen eignen, um die physikalischen 
Eigenschaften auf makroskopischen Skalen, also großen Längenskalen und großen Zeitskalen 
zu beschreiben. Für eine normale Flüssigkeit sind dies die Massendichte 
$\hat{\rho}(\mathbf{r},t)$, die Impulsdichte $\hat{\mathbf{j}}(\mathbf{r},t)$ und die 
Energiedichte $\hat{\varepsilon}(\mathbf{r},t)$.

Explizite Formeln für die Erhaltungsgrößen \eqref{equation::B_010}-\eqref{equation::B_030} 
erhält man aus dem Noether-Theorem. Man leitet diese aus den kontinuierlichen Symmetrien 
des physikalischen Systems her, in dem man die Lagrange-Dichte $\mathcal{L}$ betrachtet. 
Für unsere Berechnungen und Untersuchungen ist es sehr wichtig, dass die zugrunde 
liegende mikroskopische Theorie \emph{lokal in Raum und Zeit} ist. Das Bedeutet, es 
gibt eine \emph{lokale} Lagrange-Dichte $\mathcal{L}$, die über das Noether-Theorem 
lokale Ausdrücke für die Dichten der Erhaltungsgrößen $\hat{\rho}(\mathbf{r},t)$, 
$\hat{\mathbf{j}}(\mathbf{r},t)$ und $\hat{\varepsilon}(\mathbf{r},t)$ liefert. Da die 
makroskopischen relevanten Variablen als Mittelwerte oder Erwartungswerte dieser 
Dichten definiert werden, garantiert diese Forderung wohldefinierte hydrodynamische 
Variablen.

Ein Gegenbeispiel dazu ist die häufig verwendete Theorie für ein System aus vielen 
Teilchen, die miteinander über ein Zweiteilchen-Potential $V(\mathbf{r}_1-\mathbf{r}_2)$ 
wechselwirken. Der zugehörige Hamilton-Operator $\hat H$ ist räumlich nicht lokal. 
Dies führt zu nichtlokalen Ausdrücken für die Energiedichte $\hat{\varepsilon}(\mathbf{r},t)$. 
In Folge davon wird es schwierig, wohldefinierte Formeln für die zugehörigen Stromdichten 
in den Kontinuitätsgleichungen zu finden. In diesem Fall empfehlen wir, das 
mikroskopische Modell für das physikalisch System so zu erweitern, dass die 
Wechselwirkung über irgend ein lokales Feld vermittelt wird. Das Feld und die 
Wechselwirkung sollten so gewählt werden, dass das Zweiteilchen-Potential 
$V(\mathbf{r}_1-\mathbf{r}_2)$ als effektive Wechselwirkung herauskommt, wenn 
man das lokale Feld ausintegriert.

Eine weitere Erhaltungsgröße ist der Drehimpuls
\begin{equation}
 \hat{\mathbf{L}}(t) = \int d^d r\ \hat{\mathbf{l}}(\mathbf{r},t) \ ,
 \label{equation::B_040}
\end{equation}
welcher mit der Symmetrie des Systems unter Drehungen zusammenhängt. Die 
Drehimpuls-\emph{Dichte} lässt sich in der Form
\begin{equation}
 \hat{\mathbf{l}}(\mathbf{r},t) = \mathbf{r} \times \hat{\mathbf{j}}(\mathbf{r},t) 
 + \hat{\mathbf{s}}(\mathbf{r},t)
 \label{equation::B_050}
\end{equation}
schreiben, wobei der erste Term den Bahndrehimpuls und der zweite Term den Spindrehimpuls 
der Teilchen darstellt. Die Formel \eqref{equation::B_050} erhält man aus dem Noether-Theorem, 
zusammen mit einem expliziten Ausdruck für die Spindichte $\hat{\mathbf{s}}(\mathbf{r},t)$. 
Wie üblich lässt sich der Bahndrehimpuls durch die Radialkoordinate $\mathbf{r}$ und 
den linearen Impuls $\hat{\mathbf{j}}(\mathbf{r},t)$ darstellen.

Nach Belinfante \cite{Be39} und nach Martin \emph{et al.}\ \cite{MPP72} können wir die 
Impulsdichte $\hat{\mathbf{j}}(\mathbf{r},t)$ modifizieren, indem wir einen Beitrag 
hinzufügen, welcher mit der Spindichte zusammenhängt. Die Impulsdichte bleibt dabei 
die Dichte einer Erhaltungsgröße. Die zugehörige Drehimpulsdichte wird dann einfach 
durch die Formel 
$\hat{\mathbf{l}}(\mathbf{r},t) = \mathbf{r} \times \hat{\mathbf{j}}(\mathbf{r},t)$ 
gegeben und hängt ausschließlich von der modifizierten linearen Impulsdichte 
$\hat{\mathbf{j}}(\mathbf{r},t)$ ab. Ursprünglich wurde dieses Konzept von Belinfante 
\cite{Be39} für relativistische Feldtheorien mit einer Lagrange-Dichte 
$\mathcal{L}$ entwickelt, wobei die Dichten der Erhaltungsgrößen durch das Noether-Theorem 
definiert werden. Später wurde es von Martin \emph{et al.}\ \cite{MPP72} auf die 
Hydrodynamik übertragen. In Folge ist die Drehimpulsdichte $\hat{\mathbf{l}}(\mathbf{r},t)$ 
keine unabhängige Größe. Aus diesem Grunde werden wir die Drehimpulsdichte im Folgenden 
dieser Arbeit nicht weiter betrachten.

\subsection{Quantendynamik}
\label{section::2A}
Nachdem wir die relevanten Variablen für eine normale Flüssigkeit durch einige wenige 
Dichten von Erhaltungsgrößen identifiziert haben, benötigen wir eine Bewegungsgleichung 
für die zeitliche Entwicklung dieser Größen. Wenn wir eine solche Dichte allgemein 
bezeichnen durch den lokalen quantenmechanischen Operator $\hat{a}(\mathbf{r},t)$, 
so wird ihre zeitlichen Entwicklung durch die Heisenberg-Bewegungsgleichung 
\begin{equation}
 i\hbar \, \partial_t \, \hat{a}(\mathbf{r},t) = [ \hat{a}(\mathbf{r},t), \hat{H}(t) ]
 \label{equation::B_060}
\end{equation}
beschrieben, wobei der Hamilton-Operator $\hat{H}(t)=\hat{E}(t)$ durch die Energie gegeben 
ist, die in \eqref{equation::B_030} definiert wird. Der quantenphysikalischen Zustand des 
Systems wird beschrieben durch einen Zustandsvektor im Hilbertraum $\vert\Psi\rangle$, 
manchmal auch kurz Wellenfunktion genannt. Der Erwartungswert einer Dichte 
$\hat{a}(\mathbf{r},t)$ wird damit definiert über das Skalarprodukt
\begin{equation}
 \langle a(\mathbf{r}) \rangle_t = \langle\Psi\vert \hat{a}(\mathbf{r},t) \vert\Psi\rangle \ .
 \label{equation::B_070}
\end{equation}
In der Quantenstatistik befindet sich das physikalische System nicht in einem reinen Zustand 
$\vert\Psi\rangle$. Vielmehr nimmt man an, dass sich das System in bestimmten zueinander 
orthogonalen Zuständen $\vert\Psi_i\rangle$ mit Wahrscheinlichkeiten $w_i$ befindet. Der 
Erwartungswert ist dann gegeben durch
\begin{equation}
 \langle a(\mathbf{r}) \rangle_t = \sum_i w_i \, \langle\Psi_i\vert \hat{a}(\mathbf{r},t) 
\vert\Psi_i\rangle \ .
 \label{equation::B_080}
\end{equation}
Es ist zweckmäßig, die Dichtematrix
\begin{equation}
 \hat{\varrho} = \sum_i w_i \, \vert\Psi_i\rangle \, \langle\Psi_i\vert
 \label{equation::B_090}
\end{equation}
einzuführen. Damit bekommt der Erwartungswert die Form
\begin{equation}
 \langle a(\mathbf{r}) \rangle_t = \mathrm{Sp}\{ \hat{\varrho} \, \hat{a}(\mathbf{r},t) \} \ .
 \label{equation::B_100}
\end{equation}
Unglücklicherweise werden die Massendichte $\hat{\rho}$ in \eqref{equation::B_010} und 
die Dichtematrix $\hat{\varrho}$ in \eqref{equation::B_090} und \eqref{equation::B_100} 
mit nahezu demselben Buchstaben definiert. Das kann zu Verwechslungen führen. Man muss 
daher entweder genau hinsehen oder aus dem Kontext schließen, was gerade 
gemeint ist.

Die Gleichungen \eqref{equation::B_060}-\eqref{equation::B_100} beschreiben die Quantendynamik 
im \emph{Heisenberg-Bild}. Hier hängen die Operatoren der beobachtbaren Größen 
$\hat{a}(\mathbf{r},t)$ von der Zeit ab, der Zustandsvektor $\vert\Psi\rangle$ oder die 
Dichtematrix $\hat{\varrho}$ sind jedoch konstant. Für unsere Zwecke besser geeignet ist das 
\emph{Schrödinger-Bild}. Man erhält es durch die Transformationen 
\begin{eqnarray}
 \hat{a}(\mathbf{r}) &=& \exp(-i\hat{H}t/\hbar) \, \hat{a}(\mathbf{r},t) \, \exp(i\hat{H}t/\hbar) \ ,
 \label{equation::B_110} \\
 \vert\Psi(t)\rangle &=& \exp(-i\hat{H}t/\hbar) \, \vert\Psi\rangle \ ,
 \label{equation::B_120} \\
 \hat{\varrho}(t) &=& \exp(-i\hat{H}t/\hbar) \, \hat{\varrho} \, \exp(i\hat{H}t/\hbar) \ .
 \label{equation::B_130}
\end{eqnarray}
Links vom Gleichheitszeichen stehen die Größen im Schrödinger-Bild, rechts im Heisenberg-Bild. 
Für den Zustandsvektor gilt hier die Schrödinger-Gleichung
\begin{equation}
 i\hbar \, \partial_t \vert\Psi(t)\rangle = \hat{H} \vert\Psi(t)\rangle \ ,
 \label{equation::B_140}
\end{equation}
und für die Dichtematrix die Liouville-von-Neumann-Gleichung
\begin{equation}
 i\hbar \, \partial_t \hat{\varrho}(t) = [ \hat{H}, \hat{\varrho}(t) ] \ .
 \label{equation::B_150}
\end{equation}
Für die Erwartungswerte schreibt man im Schrödinger-Bild entsprechend
\begin{equation}
 \langle a(\mathbf{r}) \rangle_t = \langle\Psi(t)\vert \hat{a}(\mathbf{r}) \vert\Psi(t)\rangle
 \label{equation::B_160}
\end{equation}
für einen reinen Quantenzustand und
\begin{equation}
 \langle a(\mathbf{r}) \rangle_t = \mathrm{Sp}\{ \hat{\varrho}(t) \, \hat{a}(\mathbf{r}) \}
 \label{equation::B_170}
\end{equation}
in der Quantenstatistik. Im Folgenden verwenden wir immer das Schrödinger-Bild. Zur Vereinfachung 
der Schreibweise lassen wir ab jetzt das Dach über einer Variablen zur Kennzeichnung eines 
Operators weg.

\subsection{Quantenstatistik}
\label{section::2B}
In der Quantenstatistik definiert man die Entropie durch
\begin{equation}
 S(t) = - k_B\, \mathrm{Sp}\{ \varrho(t) \ln\varrho(t) \} \ .
 \label{equation::B_180}
\end{equation}
Hierbei ist $k_B$ die Boltzmann-Konstante, welche die Einheit der Entropie festlegt. 
Für einen reinen Quantenzustand $\varrho(t)=\vert\Psi(t)\rangle \langle\Psi(t)\vert$ gilt 
bekanntlich
\begin{equation}
 S(t) = - k_B\, \langle\Psi(t)\vert \ln\varrho(t) \vert\Psi(t)\rangle = - k_B\, \ln 1 = 0 \ .
 \label{equation::B_190}
\end{equation}
Die Entropie ist also null und konstant für alle Zeiten $t$. Wir greifen eine bestimmte Menge 
von \emph{relevanten} Variablen $a_i(\mathbf{r})$ heraus, welche die wesentlichen Eigenschaften 
des physikalischen Systems beschreiben. Für eine normale Flüssigkeit sind das die Dichten von Masse, 
Impuls und Energie. Wir nehmen an, dass die exakte Lösung der Schrödinger-Gleichung $\vert\Psi(t)\rangle$ 
und die exakte Lösung der Liouville-von-Neumann-Gleichung $\varrho(t)$ bekannt sind. Dann 
ist auch die zeitliche Entwicklung der Erwartungswerte der relevanten Variablen
\begin{equation}
 x_i(\mathbf{r},t) = \langle a_i(\mathbf{r}) \rangle_t = \mathrm{Sp}\{ \varrho(t) \, a_i(\mathbf{r}) \}
 \label{equation::B_200}
\end{equation}
exakt bekannt.

In der Quantenstatistik betrachtet man nun einen gemischten Zustand, der durch eine Dichtematrix 
$\tilde{\varrho}(t)$ beschrieben wird, welche die Entropie \eqref{equation::B_180} maximiert unter 
der Nebenbedingung dass die Erwartungswerte der relevanten Variablen die exakten Werte 
\eqref{equation::B_200} annehmen. Wir bekommen somit eine Maximierungsaufgabe mit Nebenbedingungen
\begin{eqnarray}
 S(t) = - k_B\, \mathrm{Sp}\{ \tilde{\varrho}(t) \ln\tilde{\varrho}(t) \} &=& \mathrm{maximum} \ ,
 \label{equation::B_210} \\
 \mathrm{Sp}\{ \tilde{\varrho}(t) \, a_i(\mathbf{r}) \} &=& x_i(\mathbf{r},t) \ ,
 \label{equation::B_220} \\
 \mathrm{Sp}\{ \tilde{\varrho}(t) \} &=& 1 \ .
 \label{equation::B_230}
\end{eqnarray}
Die letzte Nebenbedingung garantiert die Normierung der Dichtematrix. Zur Berechnung der 
Lösung definieren wir das Funktional
\begin{eqnarray}
 \Phi[\tilde{\varrho}(t)] &=& S(t) - \sum_i \int d^d r\ \lambda_i(\mathbf{r},t) \, x_i(\mathbf{r},t)
 - \mu(t) \, \langle 1 \rangle_t \nonumber\\
 &=& - k_B\, \mathrm{Sp}\{ \tilde{\varrho}(t) \ln\tilde{\varrho}(t) \} \nonumber\\
 &&- \sum_i \int d^d r\ \lambda_i(\mathbf{r},t) \, \mathrm{Sp}\{ \tilde{\varrho}(t) \, a_i(\mathbf{r}) \} \nonumber\\
 &&- \mu(t) \ \mathrm{Sp}\{ \tilde{\varrho}(t) \}
 \label{equation::B_240}
\end{eqnarray}
mit den Langrange-Parametern $\lambda_i(\mathbf{r},t)$ und $\mu(t)$. Als notwendige Bedingung 
für das Maximum muss die Variation dieses Funktional null ergeben, also
\begin{eqnarray}
 \hspace{-5mm}\delta \Phi[\tilde{\varrho}(t)] &=& - \mathrm{Sp}\biggl\{ \delta \tilde{\varrho}(t) \, 
 \biggl[ k_B ( \ln\tilde{\varrho}(t) + 1 ) \nonumber\\
 &&+ \int d^d r\ \lambda_i(\mathbf{r},t) \, a_i(\mathbf{r}) + \mu(t) \biggr] \biggr\} = 0 \ . \quad
 \label{equation::B_250}
\end{eqnarray}
Die Lösung dieser Gleichung ist die \emph{relevante} Dichtematrix
\begin{equation}
 \tilde{\varrho}(t) = [Z(t)]^{-1} \, \exp \left( - k_B^{-1} \sum_i \int d^d r\ \lambda_i(\mathbf{r},t) 
 \, a_i(\mathbf{r}) \right) \ .
 \label{equation::B_260}
\end{equation}
Anstelle des Lagrange-Parameters $\mu(t)$ verwenden wir den Normierungsfaktor 
$Z(t) = \exp( 1 + k_B^{-1} \mu(t) )$. Die Lagrange-Parameter $\lambda_i(\mathbf{r},t)$ und der 
Normierungsfaktor $Z(t)$ werden durch Einsetzen von \eqref{equation::B_260} in die Nebenbedingungen 
\eqref{equation::B_220} und \eqref{equation::B_230} bestimmt.

Die einzigen Parameter, welche in die Maximierungsaufgabe \eqref{equation::B_210}-\eqref{equation::B_230} 
eingehen, sind die Erwartungswerte $x_i(\mathbf{r},t)$. Folglich definiert die Formel 
\eqref{equation::B_260} die relevante Dichtematrix in Abhängigkeit der Erwartungswerte 
$x_i(\mathbf{r},t)$. Diese Abhängigkeit ist jedoch implizit über die Lagrange-Parameter 
$\lambda_i(\mathbf{r},t)$ und den Normierungsfaktor $Z(t)$.

\subsection{Thermisches Gleichgewicht}
\label{section::2C}
Im thermischen Gleichgewicht ist eine Flüssigkeit räumlich und zeitlich homogen. Folglich sind alle 
Erwartungswerte der Dichten $x_i(\mathbf{r},t)=x_i$, die Lagrange-Parameter $\lambda_i(\mathbf{r},t)=\lambda_i$ 
und der Normierungsfaktor $Z(t)=Z$ räumlich und zeitlich konstant. Die relevante Dichtematrix 
\eqref{equation::B_260} vereinfacht sich somit auf
\begin{equation}
 \tilde{\varrho}_\mathrm{eq} = Z^{-1} \, \exp \left( - k_B^{-1} \sum_i \lambda_i \int d^d r\ 
 a_i(\mathbf{r}) \right) \ .
 \label{equation::B_270}
\end{equation}
Setzen wir für $a_i(\mathbf{r})$ die Massendichte $\rho(\mathbf{r})$, die Impulsdichte $\mathbf{j}(\mathbf{r})$ 
und die Energiedichte $\varepsilon(\mathbf{r})$ ein und verwenden wir die integralen Erhaltungsgrößen
\eqref{equation::B_010}-\eqref{equation::B_030}, so folgt
\begin{equation}
 \tilde{\varrho}_\mathrm{eq} = Z^{-1} \, \exp \left( - k_B^{-1} [ \lambda_\rho M + \boldsymbol{\lambda}_j \cdot 
 \mathbf{P} + \lambda_\varepsilon E ] \right) \ .
 \label{equation::B_280}
\end{equation}
Wir führen die neuen Lagrange-Parameter Temperatur $T$, chemisches Potential $\mu$ und Geschwindigkeit 
$\mathbf{v}$ ein über die Beziehungen
\begin{equation}
 \lambda_\rho = - \mu/mT \ , \quad \boldsymbol{\lambda}_j = - \mathbf{v}/T \ , \quad 
 \lambda_\varepsilon = 1/T
 \label{equation::B_290}
\end{equation}
und definieren den Teilchenzahloperator $N=M/m$, wobei $m$ die Masse eines einzelnen Teilchens 
ist. Dann finden wir im Ergebnis die großkanonische Boltzmann-Verteilung für eine mit konstanter 
Geschwindigkeit $\mathbf{v}$ bewegte Flüssigkeit
\begin{equation}
 \tilde{\varrho}_\mathrm{eq} = Z^{-1} \, \exp \left( - (k_BT)^{-1} [ H - \mathbf{v} \cdot \mathbf{P} 
 - \mu N ] \right) \ .
 \label{equation::B_300}
\end{equation}
Wir schließen daraus, dass die in \eqref{equation::B_260} definierte relevante Dichtematrix 
kompatibel mit der Quantenstatistik des thermischen Gleichgewichts ist.

Die Energie $H$, der Impuls $\mathbf{P}$ und die Teilchenzahl $N$ sind Erhaltungsgrößen einer 
normalen Flüssigkeit. Sie vertauschen daher mit dem Hamilton-Operator $H$. Folglich vertauscht 
die Dichtematrix des thermischen Gleichgewichts \eqref{equation::B_300} ebenfalls mit dem 
Hamilton-Operator, also $[ H, \tilde{\varrho}_\mathrm{eq} ] = 0$. Weil die Dichtematrix 
$\tilde{\varrho}_\mathrm{eq}$ nicht von der Zeit $t$ abhängt, erfüllt sie offenbar die 
Liouville-von-Neumann-Gleichung \eqref{equation::B_150} und ist somit eine exakte Lösung.

\subsection{Thermodynamische Potentiale im Nichtgleichgewicht}
\label{section::2D}
Die relevante Dichtematrix \eqref{equation::B_260} hat die Struktur einer verallgemeinerten 
Boltzmann-Verteilung, wobei die Lagrange-Parameter $\lambda_i(\mathbf{r},t)$ vom Ort 
$\mathbf{r}$ und von der Zeit $t$ abhängen. Wenn immer diese Lagrange-Parameter nur 
\emph{langsam} mit dem Ort und der Zeit variieren, dann beschreibt die relevante 
Dichtematrix \eqref{equation::B_260} ein \emph{lokales thermisches Gleichgewicht}.
Auf diese Weise wird die grundlegende Annahme der Hydrodynamik einer normalen 
Flüssigkeit vorweg genommen. Das System ist \emph{global im Nichtgleichgewicht} aber 
\emph{lokal im Gleichgewicht}. Nichtsdestotrotz, wenn immer die Lagrange-Parameter 
$\lambda_i(\mathbf{r},t)$ nicht konstant sind sondern von Ort und Zeit abhängen, ist 
der Zustand grundsätzlich ein Nichtgleichgewicht. 

Somit kommen wir zu dem Schluss: Die relevante Dichtematrix \eqref{equation::B_260} 
eignet sich für die Definition von thermodynamischen Potentialen des Nichtgleichgewichts. 
Aus der Normierungsbedingung \eqref{equation::B_230} erhalten wir die Zustandssumme
\begin{equation}
 Z(t) = \mathrm{Sp} \left\{ \exp \left( - k_B^{-1} \sum_i \int d^d r
 \ \lambda_i(\mathbf{r},t)  \, a_i(\mathbf{r}) \right) \right\} \ .
 \label{equation::B_310}
\end{equation}
Sie ist offensichtlich ein Funktional $Z(t)=Z[\lambda(t)]$ der Lagrange-Parameter 
$\lambda_i(\mathbf{r},t)$. Die Abhängigkeit von der Zeit $t$ ist hier nur implizit und 
spielt daher eine untergeordnete Rolle. Wie in der Thermodynamik üblich definieren wir 
über den Logarithmus das thermodynamische Potential
\begin{equation}
 F[\lambda(t)] = - k_B \ln Z[\lambda(t)] \ .
 \label{equation::B_320}
\end{equation}
Bis auf einen Faktor Temperatur $T$ ist das die Verallgemeinerung des großkanonischen 
thermodynamischen Potentials auf das Nichtgleichgewicht. Wir bilden die Variation
\begin{eqnarray}
 \delta F[\lambda(t)] &=& - k_B ( Z[\lambda(t)] ) ^{-1} \delta Z[\lambda(t)] \nonumber \\
 &=& \mathrm{Sp} \left\{ \tilde{\varrho}(t) \left( \sum_i \int d^d r
 \ \delta \lambda_i(\mathbf{r},t)  \, a_i(\mathbf{r}) \right) \right\} \nonumber \\
 &=& \sum_i \int d^d r\ \delta \lambda_i(\mathbf{r},t) \, \mathrm{Sp} \left\{ \tilde{\varrho}(t) 
 \, a_i(\mathbf{r}) \right\} \nonumber \\
 &=& \sum_i \int d^d r\ \delta \lambda_i(\mathbf{r},t) \, x_i(\mathbf{r},t)
 \label{equation::B_330}
\end{eqnarray}
und finden die Erwartungswerte der relevanten Variablen $x_i(\mathbf{r},t)$ als 
Funktional-Ableitung
\begin{equation}
 \frac{ \delta F[\lambda(t)] }{ \delta \lambda_i(\mathbf{r},t) } = x_i(\mathbf{r},t) \ .
 \label{equation::B_340}
\end{equation}
Man beachte, dass bei der Definition der Funktional-Ableitung in \eqref{equation::B_330} nur 
über den Index $i$ summiert und die Ortsvariable $\mathbf{r}$ integriert wird. Die Zeit 
$t$ spielt eine untergeordnete Rolle als impliziter konstanter Parameter.

Als nächstes setzen wir die relevante Dichtematrix \eqref{equation::B_260} in die Formel 
für die Entropie \eqref{equation::B_180} ein. Wegen
\begin{equation}
 \ln \tilde{\varrho}(t) = k_B^{-1} \left( F[\lambda(t)] - \sum_i \int d^d r\ \lambda_i(\mathbf{r},t) 
 \, a_i(\mathbf{r}) \right)
 \label{equation::B_350}
\end{equation}
folgt
\begin{eqnarray}
 S[x(t)] &=& - F[\lambda(t)] + \sum_i \int d^d r\ \lambda_i(\mathbf{r},t) 
 \ \mathrm{Sp}\{ \tilde{\varrho}(t) \, a_i(\mathbf{r}) \} \nonumber\\
 &=& - F[\lambda(t)] + \sum_i \int d^d r\ \lambda_i(\mathbf{r},t) 
 \, x_i(\mathbf{r},t) \ .
 \label{equation::B_360}
\end{eqnarray}
Dies ist die klassische Formel einer Legendre-Transformation. Folglich ist die Entropie 
$S(t)=S[x(t)]$ ein Funktional der Erwartungswerte $x_i(\mathbf{r},t)$. Die Zeitabhängigkeit 
ist wiederum implizit. Aus der Variation
\begin{equation}
 \delta S[x(t)] = \sum_i \int d^d r\ \lambda_i(\mathbf{r},t) \, \delta x_i(\mathbf{r},t)
 \label{equation::B_370}
\end{equation}
erhalten wir die Lagrange-Parameter $\lambda_i(\mathbf{r},t)$ als Funktional-Ableitung
\begin{equation}
 \frac{ \delta S[x(t)] }{ \delta x_i(\mathbf{r},t) } = \lambda_i(\mathbf{r},t) \ .
 \label{equation::B_380}
\end{equation}
Aus den Überlegungen schließen wir, dass die Erwartungswerte $x_i(\mathbf{r},t)$ und die 
Lagrange-Parameter $\lambda_i(\mathbf{r},t)$ im Sinne einer Legendre-Transformation zueinander 
konjugierte Variablen sind.

Ein weiteres Funktional, das wir im Folgenden benötigen, ist die Energie des Systems
\begin{equation}
 E(t) = E[x(t)] = \mathrm{Sp}\{ \tilde{\varrho}(t) H \} = \langle H \rangle_t \ .
 \label{equation::B_390}
\end{equation}
Sie setzt sich zusammen aus der inneren Energie und der kinetischen Energie des Systems. 
Wegen der Definition der relevanten Dichtematrix \eqref{equation::B_260} ist die Energie 
$E(t)=E[\lambda(t)]$ eigentlich ein Funktional der Lagrange-Parameter $\lambda_i(\mathbf{r},t)$. 
Die Legendre-Transformation liefert jedoch eine umkehrbare Abbildung zwischen den Variablen 
$\lambda_i(\mathbf{r},t)$ und $x_i(\mathbf{r},t)$. Daher ist es möglich, die Energie 
$E(t)=E[x(t)]$ alternativ als Funktional der Erwartungswerte $x_i(\mathbf{r},t)$ 
darzustellen. Letzteres Funktional werden wir später verwenden.

\subsection{Projektionsoperatoren}
\label{section::2E}
Für einen reinen Quantenzustand ist die Dichtematrix gegeben durch 
$\varrho(t)=\vert\Psi(t)\rangle \langle\Psi(t)\vert$ wobei $\vert\Psi(t)\rangle$ eine Lösung 
der Schrödinger-Gleichung \eqref{equation::B_140} ist. Folglich ist $\varrho(t)$ eine Lösung der 
Liouville-von-Neumann-Gleichung \eqref{equation::B_150}. Wenn die exakte Lösung bekannt ist,
können wir sagen, die Dichtematrix $\varrho(t)$ ist \emph{exakt}.

Demgegenüber ist die relevante Dichtematrix $\tilde{\varrho}(t)$ definiert in 
\eqref{equation::B_260} eine \emph{Näherung}. Sie ist jedoch exakt im Unterraum der 
relevanten Variablen $a_i(\mathbf{r})$ und der $1$ in dem Sinne dass die Erwartungswerte 
$x_i(\mathbf{r},t)= \mathrm{Sp}\{ \tilde{\varrho}(t) a_i(\mathbf{r}) \}$ und 
$\langle 1 \rangle_t = \mathrm{Sp}\{ \tilde{\varrho}(t) \}$ die exakten Werte haben, denn es gilt
\begin{eqnarray}
 \mathrm{Sp}\{ \tilde{\varrho}(t) a_i(\mathbf{r}) \} &=& x_i(\mathbf{r},t) 
 \ =\ \mathrm{Sp}\{ \varrho(t) a_i(\mathbf{r}) \} \ , \qquad
 \label{equation::B_400} \\
 \mathrm{Sp}\{ \tilde{\varrho}(t) \} &=& 1 \ =\ \mathrm{Sp}\{ \varrho(t) \} \ .
 \label{equation::B_410}
\end{eqnarray}
Der Übergang von der exakten Dichtematrix $\varrho(t)$ zur relevanten Dichtematrix 
$\tilde{\varrho}(t)$ stellt eine Abbildung dar, welche sich schreiben lässt als
\begin{equation}
 \tilde{\varrho}(t) = f[ \varrho(t) ] \ .
 \label{equation::B_420}
\end{equation}
Man kann sich leicht davon überzeugen, dass die Abbildung eine Projektion ist. Weil 
das Bild, die relevante Dichtematrix $\tilde{\varrho}(t)$, nur von den Erwartungswerten 
$x_i(\mathbf{r},t)$ abhängt. Bei zweimaliger Anwendung der Abbildung kommt also wieder 
die relevante Dichtematrix heraus. Es gilt also
\begin{equation}
 f[ f[ \varrho(t) ] ] = f[ \varrho(t) ] = \tilde{\varrho}(t) \ .
 \label{equation::B_430}
\end{equation}
Die Abbildung \eqref{equation::B_420} ist zunächst nichtlinear. Durch infinitesimale 
Variation lässt sich daraus ein linearer Projektionsoperator $\mathsf{P}[x(t)]$ ableiten. Die 
Abbildung \eqref{equation::B_420} und der zugehörige lineare Projektionsoperator wurde 
von Robertson \cite{Ro66} verwendet, um eine Mastergleichung für die relevante 
Dichtematrix $\tilde{\varrho}(t)$ und eine Bewegungsgleichung für die Erwartungswerte 
$x_i(\mathbf{r},t)$ herzuleiten. Eine detaillierte Beschreibung dieser Herleitung 
findet man in den Kapiteln 17 und 18 des Buches von Fick und Sauermann \cite{FS90}.

Wir wollen hier einen etwas anderen Projektionsoperator verwenden, der auf Grabert \cite{Gr82} 
zurückgeht. Dieser wirkt nicht auf die Dichtematrix sondern auf die relevanten Variablen 
und ist für eine beliebige Variable $Y$ definiert durch
\begin{eqnarray}
 \mathsf{P}[x(t)] \, Y &=& \left( 1 + \sum_i \int d^d r \, [ a_i(\mathbf{r}) - x_i(\mathbf{r},t) ]
 \, \frac{ \delta }{ \delta x_i(\mathbf{r},t) } \right) \nonumber \\
 &&\times \, \mathrm{Sp}\{ \tilde{\varrho}(t) \, Y \} \ .
 \label{equation::B_440}
\end{eqnarray}
Man kann diesen Projektionsoperator betrachten als eine Taylorreihen-Entwicklung nach 
Potenzen in den Fluktuationen der relevanten Variablen $[ a_i(\mathbf{r}) - x_i(\mathbf{r},t) ]$ 
bis zur linearen Ordnung. Wir stellen fest, dass wegen \eqref{equation::B_400} und 
\eqref{equation::B_410} der Erwartungswert dieser Fluktuationen sowohl mit der exakten 
Dichtematrix $\varrho(t)$ als auch mit der relevanten Dichtematrix $\tilde{\varrho}(t)$ null 
ergibt gemäß
\begin{eqnarray}
 &&\mathrm{Sp}\{ \varrho(t) \, [ a_i(\mathbf{r}) - x_i(\mathbf{r},t) ] \} = \nonumber\\
 &&= \mathrm{Sp}\{ \tilde{\varrho}(t) \, [ a_i(\mathbf{r}) - x_i(\mathbf{r},t) ] \} = 0 \ .
 \label{equation::B_450}
\end{eqnarray}
Wenn wir also den Erwartungswert von der Gleichung \eqref{equation::B_440} mit der exakten 
Dichtematrix bilden, dann vereinfacht sich diese auf
\begin{equation}
 \mathrm{Sp}\{ \varrho(t) \, \mathsf{P}[x(t)] \, Y \} = \mathrm{Sp}\{ \tilde{\varrho}(t) \, Y \} 
 = \mathrm{Sp}\{ f[\varrho(t)] \, Y \} \ .
 \label{equation::B_460}
\end{equation}
Wenden wir hier zusätzlich \eqref{equation::B_420} an, um das zweite Gleichheitszeichen 
und den Term auf der rechten Seite zu erklären, so kommen wir zu der Erkenntnis, dass bei 
der Berechnung des Erwartungswertes einer beliebigen Variablen $Y$ der Projektionsoperator 
von Grabert \eqref{equation::B_440} äquivalent ist zu der nichtlinearen Projektion der 
Dichtematrix \eqref{equation::B_420}.

Der Projektionsoperator \eqref{equation::B_440} ist so definiert, dass er nach rechts auf eine 
physikalische Variable $Y$ wirkt. Es ist auch möglich eine andere Variante des 
Projektionsoperators zu definieren, der nach links auf eine Dichtematrix wirkt. Diese Variante 
ist bekannt unter dem Namen Kawasaki-Gunton-Projektionsoperator \cite{KG73} und führt 
zu denselben Ergebnissen.

Die rechte Seite von \eqref{equation::B_440} ist eine Linearkombination der relevanten 
Variablen $a_i(\mathbf{r})$ und der $1$. Folglich projiziert der Operator $\mathsf{P}[x(t)]$ 
eine beliebige Variable $Y$ in der Unterraum der relevanten Variablen $a_i(\mathbf{r})$ und 
der $1$. Speziell gilt
\begin{equation}
 \mathsf{P}[x(t)] \, a_i(\mathbf{r}) = a_i(\mathbf{r}) \ , \qquad
 \mathsf{P}[x(t)] \, 1 = 1 \ .
 \label{equation::B_470}
\end{equation}
Man kann dies explizit nachprüfen durch Einsetzen in die Formel \eqref{equation::B_440}. 
Der Projektionsoperator $\mathsf{P}[x(t)]$ ist weiterhin linear. Ist $Y^\prime$ eine 
beliebige Linearkombination der relevanten Variablen $a_i(\mathbf{r})$ und der $1$, so folgt
$\mathsf{P}[x(t)] \, Y^\prime = Y^\prime$. Wählen wir speziell 
$Y^\prime = \mathsf{P}[x(t^\prime)] \, Y$, so finden wir
\begin{equation}
 \mathsf{P}[x(t)] \, \mathsf{P}[x(t^\prime)] \, Y = \mathsf{P}[x(t^\prime)] \, Y \ .
 \label{equation::B_480}
\end{equation}
Da die Variable $Y$ beliebig ist, können wir diese auch weglassen und finden formal einfach 
\begin{equation}
 \mathsf{P}[x(t)] \, \mathsf{P}[x(t^\prime)] = \mathsf{P}[x(t^\prime)] \ .
 \label{equation::B_490}
\end{equation}
Diese Gleichung ist eine Verallgemeinerung der Eigenschaft eines Projektionsoperators, dass 
zweimaliges Anwenden hintereinander nichts Neues bewirkt. Die Verallgemeinerung besteht darin, 
dass die Erwartungswerte $x_i(\mathbf{r},t)$ und $x_i(\mathbf{r},t^\prime)$ zu unterschiedlichen 
Zeiten $t$ und $t^\prime$ genommen werden dürfen.

Während der Operator $\mathsf{P}[x(t)]$ die relevanten Variablen heraus projiziert, ist es 
zweckmäßig, den orthogonalen Operator
\begin{equation}
 \mathsf{Q}[x(t)] = 1 - \mathsf{P}[x(t)]
 \label{equation::B_500}
\end{equation}
zu definieren, welcher alle übrigen nicht relevanten Variablen heraus projiziert. Durch 
zweimaliges Anwenden dieses Projektionsoperators zu unterschiedlichen Zeiten $t$ und $t^\prime$ 
und explizites Nachrechnen unter Verwendung von \eqref{equation::B_500} und 
\eqref{equation::B_490} finden wir 
\begin{equation}
 \mathsf{Q}[x(t)] \, \mathsf{Q}[x(t^\prime)] = \mathsf{Q}[x(t)] \ .
 \label{equation::B_510}
\end{equation}
Die Gleichungen \eqref{equation::B_490} und \eqref{equation::B_510} sind ähnlich zueinander. 
Man beachte jedoch auf der rechten Seite den Unterschied in der Abhängigkeit von den 
Zeiten $t^\prime$ und $t$.

Da der Projektionsoperator $\mathsf{P}[x(t)]$ über die Erwartungswerte $x_i(\mathbf{r},t)$ 
implizit von der Zeit $t$ abhängt, kann man erwarten, dass bei der Herleitung der 
Mastergleichung die Zeitableitungen $\partial_t \mathsf{P}[x(t)]$ des Projektionsoperators 
auftreten. Man kann jedoch zeigen dass solche Terme in der Mastergleichung null sind und 
folglich herausfallen. Dazu berechnen wir den Erwartungswert einer beliebigen Variablen 
$Y$ mit der exakten Dichtematrix $\varrho(t)$ unter Anwendung des zeitlich abgeleiteten 
Projektionsoperators $\partial_t \mathsf{P}[x(t)]$. Wir finden
\begin{widetext}
\begin{eqnarray}
 \mathrm{Sp}\{ \varrho(t) \, \partial_t \mathsf{P}[x(t)] \, Y \} &=& \mathrm{Sp}\left\{ \varrho(t)
 \, \partial_t \left( 1 + \sum_i \int d^d r \, [ a_i(\mathbf{r}) - x_i(\mathbf{r},t) ]
 \, \frac{ \delta }{ \delta x_i(\mathbf{r},t) } \right)
 \, \mathrm{Sp}\{ \tilde{\varrho}(t) \, Y \} \right\} \nonumber \\
 &=& \partial_t \, \mathrm{Sp}\{ \tilde{\varrho}(t) \, Y \} - \sum_i \int d^d r \, 
 ( \partial_t x_i(\mathbf{r},t) ) \, \frac{ \delta }{ \delta x_i(\mathbf{r},t) }
 \, \mathrm{Sp}\{ \tilde{\varrho}(t) \, Y \} \nonumber \\
 &=& \partial_t \, \mathrm{Sp}\{ \tilde{\varrho}(t) \, Y \} 
 - \partial_t \, \mathrm{Sp}\{ \tilde{\varrho}(t) \, Y \} = 0 \ .
 \label{equation::B_520}
\end{eqnarray}
\end{widetext}
Das vorletzte Gleichheitszeichen folgt aus der Kettenregel der Differentialrechnung, weil 
die relevante Dichtematrix $\tilde{\varrho}(t)$ implizit über die Erwartungswerte 
$x_i(\mathbf{r},t)$ von der Zeit $t$ abhängt. Die Gleichung \eqref{equation::B_520} wurde 
für eine beliebige Variable $Y$ hergeleitet. Wir dürfen daher $Y$ und die Spur weglassen, 
und es gilt ebenso
\begin{equation}
 \varrho(t) \, \partial_t \mathsf{P}[x(t)] = 0 \ .
 \label{equation::B_530}
\end{equation}
Diese Gleichung werden wir später verwenden um zu zeigen, dass die Zeitableitung des 
Projektionsoperators $\partial_t \mathsf{P}[x(t)]$ aus der Mastergleichung herausfällt.

\section{Mastergleichung}
\label{section::3}
Unser Ziel ist die Herleitung von hydrodynamischen Gleichungen für die Erwartungswerte 
$x_i(\mathbf{r},t)$. Mit Hilfe der Definition \eqref{equation::B_200} und der 
Liouville-von-Neumann-Gleichung \eqref{equation::B_150} finden wir
\begin{eqnarray}
 \partial_t x_i(\mathbf{r},t) &=& \mathrm{Sp}\{ \partial_t\varrho(t) \, a_i(\mathbf{r}) \}
 \nonumber \\
 &=& (i\hbar)^{-1} \, \mathrm{Sp}\{ [ H, \varrho(t) ] \, a_i(\mathbf{r}) \} \nonumber \\
 &=& (i\hbar)^{-1} \, \mathrm{Sp}\{ H \, \varrho(t) \, a_i(\mathbf{r}) 
 - \varrho(t) \, H \, a_i(\mathbf{r}) \} \nonumber \\
 &=& (i/\hbar) \ \mathrm{Sp}\{ \varrho(t) \, H \, a_i(\mathbf{r}) 
 - \varrho(t) \, a_i(\mathbf{r}) \, H \} \nonumber \\
 &=& (i/\hbar) \ \mathrm{Sp}\{ \varrho(t) \, [ H, a_i(\mathbf{r}) ] \} \nonumber \\
 &=& i \ \mathrm{Sp}\{ \varrho(t) \, \mathsf{L} \, a_i(\mathbf{r}) \} \ .
 \label{equation::C_010}
\end{eqnarray}
Zur Vereinfachung der Schreibweise definieren wir den Liouville-Operator $\mathsf{L}$, 
der auf eine beliebige Variable $Y$ wirkt, durch
\begin{equation}
 \mathsf{L} \, Y = \hbar^{-1} \, [ H, Y ] \ .
 \label{equation::C_020}
\end{equation}
Vergleichen wir miteinander die erste und letzte Zeile von \eqref{equation::C_010}, 
so finden wir, dass sich die Liouville-von-Neumann-Gleichung formal schreiben lässt als 
\begin{equation}
 \partial_t\varrho(t) = \varrho(t) \, i \, \mathsf{L} \ .
 \label{equation::C_030}
\end{equation}
Die Bewegungsgleichung für die Erwartungswerte \eqref{equation::C_010} wollen wir auf 
eine \emph{geschlossene} Form bringen, so dass die rechte Seite möglichst ein Funktional 
von den $x_i(\mathbf{r},t)$ ist. In der letzten Zeile von \eqref{equation::C_010} ist 
das leider nicht erkennbar. Wir wissen jedoch, dass die relevante Dichtematrix 
$\tilde{\varrho}(t)$ über die Lagrange-Parameter $\lambda_i(\mathbf{r},t)$ ein Funktional 
von den Erwartungswerten $x_i(\mathbf{r},t)$ ist. Schreiben wir die Erwartungswerte 
in der Form mit der relevanten Dichtematrix, wie auf der linken Seite von 
\eqref{equation::B_400} dargestellt, so kommen wir dem Ziel schon näher mit
\begin{equation}
 \partial_t x_i(\mathbf{r},t) = \mathrm{Sp}\{ \partial_t\tilde{\varrho}(t) 
 \, a_i(\mathbf{r}) \} \ .
 \label{equation::C_040}
\end{equation}
Wir benötigen dazu eine Bewegungsgleichung für die relevante Dichtematrix $\tilde{\varrho}(t)$, 
welche der Liouville-von-Neumann-Gleichung äquivalent ist. Eine solche Gleichung heißt 
\emph{Mastergleichung}. Wir wollen sie im Folgenden herleiten.

\subsection{Mastergleichung für die relevante Dichtematrix}
\label{section::3A}
Mit den Projektionsoperatoren 
\begin{equation}
 \mathsf{P}(t)=\mathsf{P}[x(t)] \ , \quad \mathsf{Q}(t)=\mathsf{Q}[x(t)]
 \label{equation::C_050}
\end{equation}
zerlegen wir die exakte Dichtematrix $\varrho(t)$ in einen relevanten Anteil $\tilde{\varrho}(t)$ 
und einen Restanteil $\varrho^\prime(t)$ gemäß
\begin{eqnarray}
 \varrho(t) &=& \varrho(t) \, [ \mathsf{P}(t) + \mathsf{Q}(t) ] \nonumber \\
 &=& \varrho(t) \, \mathsf{P}(t) + \varrho(t) \, \mathsf{Q}(t) \nonumber \\
 &=& \tilde{\varrho}(t) + \varrho^\prime(t) \ ,
 \label{equation::C_060}
\end{eqnarray}
so dass
\begin{equation}
 \tilde{\varrho}(t) = \varrho(t) \, \mathsf{P}(t) \ , \quad 
 \varrho^\prime(t) = \varrho(t) \, \mathsf{Q}(t) \ .
 \label{equation::C_070}
\end{equation}
Für die relevante Dichtematrix finden wir dann mit \eqref{equation::C_030} die 
Bewegungsgleichung 
\begin{eqnarray}
 \partial_t \tilde{\varrho}(t) &=& \partial_t ( \varrho(t) \, \mathsf{P}(t) ) \nonumber \\
 &=& ( \partial_t \varrho(t) ) \, \mathsf{P}(t) + \varrho(t) \, ( \partial_t \mathsf{P}(t) )
 \nonumber \\
 &=& \varrho(t) \, i \, \mathsf{L} \, \mathsf{P}(t) + \varrho(t) \, ( \partial_t \mathsf{P}(t) ) \ .
 \label{equation::C_080}
\end{eqnarray}
Analog finden wir für den Restanteil der Dichtematrix die Bewegungsgleichung 
\begin{eqnarray}
 \partial_t \varrho^\prime(t) &=& \partial_t ( \varrho(t) \, \mathsf{Q}(t) ) \nonumber \\
 &=& ( \partial_t \varrho(t) ) \, \mathsf{Q}(t) + \varrho(t) \, ( \partial_t \mathsf{Q}(t) )
 \nonumber \\
 &=& \varrho(t) \, i \, \mathsf{L} \, \mathsf{Q}(t) - \varrho(t) \, ( \partial_t \mathsf{P}(t) ) \ .
 \label{equation::C_090}
\end{eqnarray}
Die zweiten Terme in den beiden Gleichungen mit der Zeitableitung des Projektionsoperators 
$\partial_t \mathsf{P}(t)$ fallen offensichtlich weg wegen \eqref{equation::B_530}. In den 
ersten Termen setzen wir die Zerlegung \eqref{equation::C_060} ein. Wir erhalten dann zwei 
gekoppelte Gleichungen 
\begin{eqnarray}
 \partial_t \tilde{\varrho}(t) &=& \tilde{\varrho}(t) \, i \, \mathsf{L} \, \mathsf{P}(t)
 + \varrho^\prime(t) \, i \, \mathsf{L} \, \mathsf{P}(t) \ ,
 \label{equation::C_100} \\
 \partial_t \varrho^\prime(t) &=& \tilde{\varrho}(t) \, i \, \mathsf{L} \, \mathsf{Q}(t) 
 + \varrho^\prime(t) \, i \, \mathsf{L} \, \mathsf{Q}(t)
 \label{equation::C_110}
\end{eqnarray}
für die zwei Anteile der Dichtematrix. Die Mastergleichung für die relevante Dichtematrix 
$\tilde{\varrho}(t)$ bekommen wir nun, indem wir den Restanteil $\varrho^\prime(t)$ eliminieren. 
Dazu lösen wir formal die zweite Gleichung. Diese ist eine inhomogene Differentialgleichung 
für $\varrho^\prime(t)$. Daher lösen wir zuerst den homogenen Anteil der Gleichung
\begin{equation}
 \partial_t U(t_0,t) = U(t_0,t) \, i \, \mathsf{L} \, \mathsf{Q}(t)
 \label{equation::C_120}
\end{equation}
mit der Anfangsbedingung $U(t_0,t_0)=1$. Wir finden
\begin{equation}
 U(t_0,t) = \mathsf{T} \, \exp \left\{ i \int_{t_0}^t dt^\prime \ \mathsf{L} 
 \, \mathsf{Q}(t^\prime) \right\} \ .
 \label{equation::C_130}
\end{equation}
Hierbei ist $\mathsf{T}$ der aus der Quantenfeldtheorie bekannte Zeitordnungs-Operator. 
Für $t>t_0$ ordnet er die Zeiten aufsteigend von links nach rechts, das ist hier genau 
umgekehrt wie sonst in der Quantenfeldtheorie. Mit einem geeigneten Ansatz berechnen wir 
darauf folgend auch die Lösung der inhomogenen Gleichung und bekommen den Restanteil der 
Dichtematrix
\begin{equation}
 \varrho^\prime(t) = \varrho^\prime(t_0) \, U(t_0,t) + \int_{t_0}^t dt^\prime 
 \ \tilde{\varrho}(t^\prime) \, i \, \mathsf{L} \, \mathsf{Q}(t^\prime) 
 \, U(t^\prime,t) \ .
 \label{equation::C_140} 
\end{equation}
Wir setzen nun diese Formel in die Bewegungsgleichung für die relevante Dichtematrix 
\eqref{equation::C_100} ein, ersetzen $\varrho^\prime(t_0) = \varrho(t_0) \, \mathsf{Q}(t_0)$ mit 
Hilfe von \eqref{equation::C_070} und ordnen die Reihenfolge einiger Terme um. Als Ergebnis 
erhalten wir die \emph{Mastergleichung} für die relevante Dichtematrix
\begin{eqnarray}
 \partial_t \tilde{\varrho}(t) &=& \tilde{\varrho}(t) \, i \, \mathsf{L} \, \mathsf{P}(t) \nonumber \\
 &&+ \, \int_{t_0}^t dt^\prime \ \tilde{\varrho}(t^\prime) \, i \, \mathsf{L} 
 \, \mathsf{Q}(t^\prime) \, U(t^\prime,t) \, i \, \mathsf{L} \, \mathsf{P}(t) \nonumber \\
 &&+ \, \varrho(t_0) \, \mathsf{Q}(t_0) \, U(t_0,t) \, i \, \mathsf{L} 
 \, \mathsf{P}(t) \ .
 \label{equation::C_150}
\end{eqnarray}
Diese Gleichung ist das zentrale Ergebnis von diesem Abschnitt. Man beachte, dass die 
relevante Dichtematrix $\tilde{\varrho}(t)$ und die Projektionsoperatoren $\mathsf{P}(t)$ 
und $\mathsf{Q}(t)$ eine spezielle Form haben. Sie sind definiert durch \eqref{equation::B_260}, 
\eqref{equation::B_440} und \eqref{equation::B_500}. Die Mastergleichung in der speziellen 
Form \eqref{equation::C_150} wurde zuerst von Robertson \cite{Ro66} hergeleitet und ist bekannt 
unter dem Namen \emph{Robertson-Gleichung}.

Eine einfachere Version der Mastergleichung wurde zuvor von Nakajima und Zwanzig 
\cite{Na58,Zw60} hergeleitet, wobei die Projektionsoperatoren $\mathsf{P}$ und 
$\mathsf{Q}$ konstant in der Zeit sind. In diesem Fall ist die Projektion 
$\tilde{\varrho}(t) = \varrho(t) \, \mathsf{P}$ eine lineare Abbildung im Raum der 
Quanten-Operatoren oder im Raum der Funktionen im klassischen Phasenraum. Folglich 
darf in der Gleichung \eqref{equation::C_130} der Zeitordnungs-Operator weggelassen 
werden, so dass die Zeitentwicklung durch eine einfache operatorwertige Exponentialfunktion 
$U(t_0,t) = \exp\{ i \mathsf{L} \mathsf{Q} ( t - t_0 ) \}$ beschrieben wird. Im Ergebnis 
wurde eine lineare Antworttheorie hergeleitet, um kleine Abweichungen vom thermischen 
Gleichgewicht zu beschreiben. Diese Theorie ist wohl bekannt unter dem Namen 
\emph{Zwanzig-Mori-Formalismus} \cite{Zw60,Zw61,Zw01,Mo65a,Mo65b}. In der vorliegenden 
Arbeit wollen wir jedoch Nichtgleichgewichtszustände weit entfernt vom Gleichgewicht 
in einer vollständig \emph{nichtlinearen} Weise betrachten. In der Robertson-Gleichung 
\eqref{equation::C_150} hängen die Projektionsoperatoren \eqref{equation::C_050} 
\emph{nichtlinear} von den hydrodynamischen Variablen $x_i(t)$ und folglich implizit 
auch von der Zeit $t$ ab.

Die einzelnen Zeilen der Formel \eqref{equation::C_150} lassen sich folgendermaßen 
interpretieren. Die erste Zeile ist der Beitrag zur Dynamik von den relevanten Variablen. 
Die zweite Zeile enthält die Gedächtniseffekte, welche dadurch entstehen, dass die 
irrelevanten Variablen eliminiert werden. Die dritte Zeile enthält die restlichen Effekte 
der irrelevanten Variablen. Hier handelt es sich um fluktuierende Kräfte, die im 
wesentlichen Rauschen darstellen.

\subsection{Bewegungsgleichungen für die Erwartungswerte}
\label{section::3B}
Für die Erwartungswerte $x_i(\mathbf{r},t)$ ist die Bewegungsgleichung definiert durch 
\eqref{equation::C_040}. Auf der rechten Seite dieser Gleichung setzen wir die Mastergleichung 
\eqref{equation::C_150} für die Zeitableitung der relevanten Dichtematrix ein. Weil die 
rechtesten Projektionsoperatoren in den einzelnen Termen von \eqref{equation::C_150} nun immer 
auf die relevante Variable $a_i(\mathbf{r})$ wirken, bewirkt \eqref{equation::B_470} das wir 
diese Projektionsoperatoren weglassen dürfen. Wir erhalten also
\begin{eqnarray}
 \partial_t x_i(\mathbf{r},t) &=& \mathrm{Sp}\{ \tilde{\varrho}(t) \, i \, \mathsf{L} 
 \, a_i(\mathbf{r}) \} \nonumber \\
 &&+ \, \int_{t_0}^t dt^\prime \ \mathrm{Sp}\{ \tilde{\varrho}(t^\prime) \, i \, \mathsf{L} 
 \, \mathsf{Q}(t^\prime) \, U(t^\prime,t) \, i \, \mathsf{L} \, a_i(\mathbf{r}) \} \nonumber \\
 &&+ \, \mathrm{Sp}\{ \varrho(t_0) \, \mathsf{Q}(t_0) \, U(t_0,t) \, i \, \mathsf{L} 
 \, a_i(\mathbf{r}) \} \ .
 \label{equation::C_160}
\end{eqnarray}
Der Liouville-Operator $\mathsf{L}$ wurde in \eqref{equation::C_020} so definiert, dass 
er nach rechts auf eine Variable $Y$ wirkt. Betrachtet man die Umformungen in 
\eqref{equation::C_010} genauer, so stellt man fest, dass man den Liouville-Operator 
auch nach links auf die Dichtematrix wirken lassen kann. Wir finden also
\begin{widetext}
\begin{eqnarray}
 \tilde{\varrho}(t) \, i \, \mathsf{L} &=& (i\hbar)^{-1} [ H, \tilde{\varrho}(t) ] \nonumber \\
 &=& (i\hbar)^{-1} \int_0^1 d\alpha \ (\tilde{\varrho}(t))^\alpha \, \biggl[ H, - k_B^{-1} \sum_k 
 \int d^d r\ \lambda_k(\mathbf{r},t) \, a_k(\mathbf{r}) \biggr] \, (\tilde{\varrho}(t))^{1-\alpha}
 \nonumber \\
 &=& - k_B^{-1} \sum_k \int d^d r\ \lambda_k(\mathbf{r},t) \, \int_0^1 d\alpha 
 \ (\tilde{\varrho}(t))^\alpha \, (i\hbar)^{-1} [ H, a_k(\mathbf{r}) ] 
 \, (\tilde{\varrho}(t))^{1-\alpha} \nonumber \\
 &=& k_B^{-1} \sum_k \int d^d r\ \lambda_k(\mathbf{r},t) \, \int_0^1 d\alpha 
 \ (\tilde{\varrho}(t))^\alpha \, ( i \, \mathsf{L} \, a_k(\mathbf{r}) )
 \, (\tilde{\varrho}(t))^{1-\alpha} \ .
 \label{equation::C_170}
\end{eqnarray}
Der Kommutator auf der rechten Seite von der ersten Zeile lässt sich auswerten, indem 
wir für die relevante Dichtematrix die Formel \eqref{equation::B_260} einsetzen und 
die Exponentialfunktion durch die entsprechende Produktentwicklung ersetzen. Auf 
diese Weise erhalten wir die Integralformel mit dem Kommutator in der zweiten Zeile. 
Mit dem Ergebnis von \eqref{equation::C_170} lassen sich die ersten beiden Terme in 
der Bewegungsgleichung \eqref{equation::C_160} umformen. Definieren wir die Frequenzmatrix
\begin{equation}
 \Omega_{ik}(\mathbf{r},\mathbf{r}^\prime;t) = k_B^{-1} \int_0^1 d\alpha \ \mathrm{Sp}\{ 
 (\tilde{\varrho}(t))^\alpha \, ( i \, \mathsf{L} \, a_k(\mathbf{r}^\prime) ) 
 \, (\tilde{\varrho}(t))^{1-\alpha} \, a_i(\mathbf{r}) \} \ ,
 \label{equation::C_180}
\end{equation}
die Gedächtnismatrix
\begin{equation}
 M_{ik}(\mathbf{r},t;\mathbf{r}^\prime,t^\prime) = k_B^{-1} \int_0^1 d\alpha \ \mathrm{Sp}\{ 
 (\tilde{\varrho}(t^\prime))^\alpha \, ( i \, \mathsf{L} \, a_k(\mathbf{r}^\prime) )
 \, (\tilde{\varrho}(t^\prime))^{1-\alpha} \, \mathsf{Q}(t^\prime) \, U(t^\prime,t) 
 \, ( i \, \mathsf{L} \, a_i(\mathbf{r}) ) \}
 \label{equation::C_190}
\end{equation}
\end{widetext}
und die fluktuierende Kraft
\begin{equation}
 f_i(\mathbf{r},t) = \mathrm{Sp}\{ \varrho(t_0) \, \mathsf{Q}(t_0) \, U(t_0,t) \, i \, \mathsf{L} 
 \, a_i(\mathbf{r}) \} \ ,
 \label{equation::C_200}
\end{equation}
so erhalten wir die Bewegungsgleichung
\begin{eqnarray}
 \partial_t x_i(\mathbf{r},t) &=& \sum_k \int d^d r^\prime \ \Omega_{ik}(\mathbf{r},\mathbf{r}^\prime;t)
 \, \lambda_k(\mathbf{r}^\prime,t) \nonumber \\
 &&+ \, \sum_k \int d^d r^\prime \int_{t_0}^t dt^\prime \ M_{ik}(\mathbf{r},t;\mathbf{r}^\prime,t^\prime)
 \, \lambda_k(\mathbf{r}^\prime,t^\prime) \nonumber \\
 &&+ \, f_i(\mathbf{r},t) \ .
 \label{equation::C_210}
\end{eqnarray}
Man beachte, dass die ersten beiden Terme auf der rechten Seite die Lagrange-Parameter 
$\lambda_i(\mathbf{r},t)$ enthalten. Die drei Terme auf der rechten Seite kann man 
folgendermaßen interpretieren. Der erste Term beschreibt die Kopplungen innerhalb des 
Unterraums der relevanten Variablen. Der zweite Term beschreibt die Gedächtniseffekte, 
die entstehen, wenn die relevanten Variablen mit den restlichen ausintegrierten Variablen 
wechselwirken. Der letzte Term enthält die restlichen Kräfte der ausintegrierten Variablen. 
Diese sind meist Fluktuationen auf kurzen räumlichen und zeitlichen Skalen.

Wählt man zum Anfangszeitpunkt die Dichtematrix $\varrho(t_0)=\tilde{\varrho}(t_0)$, so gilt
\begin{equation}
 \varrho(t_0) \, \mathsf{Q}(t_0) =\tilde{\varrho}(t_0) \, \mathsf{Q}(t_0) = \tilde{\varrho}(t_0) 
 - \tilde{\varrho}(t_0) \, \mathsf{P}(t_0) = 0 \ ,
 \label{equation::C_220}
\end{equation}
und die fluktuierende Kraft \eqref{equation::C_200} ist $f_i(\mathbf{r},t)=0$. Die 
Bewegungsgleichung \eqref{equation::C_210} zusammen mit der Frequenzmatrix 
\eqref{equation::C_180} und der Gedächtnismatrix \eqref{equation::C_190} jedoch ohne die 
fluktuierende Kraft \eqref{equation::C_200} wurde bereits von Robertson \cite{Ro66} 
hergeleitet und ist beschrieben in Kapitel 18 im Buch von Fick und Sauermann \cite{FS90}.

\subsection{Mori-Skalarprodukt}
\label{section::3C}
Das Mori-Skalarprodukt ist ein hermitesches Skalarprodukt für zwei quantenmechanische Variablen 
$Y_1$ und $Y_2$, mit dem sich die Formeln für die Frequenzmatrix \eqref{equation::C_180} und die 
Gedächtnismatrix \eqref{equation::C_190} vereinfachen lassen. Ursprünglich \cite{Mo65a,Mo65b} 
wurde es für das thermische Gleichgewicht mit einer Dichtematrix $\tilde{\varrho}_\mathrm{eq}$ 
definiert, welche eine Boltzmannstruktur wie \eqref{equation::B_300} hat. Man kann es jedoch 
auch allgemeiner definieren für das Nichtgleichgewicht, in dem man die relevante Dichtematrix 
\eqref{equation::B_260} einsetzt. Wir verwenden hier das verallgemeinerte Mori-Skalarprodukt 
in der Form
\begin{equation}
 ( Y_1 | Y_2 )_t = \int_0^1 d\alpha \ \mathrm{Sp}\{ (\tilde{\varrho}(t))^\alpha \, Y_1^+ 
 \, (\tilde{\varrho}(t))^{1-\alpha} \, Y_2 \} \ .
 \label{equation::C_230}
\end{equation}
Weil die relevante Dichtematrix $\tilde{\varrho}(t)$ über die Erwartungswerte $x_i(\mathbf{r},t)$ 
implizit von der Zeit abhängt, gilt dasselbe ebenso für das Mori-Skalarprodukt. Es gelten die 
üblichen Regeln für Skalarprodukte, welche in der Quantentheorie verwendete werden. Es ist 
bilinear, positiv definit und hermitesch. Letztere Eigenschaft bewirkt z.B.\ die Gleichung 
$( Y_1 | Y_2 )_t = ( Y_2 | Y_1 )_t^*$.

Wir haben bisher drei Operatoren definiert, welche auf die Variablen $Y$ wirken. Dies 
sind die Projektionsoperatoren $\mathsf{P}(t)$, $\mathsf{Q}(t)$ und der Liouville-Operator 
$\mathsf{L}$, definiert in \eqref{equation::B_440}, \eqref{equation::B_500} und 
\eqref{equation::C_020}. Wir wollen untersuchen, in wieweit diese Operatoren selbstadjungiert 
oder hermitesch sind bezüglich dem Mori-Skalarprodukt. Setzen wir den Projektionsoperator 
\eqref{equation::B_440} einmal hinten und einmal vorne in das Mori-Skalarprodukt ein, so 
finden wir nach einigen Umformungen die symmetrische Formel
\begin{widetext}
\begin{eqnarray}
 ( Y_1 | \mathsf{P}(t) | Y_2 )_t &=& ( Y_1 | \mathsf{P}(t) Y_2 )_t = ( \mathsf{P}(t) Y_1 | Y_2 )_t 
 \nonumber \\
 &=& \mathrm{Sp}\{ (\tilde{\varrho}(t) Y_1^+ \} \, \mathrm{Sp}\{ (\tilde{\varrho}(t) Y_2 \}
 - \sum_{ik} \int d^d r \int d^d r^\prime \left( \frac{\delta}{\delta x_i(\mathbf{r},t)}
 \mathrm{Sp}\{ (\tilde{\varrho}(t) Y_1^+ \} \right) \nonumber \\
 &&\hspace{50mm} \times \chi_{ik}( \mathbf{r}, \mathbf{r}^\prime; t )
 \left( \frac{\delta}{\delta x_k(\mathbf{r}^\prime,t)} \mathrm{Sp}\{ (\tilde{\varrho}(t) Y_2 \} \right)
 \label{equation::C_240}
\end{eqnarray}
mit der Suszeptibilität
\begin{equation}
 \chi_{ik}( \mathbf{r}, \mathbf{r}^\prime; t ) = 
 \frac{\delta x_i(\mathbf{r},t)}{\delta \lambda_k(\mathbf{r}^\prime,t)} =
 \frac{\delta x_k(\mathbf{r}^\prime,t)}{\delta \lambda_i(\mathbf{r},t)} =
 \frac{\delta^2 F[\lambda(t)]}{\delta \lambda_i(\mathbf{r},t) \, \delta \lambda_k(\mathbf{r}^\prime,t)} \ .
 \label{equation::C_250}
\end{equation}
\end{widetext}
Eine analoge Formel finden wir ebenso für den orthogonalen Projektionsoperator \eqref{equation::B_500},
nämlich
\begin{equation}
 ( Y_1 | \mathsf{Q}(t) | Y_2 )_t = ( Y_1 | \mathsf{Q}(t) Y_2 )_t = ( \mathsf{Q}(t) Y_1 | Y_2 )_t \ .
 \label{equation::C_260}
\end{equation}
Wir stellen somit fest, dass die Projektionsoperatoren $\mathsf{P}(t)$ und $\mathsf{Q}(t)$ beide 
selbstadjungiert oder hermitesch sind. Diese Eigenschaft garantiert, dass die Projektionsoperatoren 
in einem besonderen Sinne kompatibel mit dem Mori-Skalarprodukt sind.

Anders verhält sich der Liouville-Operator $\mathsf{L}$. Setzen wir diesen hinten und vorne 
in das Mori-Skalarprodukt ein, so finden wir
\begin{equation}
 ( Y_1 | \mathsf{L} Y_2 )_t = ( \mathsf{L} Y_1 | Y_2 )_t 
 \quad \Longleftrightarrow \quad [\tilde{\varrho}(t), H] = 0 \ .
 \label{equation::C_270}
\end{equation}
Das bedeutet, der Liouville-Operator $\mathsf{L}$ ist dann und nur dann hermitesch, wenn
die relevante Dichtematrix $\tilde{\varrho}(t)$ mit dem Hamilton-Operator $H$ vertauscht. Dies 
ist im thermischen Gleichgewicht für die Dichtematrix \eqref{equation::B_300} des großkanonischen 
Ensembles erfüllt, weil der Impuls $\mathbf{P}$ und die Teilchenzahl $N$ Erhaltungsgrößen sind. 
Im Nichtgleichgewicht gilt das jedoch im allgemeinen nicht. Der Liouville-Operator $\mathsf{L}$ 
ist also im allgemeinen nicht selbstadjungiert oder hermitesch.

Wir schreiben nun die Frequenzmatrix \eqref{equation::C_180} mit dem Mori-Skalarprodukt um 
und erhalten
\begin{equation}
 \Omega_{ik}(\mathbf{r},\mathbf{r}^\prime;t) = -i \, k_B^{-1}
 ( \mathsf{L} \, a_k(\mathbf{r}^\prime) | a_i(\mathbf{r}) )_t \ .
 \label{equation::C_280}
\end{equation}
Ebenso schreiben wir die Gedächtnismatrix \eqref{equation::C_190} um und erhalten
\begin{equation}
 M_{ik}(\mathbf{r},t;\mathbf{r}^\prime,t^\prime) = k_B^{-1} ( \mathsf{L} \, a_k(\mathbf{r}^\prime)
 | \mathsf{Q}(t^\prime) \, U(t^\prime,t) 
 \, \mathsf{L} \, a_i(\mathbf{r}) )_{t^\prime} \ .
 \label{equation::C_290}
\end{equation}
Die imaginären Faktoren $i$ haben wir aus dem Mori-Skalarprodukt nach vorne herausgezogen. 
Die neuen Formeln \eqref{equation::C_280} und \eqref{equation::C_290} haben eine erheblich 
einfachere Struktur. Sie sehen jedoch nicht gerade symmetrisch aus.

Eine Symmetrisierung bezügliche dem orthogonalen Projektionsoperator $\mathsf{Q}(t)$ ist 
möglich, weil dieser gemäß \eqref{equation::C_260} hermitesch ist und die allgemeine Formel 
\eqref{equation::B_510} erfüllt. Wir stellen fest, dass der Zeitentwicklungsoperator 
$U(t_0,t)$, definiert in \eqref{equation::C_130}, immer nur mit einem vorangestelltem 
Projektionsoperator $\mathsf{Q}(t_0)$ auftritt. Wegen \eqref{equation::B_510} dürfen wir 
daher zwei Dinge tun. Zum einen gilt
\begin{equation}
 \mathsf{Q}(t_0) \, U(t_0,t) = \mathsf{Q}(t_0) \, U(t_0,t) \, \mathsf{Q}(t) \ ,
 \label{equation::C_300}
\end{equation}
und zum anderen dürfen wir den Zeitentwicklungsoperator durch einen Ausdruck mit symmetrischem 
Exponenten und zwei orthogonalen Projektionsoperatoren darin ersetzen gemäß
\begin{equation}
 U(t_0,t) = \mathsf{T} \, \exp \left\{ i \int_{t_0}^t dt^\prime \ \mathsf{Q}(t^\prime) 
 \, \mathsf{L} \, \mathsf{Q}(t^\prime) \right\} \ .
 \label{equation::C_310}
\end{equation}
Setzen wir nun \eqref{equation::C_300} in \eqref{equation::C_290} ein und verwenden wir die 
Hermitezität des orthogonalen Projektionsoperators \eqref{equation::C_260}, so bekommen wir 
die Gedächtnismatrix
\begin{equation}
 M_{ik}(\mathbf{r},t;\mathbf{r}^\prime,t^\prime) = k_B^{-1} 
 ( \mathsf{Q}(t^\prime) \, \mathsf{L} \, a_k(\mathbf{r}^\prime) | 
 U(t^\prime,t) \, \mathsf{Q}(t) \, \mathsf{L} \, a_i(\mathbf{r}) )_{t^\prime} \ .
 \label{equation::C_320}
\end{equation}
Diese Formel ist symmetrisch bis auf die Position des Zeitentwicklungsoperators $U(t^\prime,t)$. 

Eine weitere Symmetrisierung von Frequenzmatrix \eqref{equation::C_280} und Gedächtnismatrix 
\eqref{equation::C_320} ist nur möglich, wenn auch der Liouville-Operator $\mathsf{L}$ 
hermitesch ist. Dies ist im thermischen Gleichgewicht der Fall. Hier finden wir für die 
Frequenzmatrix die symmetrische Formel
\begin{equation}
 \Omega^{(\mathrm{eq})}_{ik}(\mathbf{r},\mathbf{r}^\prime) = -i \, k_B^{-1}
 ( a_k(\mathbf{r}^\prime) ) \, | \, \mathsf{L} \, | \, a_i(\mathbf{r}) )_\mathrm{eq} \ .
 \label{equation::C_330}
\end{equation}
Ebenso finden wir für die Gedächtnismatrix die symmetrische Formel
\begin{eqnarray}
 M^{(\mathrm{eq})}_{ik}(\mathbf{r},t;\mathbf{r}^\prime,t^\prime) &=& k_B^{-1} 
 ( \mathsf{Q}(t^\prime) \, \mathsf{L} \, a_k(\mathbf{r}^\prime) \, | \nonumber \\
 &&\times \, U(t^\prime,t) \, | \, \mathsf{Q}(t) \, \mathsf{L} \, a_i(\mathbf{r}) )_\mathrm{eq} \ .
 \hspace{10mm}
 \label{equation::C_340}
\end{eqnarray}
Weil wir im thermischen Gleichgewicht den Liouville-Operator $\mathsf{L}$ und den 
Zeitentwicklungsoperator $U$ entweder vorne oder hinten in das Mori-Skalarprodukt schreiben 
dürfen, setzen wir diese in die Mitte zwischen zwei senkrechte Striche.

Wir stellen fest, dass die Indizes, die Ortsvariablen und die Zeitvariablen auf den linken 
Seiten der Gleichungen für die Frequenzmatrix und die Gedächtnismatrix die umgekehrte 
Reihenfolge haben als auf den rechten Seiten. Das liegt an unserer Definition der Erwartungswerte 
\eqref{equation::B_170}, wo die Dichtematrix links und die physikalische Variable rechts stehen. 
Ebenso liegt das an unserer Definition des Projektionsoperators \eqref{equation::B_440}, der 
nach rechts auf die physikalische Variable wirkt. Man kann die Faktoren in den Spuren auch 
umordnen und die Richtungen und Reihenfolgen umkehren. Dann erhält man für die Frequenzmatrix 
und die Gedächtnismatrix Ergebnisse mit gleichen Reihenfolgen von Indizes, Ortsvariablen und 
Zeitvariablen auf beiden Seiten der Gleichungen. Der Unterschied zwischen unserer und letzter 
Schreibweise ist jedoch nur formaler Natur. Im Ergebnis gibt es keinen Unterschied.

\subsection{GENERIC Formalismus}
\label{section::3D}
Unsere Herleitung der Bewegungsgleichungen \eqref{equation::C_210} mit der Frequenzmatrix 
\eqref{equation::C_280}, der Gedächtnismatrix \eqref{equation::C_320} und den fluktuierenden 
Kräften \eqref{equation::C_200} folgte der Originalarbeit von Robertson \cite{Ro66} und der 
Darstellung von Fick und Sauermann \cite{FS90}. Eine alternative Formulierung von solchen 
Bewegungsgleichungen wurde von Öttinger und Grmela \cite{GO97A,GO97B,Ot05} gegeben im 
Rahmen eines allgemeinen Konzeptes, das GENERIC-Formalismus genannt wird. Wenn wir unsere 
Bewegungsgleichung \eqref{equation::C_210} in die GENERIC-Form umschreiben wollen, dann 
muss diese eine Struktur haben wie
\begin{eqnarray}
 \partial_t x_i(\mathbf{r},t) &=& \sum_k \int d^d r^\prime \ L_{ik}(\mathbf{r},\mathbf{r}^\prime;t)
 \, \frac{ \delta E[x(t)] }{ \delta x_k(\mathbf{r}^\prime,t) } \nonumber \\
 &+& \sum_k \int d^d r^\prime \int_{t_0}^t dt^\prime \ M_{ik}(\mathbf{r},t;\mathbf{r}^\prime,t^\prime)
 \, \frac{ \delta S[x(t^\prime)] }{ \delta x_k(\mathbf{r}^\prime,t^\prime) } \nonumber \\
 &+& f_i(\mathbf{r},t) \ .
 \label{equation::C_350}
\end{eqnarray}
Die ersten beiden Terme werden dargestellt durch Funktional-Ableitungen von der Energie 
$E[x(t)]$ und der Entropie $S[x(t)]$ nach den Erwartungswerten der relevanten Variablen 
$x_i(\mathbf{r},t)$. Sie enthalten die direkten Kopplungen der relevanten Veriablen und 
die indirekten Kopplungen mit Gedächtniseffekten. Der dritte Term beschreibt die restlichen 
fluktuierenden Kräfte von den ausintegrierten Variablen. Wir stellen fest, dass der zweite 
und der dritte Term unserer Gleichung \eqref{equation::C_210} bereits die erforderliche Form 
haben, denn die Lagrange-Parameter $\lambda_i(\mathbf{r},t)$ sind nach \eqref{equation::B_380} 
Funktional-Ableitungen der Entropie nach den Erwartungswerten.

Der erste Term in \eqref{equation::C_210} hat nicht die gewünschte Form. Die Kopplung dieser 
Kräfte an die Funktional-Ableitungen der \emph{Entropie} müssen ersetzt werden durch die 
Kopplung an die Funktional-Ableitungen der \emph{Energie}. Der erste Term muss folglich neu 
berechnet werden. Wir gehen daher zurück zur Gleichung \eqref{equation::C_160}. Wir verlangen 
als zusätzliche Bedingung, dass der Hamilton-Operator $H$ zu den relevanten Variablen 
$a_i(\mathbf{r})$ gehört. Er soll sich als Linearkombination
\begin{equation}
 H = \int d^d r \sum_i \varepsilon_i \, a_i(\mathbf{r})
 \label{equation::C_360}
\end{equation}
mit geeigneten Koeffizienten $\varepsilon_i$ darstellen lassen. Dies ist am einfachsten 
erfüllt, wenn eine der relevanten Variablen $a_i(\mathbf{r})$ die Energiedichte ist. Aus 
\eqref{equation::B_470} bekommen wir dann die Projektion
\begin{equation}
 \mathsf{P}(t) \, H = H \ .
 \label{equation::C_370}
\end{equation}
Wir formen damit den ersten Term auf der rechten Seite von \eqref{equation::C_160} um 
und erhalten
\begin{widetext}
\begin{eqnarray}
 \mathrm{Sp}\{ \tilde{\varrho}(t) \, i \, \mathsf{L} \, a_i(\mathbf{r}) \} 
 &=& ( i / \hbar ) \ \mathrm{Sp}\{ \tilde{\varrho}(t) \, [ H, a_i(\mathbf{r}) ] \} \nonumber \\
 &=& ( i / \hbar ) \ \mathrm{Sp}\{ \tilde{\varrho}(t) 
 \, [ ( \mathsf{P}(t) \, H ), a_i(\mathbf{r}) ] \} \nonumber \\
 &=& \frac{ i }{ \hbar } \ \sum_k \int d^d r^\prime \ \mathrm{Sp}\{ \tilde{\varrho}(t) 
 \, [ a_k(\mathbf{r}^\prime), a_i(\mathbf{r}) ] \} \, \frac{ \delta }{ \delta x_k(\mathbf{r}^\prime,t) }
 \, \mathrm{Sp}\{ \tilde{\varrho}(t) \, H \} \nonumber \\
 &=& \frac{ 1 }{ i \hbar } \ \sum_k \int d^d r^\prime \ \mathrm{Sp}\{ \tilde{\varrho}(t) 
 \, [ a_i(\mathbf{r}), a_k(\mathbf{r}^\prime) ] \} 
 \, \frac{ \delta E[x(t)] }{ \delta x_k(\mathbf{r}^\prime,t) } \nonumber \\
 &=& \sum_k \int d^d r^\prime \ L_{ik}(\mathbf{r},\mathbf{r}^\prime;t)
 \, \frac{ \delta E[x(t)] }{ \delta x_k(\mathbf{r}^\prime,t) } \ .
 \label{equation::C_380}
\end{eqnarray}
\end{widetext}
In der dritten Zeile dieser Gleichung haben wir die explizite Form des Projektionsoperators 
\eqref{equation::B_440} eingesetzt. Dieser erste Term bekommt schließlich die in 
\eqref{equation::C_350} gewünschte Form mit der \emph{Poisson-Matrix}
\begin{equation}
 L_{ik}(\mathbf{r},\mathbf{r}^\prime;t) = ( i \hbar )^{-1} \, \mathrm{Sp}\{ \tilde{\varrho}(t) 
 \, [ a_i(\mathbf{r}), a_k(\mathbf{r}^\prime) ] \} \ .
 \label{equation::C_390}
\end{equation}
Ersetzen wir auf der rechten Seite den quantenmechanischen Kommutator durch die klassische 
Poisson-Klammer, so erklärt sich der Name Poisson-Matrix von selbst. Denn dieser Ausdruck 
ist offensichtlich der Erwartungswert der Poisson-Klammer von zwei relevanten Variablen. 
Die Poisson-Matrix ist antisymmetrisch gemäß
\begin{equation}
 L_{ik}(\mathbf{r},\mathbf{r}^\prime;t) = - L_{ki}(\mathbf{r}^\prime,\mathbf{r};t) \ .
 \label{equation::C_400}
\end{equation}
Dies folgt aus der entsprechenden Eigenschaft des Kommutators und der Poisson-Klammer.

Im Ergebnis finden wir also, dass sich die Bewegungsgleichung \eqref{equation::C_160},
welche ursprünglich von Robertson \cite{Ro66} hergeleitet wurde, in die GENERIC-Form 
umschreiben lässt. Der GENERIC-Formalismus ist damit jedoch noch nicht abgeschlossen. 
Er liefert als zusätzliche Elemente noch einige Nebenbedingungen. Als erstes betrachten 
wir dazu
\begin{eqnarray}
 &&\sum_k \int d^d r^\prime \ L_{ik}(\mathbf{r},\mathbf{r}^\prime;t)
 \, \frac{ \delta S[x(t)] }{ \delta x_k(\mathbf{r}^\prime,t) } \nonumber \\
 &&= \frac{ 1 }{ i \hbar } \ \sum_k \int d^d r^\prime \ \mathrm{Sp}\{ \tilde{\varrho}(t) 
 \, [ a_i(\mathbf{r}), a_k(\mathbf{r}^\prime) ] \} \, \lambda_k(\mathbf{r}^\prime,t) \nonumber \\
 &&= \frac{ 1 }{ i \hbar } \ \mathrm{Sp}\left\{ \tilde{\varrho}(t) 
 \, \left[ a_i(\mathbf{r}), \sum_k \int d^d r^\prime \ \lambda_k(\mathbf{r}^\prime,t) 
 \, a_k(\mathbf{r}^\prime) \right] \right\} \nonumber \\
 &&= ( i \hbar )^{-1} \ \mathrm{Sp}\{ \tilde{\varrho}(t) 
 \, [ a_i(\mathbf{r}), - k_B \, \ln \tilde{\varrho}(t) ] \} \nonumber \\
 &&= ( i \hbar )^{-1} \ \mathrm{Sp}\{ a_i(\mathbf{r}) 
 \, [ \ln \tilde{\varrho}(t), \tilde{\varrho}(t) ] \} \, ( - k_B ) \ = \ 0 \ .
 \label{equation::C_410}
\end{eqnarray}
Für die Umformung zwischen der dritten und der vierten Zeile verwenden wir die explizite Form 
der relevanten Dichtematrix \eqref{equation::B_260}. Das letzte Gleichheitszeichen folgt aus 
dem Kommutator $[\ln\tilde{\varrho}(t),\tilde{\varrho}(t)]=0$. Wir erhalten also die Nebenbedingung
\begin{equation}
 \sum_k \int d^d r^\prime \ L_{ik}(\mathbf{r},\mathbf{r}^\prime;t)
 \, \frac{ \delta S[x(t)] }{ \delta x_k(\mathbf{r}^\prime,t) } = 0 \ .
 \label{equation::C_420}
\end{equation}
Da die Poisson-Matrix gemäß \eqref{equation::C_400} antisymmetrisch ist, gilt die Nebenbedingung 
auch in der adjungierten Form
\begin{equation}
 \sum_i \int d^d r \ \frac{ \delta S[x(t)] }{ \delta x_i(\mathbf{r},t) } 
 \, L_{ik}(\mathbf{r},\mathbf{r}^\prime;t) = 0 \ .
 \label{equation::C_430}
\end{equation}

Als zweites betrachten wir
\begin{eqnarray}
 &&\sum_i \int d^d r \ \mathsf{L} \, a_i(\mathbf{r}) \, \frac{ \delta E[x(t)] }{ \delta x_i(\mathbf{r},t) }
 \nonumber \\
 &&= \sum_i \int d^d r \ \mathsf{L} \, a_i(\mathbf{r}) \, \frac{ \delta }{ \delta x_i(\mathbf{r},t) } 
 \, \mathrm{Sp}\{ \tilde{\varrho}(t) \, H \} \nonumber \\
 &&= \mathsf{L} \, \mathsf{P}(t) \, H = \mathsf{L} \, H = \hbar^{-1} \, [ H, H ] = 0 \ .
 \label{equation::C_440}
\end{eqnarray}
Wir multiplizieren nun die Gedächtnismatrix \eqref{equation::C_320} mit einer Funktional-Ableitung 
der Energie $E[x(t)]$. Summieren wir anschließend über den hinteren Index, und intergrieren wir über die 
hintere Ortsvariable, so bekommen wir die Nebenbedingung
\begin{equation}
 \sum_k \int d^d r^\prime \ M_{ik}(\mathbf{r},t;\mathbf{r}^\prime,t^\prime)
 \, \frac{ \delta E[x(t^\prime)] }{ \delta x_k(\mathbf{r}^\prime,t^\prime) } = 0 \ .
 \label{equation::C_450}
\end{equation}
Summieren wir andererseits über den vorderen Index, und integrieren wir über die vordere 
Ortsvariable, so bekommen wir die Nebenbedingung in der adjungierten Form
\begin{equation}
 \sum_k \int d^d r^\prime \ \frac{ \delta E[x(t)] }{ \delta x_i(\mathbf{r},t) } 
 \, M_{ik}(\mathbf{r},t;\mathbf{r}^\prime,t^\prime) = 0 \ .
 \label{equation::C_460}
\end{equation}

Weitere Nebenbedingungen lassen sich für Erhaltungsgrößen ableiten. In einer normalen 
Flüssigkeit sind neben der Energie $E$ ebenso der Impuls $\mathbf{P}$ und die Teilchenzahl 
$N$ Erhaltungsgrößen. Die Erwartungswerte dieser Erhaltungsgrößen sind wiederum Funktionale 
der Erwartungswerte der relevanten Variablen $x_i(\mathbf{r},t)$ gemäß
\begin{eqnarray}
 \mathbf{P}[x(t)] &=& \mathrm{Sp}\{ \tilde{\varrho}(t) \, \mathbf{P} \} \ ,
 \label{equation::C_470} \\
 N[x(t)] &=& \mathrm{Sp}\{ \tilde{\varrho}(t) \, N \} \ .
 \label{equation::C_480}
\end{eqnarray}
Wir nehmen an, dass sich die Operatoren der Erhaltungsgrößen ähnlich wie der Hamilton-Operator 
\eqref{equation::C_360} als Linearkombinationen der relevanten Variablen
\begin{eqnarray}
 \mathbf{P} &=& \int d^d r \sum_i \mathbf{p}_i \, a_i(\mathbf{r}) ) \ ,
 \label{equation::C_490} \\
 N &=& \int d^d r \sum_i n_i \, a_i(\mathbf{r}) )
 \label{equation::C_500}
\end{eqnarray}
darstellen lassen, wobei $\mathbf{p}_i$ und $n_i$ geeignete Koeffizienten sind. Es folgen dann 
die Projektionen
\begin{equation}
 \mathsf{P}(t) \, \mathbf{P} = \mathbf{P} \ , \qquad \mathsf{P}(t) \, N = N \ .
 \label{equation::C_510}
\end{equation}
Mit diesen Projektionen können wir nun für die Erhaltungsgrößen Überlegungen analog zu 
\eqref{equation::C_380} durchführen, allerdings in Rückwärtsrichtung von unten nach oben.
Wir finden dann für den Impuls
\begin{eqnarray}
 &&\sum_k \int d^d r^\prime \ L_{ik}(\mathbf{r},\mathbf{r}^\prime;t)
 \, \frac{ \delta \mathbf{P}[x(t)] }{ \delta x_k(\mathbf{r}^\prime,t) } \nonumber \\
 &&= ( i / \hbar ) \ \mathrm{Sp}\{ \tilde{\varrho}(t) 
 \, [ ( \mathsf{P}(t) \, \mathbf{P} ), a_i(\mathbf{r}) ] \} \nonumber \\
 &&= ( i / \hbar ) \ \mathrm{Sp}\{ \tilde{\varrho}(t) 
 \, [ \mathbf{P}, a_i(\mathbf{r}) ] \}
 \label{equation::C_520}
\end{eqnarray}
und für die Teilchenzahl
\begin{eqnarray}
 &&\sum_k \int d^d r^\prime \ L_{ik}(\mathbf{r},\mathbf{r}^\prime;t)
 \, \frac{ \delta N[x(t)] }{ \delta x_k(\mathbf{r}^\prime,t) } \nonumber \\
 &&= ( i / \hbar ) \ \mathrm{Sp}\{ \tilde{\varrho}(t) 
 \, [ ( \mathsf{P}(t) \, N ), a_i(\mathbf{r}) ] \} \nonumber \\
 &&= ( i / \hbar ) \ \mathrm{Sp}\{ \tilde{\varrho}(t) \, [ N, a_i(\mathbf{r}) ] \} \ .
 \label{equation::C_530}
\end{eqnarray}
Ob auf den rechten Seiten nun eine Null steht oder irgend etwas Anderes, hängt davon ab,
was die Kommutatoren der Erhaltungsgrößen mit den relevanten Variablen ergeben. Für 
unsere beiden Erhaltungsgrößen in einer normalen Flüssigkeit gilt
\begin{equation}
 [ a_i(\mathbf{r}), \mathbf{P} ] = - i\hbar \nabla a_i(\mathbf{r}) \ , \qquad
 [ a_i(\mathbf{r}), N ] = 0 \ .
 \label{equation::C_540}
\end{equation}
Wir finden somit die Nebenbedingungen für den Impuls
\begin{equation}
 \sum_k \int d^d r^\prime \ L_{ik}(\mathbf{r},\mathbf{r}^\prime;t)
 \, \frac{ \delta \mathbf{P}[x(t)] }{ \delta x_k(\mathbf{r}^\prime,t) }
 = - \nabla \, x_i(\mathbf{r},t)
 \label{equation::C_550}
\end{equation}
und für die Teilchenzahl
\begin{equation}
 \sum_k \int d^d r^\prime \ L_{ik}(\mathbf{r},\mathbf{r}^\prime;t)
 \, \frac{ \delta N[x(t)] }{ \delta x_k(\mathbf{r}^\prime,t) } = 0 \ .
 \label{equation::C_560}
\end{equation}

Die Überlegungen von \eqref{equation::C_440} können wir ebenso mit den Erhaltungsgrößen 
durchführen. Wir erhalten dann für den Impuls
\begin{eqnarray}
 &&\sum_i \int d^d r \ \mathsf{L} \, a_i(\mathbf{r}) 
 \, \frac{ \delta \mathbf{P}[x(t)] }{ \delta x_i(\mathbf{r},t) } \nonumber \\
 &&= \sum_i \int d^d r \ \mathsf{L} \, a_i(\mathbf{r}) \, \frac{ \delta }{ \delta x_i(\mathbf{r},t) } 
 \, \mathrm{Sp}\{ \tilde{\varrho}(t) \, \mathbf{P} \} \nonumber \\
 &&= \mathsf{L} \, \mathsf{P}(t) \, \mathbf{P} = \mathsf{L} \, \mathbf{P} = \hbar^{-1} 
 \, [ H, \mathbf{P} ] = \mathsf{0}
 \label{equation::C_570}
\end{eqnarray}
und für die Teilchenzahl
\begin{eqnarray}
 &&\sum_i \int d^d r \ \mathsf{L} \, a_i(\mathbf{r}) \, \frac{ \delta N[x(t)] }{ \delta x_i(\mathbf{r},t) }
 \nonumber \\
 &&= \sum_i \int d^d r \ \mathsf{L} \, a_i(\mathbf{r}) \, \frac{ \delta }{ \delta x_i(\mathbf{r},t) } 
 \, \mathrm{Sp}\{ \tilde{\varrho}(t) \, N \} \nonumber \\
 &&= \mathsf{L} \, \mathsf{P}(t) \, N = \mathsf{L} \, N = \hbar^{-1} \, [ H, N ] = 0 \ .
 \label{equation::C_580}
\end{eqnarray}
In diesem Falle sind die rechten Seiten immer null, weil die Operatoren der Erhaltungsgrößen 
mit dem Hamilton-Operator vertauschen. Wir bekommen somit weitere Nebenbedingungen für den Impuls 
\begin{equation}
 \sum_k \int d^d r^\prime \ M_{ik}(\mathbf{r},t;\mathbf{r}^\prime,t^\prime)
 \, \frac{ \delta \mathbf{P}[x(t^\prime)] }{ \delta x_k(\mathbf{r}^\prime,t^\prime) } = 0 \ .
 \label{equation::C_590}
\end{equation}
und für die Teilchenzahl
\begin{equation}
 \sum_k \int d^d r^\prime \ M_{ik}(\mathbf{r},t;\mathbf{r}^\prime,t^\prime)
 \, \frac{ \delta N[x(t^\prime)] }{ \delta x_k(\mathbf{r}^\prime,t^\prime) } = 0 \ .
 \label{equation::C_600}
\end{equation}
Wir haben also für die zwei Erhaltungsgrößen Impuls $\mathbf{P}[x(t)]$ und Teilchenzahl 
$N[x(t)]$ die vier Nebenbedingungen \eqref{equation::C_550}, \eqref{equation::C_560} 
und \eqref{equation::C_590}, \eqref{equation::C_600} hergeleitet. In diesen stehen die 
Funktional-Ableitungen immer auf der rechten Seite. Es gibt weitere vier Nebenbedingungen in 
der adjungierten Form, wo die Funktional-Ableitungen auf der linken Seite stehen.

Die wesentlichen Komponenten des GENERIC-For\-malismus sind somit gefunden. Sie bestehen zum 
einen aus der Bewegungsgleichungen für die Erwartungswerte der relevanten Variablen 
\eqref{equation::C_350} und zum anderen aus einer Reihe von Nebenbedingungen für die 
Funktionale von Energie $E[x(t)]$, Entropie $S[x(t)]$ und die Erhaltungsgrößen wie Impuls 
$\mathbf{P}[x(t)]$ und Teilchenzahl $N[x(t)]$. Hinzu gehören die expliziten Formeln für die 
Poisson-Matrix \eqref{equation::C_390}, die Gedächtnismatrix \eqref{equation::C_320} und 
die fluktuierenden Kräfte der nicht relevanten Freiheitsgrade \eqref{equation::C_200}.

\subsection{Fluktuierende Kräfte}
\label{section::3E}
Der dritte Term in der Bewegungsgleichung \eqref{equation::C_350} sind die fluktuierenden 
Kräfte $f_i(\mathbf{r},t)$, definiert in \eqref{equation::C_200}. Wir dürfen hier ebenfalls 
den Zeitentwicklungsoperator in der symmetrischen Form \eqref{equation::C_300} verwenden. 
Es folgt dann
\begin{equation}
 f_i(\mathbf{r},t) = i \ \mathrm{Sp}\{ \varrho(t_0) \, \mathsf{Q}(t_0) \, U(t_0,t) \, \mathsf{Q}(t) 
 \, \mathsf{L} \, a_i(\mathbf{r}) \}
 \label{equation::C_610}
\end{equation}
mit $U(t_0,t)$ definiert in \eqref{equation::C_310}. Folgen wir den Überlegungen von 
\eqref{equation::C_440}, \eqref{equation::C_570} und \eqref{equation::C_580}, so finden wir 
Nebenbedingungen für die fluktuierenden Kräfte ähnlich wie für die Gedächtnismatrix. Wir finden 
Nebenbedingungen für die Energie
\begin{equation}
 \sum_i \int d^d r \ f_i(\mathbf{r},t) \, \frac{ \delta E[x(t)] }{ \delta x_i(\mathbf{r},t) } = 0 \ ,
 \label{equation::C_620}
\end{equation}
für den Impuls
\begin{equation}
 \sum_i \int d^d r \ f_i(\mathbf{r},t) \, \frac{ \delta \mathbf{P}[x(t)] }{ \delta x_i(\mathbf{r},t) } = 0
 \label{equation::C_630}
\end{equation}
und für die Teilchenzahl
\begin{equation}
 \sum_i \int d^d r \ f_i(\mathbf{r},t) \, \frac{ \delta N[x(t)] }{ \delta x_i(\mathbf{r},t) } = 0 \ .
 \label{equation::C_640}
\end{equation}
Wir finden jedoch keine solche Nebenbedingung für die Entropie $S[x(t)]$.
Wählt man für die Anfangsbedingung einen reinen Quantenzustand 
$\varrho(t_0) = \vert\Psi(t_0)\rangle \langle\Psi(t_0)\vert$, so kann man erwarten, dass die 
fluktuierenden Kräfte auf kurzen Längenskalen und kurzen Zeitskalen variieren. Wählt man 
andererseits eine relevante Dichtematrix als Anfangsbedingung $\varrho(t_0)=\tilde{\varrho}(t_0)$, 
welche definiert ist in \eqref{equation::B_260}, so sind die fluktuierenden Kräfte immer null.

\subsection{Kontinuitätsgleichungen}
\label{section::3F}
In unseren bisherigen Überlegungen und Betrachtungen sind die quantenmechanischen Operatoren 
$a_i(\mathbf{r},t)$, welche die relevanten Variablen in Form von irgendwelchen Dichten 
darstellen, nicht weiter spezifiziert worden. Da wir am Ende eine normale Flüssigkeit 
betrachten, handelt es sich in unserem Fall speziell um Dichten von Erhaltungsgrößen. Das 
Noether-Theorem liefert explizite Ausdrücke nicht nur für die Dichten $a_i(\mathbf{r},t)$ 
sondern auch für die zugehörigen Stromdichten $b_{im}(\mathbf{r},t)$, so dass auf der Ebene 
der quantenmechanischen Operatoren die Kontinuitätsgleichung
\begin{equation}
 \partial_t a_i(\mathbf{r},t) + \partial_m b_{im}(\mathbf{r},t) = 0 
 \label{equation::C_650}
\end{equation}
gilt, wobei $\partial_m = \partial / \partial r_m$ die Differentialoperatoren für die 
räumlichen Ableitungen sind. Die quantenmechanischen Operatoren sind hier vorübergehend 
im Heisenberg-Bild definiert und hängen explizit von der Zeit ab. Daher gilt für die 
Dichten auch die Heisenberg-Bewegungsgleichung
\begin{equation}
 \partial_t a_i(\mathbf{r},t) = i \, \mathsf{L} \, a_i(\mathbf{r},t) 
 = (i\hbar)^{-1} [ a_i(\mathbf{r},t) , H(t) ] \ . 
 \label{equation::C_660}
\end{equation}
Durch Vergleich der beiden Gleichungen \eqref{equation::C_650} und \eqref{equation::C_660} 
finden wir im Heisenberg-Bild
\begin{equation}
 i \, \mathsf{L} \, a_i(\mathbf{r},t) = - \, \partial_m b_{im}(\mathbf{r},t) \ . 
 \label{equation::C_670}
\end{equation}
Entsprechend finden wir im Schrödinger-Bild 
\begin{equation}
 i \, \mathsf{L} \, a_i(\mathbf{r}) = - \, \partial_m b_{im}(\mathbf{r}) \ . 
 \label{equation::C_680}
\end{equation}
Der Liouville-Operator $\mathsf{L}$ bildet also die relevanten Variablen $a_i(\mathbf{r})$ 
in räumliche Divergenzen von Stromdichten $\partial_m b_{im}(\mathbf{r})$ ab. 

Die Erwartungswerte der relevanten Variablen $x_i(\mathbf{r},t)$ werden durch die Formel 
\eqref{equation::B_200} mit der exakten Dichtematrix $\varrho(t)$ gebildet. Entsprechend 
definieren wir die Erwartungswerte der Stromdichten durch die Formel
\begin{equation}
 J_{im}(\mathbf{r},t) = \langle b_{im}(\mathbf{r}) \rangle_t 
 = \mathrm{Sp}\{ \varrho(t) \, b_{im}(\mathbf{r}) \} \ .
 \label{equation::C_690}
\end{equation}
Aus der Kontinuitätsgleichung für die quantenmechanischen Operatoren folgt dann die 
entsprechende Kontinuitätsgleichung für die Erwartungswerte
\begin{equation}
 \partial_t x_i(\mathbf{r},t) + \partial_m J_{im}(\mathbf{r},t) = 0 \ .
 \label{equation::C_700}
\end{equation}

Weiter oben in Kapitel \ref{section::3B} haben wir mit den Projektionsoperatoren die 
Bewegungsgleichung für die relevanten Variablen \eqref{equation::C_160} hergeleitet. 
Diese hat drei Terme auf der rechten Seite, einen reversiblen, einen dissipativen und 
einen fluktuierenden. In jedem dieser drei Terme finden wir den Ausdruck 
$\mathsf{L} \, a_i(\mathbf{r})$, welchen wir mit \eqref{equation::C_680} umformen. 
Auf diese Weise lässt sich die Bewegungsgleichung für die relevanten Variablen in der 
Form der Kontinuitätsgleichung \eqref{equation::C_700} schreiben mit den Stromdichten 
\begin{eqnarray}
 J_{im}(\mathbf{r},t) &=& \mathrm{Sp}\{ \tilde{\varrho}(t) \, b_{im}(\mathbf{r}) \} \nonumber \\
 &&+ \, \int_{t_0}^t dt^\prime \ \mathrm{Sp}\{ \tilde{\varrho}(t^\prime) \, i \, \mathsf{L} 
 \, \mathsf{Q}(t^\prime) \, U(t^\prime,t) \, b_{im}(\mathbf{r}) \} \nonumber \\
 &&+ \, \mathrm{Sp}\{ \varrho(t_0) \, \mathsf{Q}(t_0) \, U(t_0,t) \, b_{im}(\mathbf{r}) \} \ .
 \label{equation::C_710}
\end{eqnarray}
Die Stromdichten haben offensichtlich ebenfalls drei Terme, einen reversiblen, einen 
dissipativen und einen fluktuierenden.

Da die Bewegungsgleichung in der GENERIC-Form \eqref{equation::C_350} eine äquivalente 
Darstellung ist, muss sich diese auch in der Form der Kontinuitätsgleichung 
\eqref{equation::C_700} schreiben lassen. Für den ersten Term auf der rechten Seite 
prüfen wir das später in Kapitel \ref{section::4C} explizit nach, in dem wir für eine 
normale Flüssigkeit die Poisson-Klammern der relevanten Variablen und somit die 
Poisson-Matrix $L_{ik}(\mathbf{r},\mathbf{r}^\prime;t)$ explizit berechnen. Für den 
zweiten Term müssen wir die Gedächtnismatrix $M_{ik}(\mathbf{r},t;\mathbf{r}^\prime,t^\prime)$ 
genauer untersuchen. Diese ist in der Formel \eqref{equation::C_320} mit dem Mori-Skalarprodukt 
definiert. Wir finden hier den Ausdruck $\mathsf{L} \, a_i(\mathbf{r})$ gleich zweimal. 
Ersetzen wir diesen durch die Divergenz der Stromdichten gemäß \eqref{equation::C_680}, 
so finden wir die Darstellung 
\begin{equation}
 M_{ik}(\mathbf{r},t;\mathbf{r}^\prime,t^\prime) = \partial^{\ }_m \partial^\prime_n 
 \, N_{im,kn}(\mathbf{r},t;\mathbf{r}^\prime,t^\prime)
 \label{equation::C_720}
\end{equation}
mit zwei räumlichen Differentialoperatoren $\partial^{\ }_m = \partial / \partial r^{\ }_m$, 
$\partial^\prime_n = \partial / \partial r^\prime_n$ und der neuen Gedächtnismatrix mit den 
Stromdichten 
\begin{eqnarray}
 N_{im,kn}(\mathbf{r},t;\mathbf{r}^\prime,t^\prime) = \hspace{35mm} \nonumber\\ 
 = k_B^{-1} ( \mathsf{Q}(t^\prime) \, b_{kn}(\mathbf{r}^\prime) | 
 U(t^\prime,t) \, \mathsf{Q}(t) \, b_{im}(\mathbf{r}) )_t \ .
 \label{equation::C_730}
\end{eqnarray}
Setzen wir nun die Gedächtnismatrix \eqref{equation::C_720} in die Bewegungsgleichung in 
GENERIC-Form \eqref{equation::C_350} ein, so finden wir, dass sich auch der zweite Term 
in Form einer Divergenz einer Stromdichte schreiben lässt. Die Gedächtnismatrix in der 
Form \eqref{equation::C_720} werden wir später in Kapitel \ref{section::4D} als Ausgangspunkt 
für unsere Näherungen verwenden.

Zum Schluss betrachten wir die fluktuierenden Kräfte $f_i(\mathbf{r},t)$, welche in 
\eqref{equation::C_200} oder äquivalent in \eqref{equation::C_610} definiert werden. 
Diese stellen den dritten Term in der Bewegungsgleichung \eqref{equation::C_350} dar. 
Wir finden hier wieder den Ausdruck $\mathsf{L} \, a_i(\mathbf{r})$ und ersetzen diesen 
durch die Divergenz der Stromdichten gemäß \eqref{equation::C_680}. Dann finden wir die 
fluktuierenden Kräfte in der Form 
\begin{eqnarray}
 f_i(\mathbf{r},t) = - \, \partial_m \, g_{im}(\mathbf{r},t) 
 \label{equation::C_740}
\end{eqnarray}
mit den fluktuierenden Stromdichten 
\begin{equation}
 g_{im}(\mathbf{r},t) = \mathrm{Sp}\{ \varrho(t_0) \, \mathsf{Q}(t_0) \, U(t_0,t) \, \mathsf{Q}(t) 
 \, b_{im}(\mathbf{r}) \} \ .
 \label{equation::C_750}
\end{equation}
Die Darstellung der fluktuierenden Kräfte \eqref{equation::C_740} durch Divergenzen von 
fluktuierenden Stromdichten werden wir später in Kapitel \ref{section::4E} verwenden.

\subsection{Allgemeine Bewegungsgleichungen für Entropie, Energie und die Erhaltungsgrößen}
\label{section::3G}
Aus der Bewegungsgleichung für die Erwartungswerte der relevanten Variablen $x_i(\mathbf{r},t)$ 
können wir Bewegungsgleichungen für die Entropie $S[x(t)]$, die Energie $E[x(t)]$ und die 
weiteren Erhaltungsgrößen wie Impuls $\mathbf{P}[x(t)]$ und Teilchenzahl $N[x(t)]$ herleiten. 
Wir beginnen mit der Entropie und erhalten
\begin{widetext}
\begin{eqnarray}
 \frac{ d }{ dt } \, S[x(t)] &=& \sum_i \int d^d r \ \frac{ \delta S[x(t)] }{ \delta x_i(\mathbf{r},t) }
 \ \frac{ \partial x_i(\mathbf{r},t) }{ \partial t } \nonumber \\
 &=& \sum_i \int d^d r \ \sum_k \int d^d r^\prime \ \frac{ \delta S[x(t)] }{ \delta x_i(\mathbf{r},t) }
 \, L_{ik}(\mathbf{r},\mathbf{r}^\prime;t) \, \frac{ \delta E[x(t)] }{ \delta x_k(\mathbf{r}^\prime,t) } 
 \nonumber \\
 &&+ \, \sum_i \int d^d r \ \sum_k \int d^d r^\prime \int_{t_0}^t dt^\prime 
 \ \frac{ \delta S[x(t)] }{ \delta x_i(\mathbf{r},t) }
 \, M_{ik}(\mathbf{r},t;\mathbf{r}^\prime,t^\prime)
 \, \frac{ \delta S[x(t^\prime)] }{ \delta x_k(\mathbf{r}^\prime,t^\prime) } \nonumber \\
 &&+ \, \sum_i \int d^d r \ \frac{ \delta S[x(t)] }{ \delta x_i(\mathbf{r},t) }
 \, f_i(\mathbf{r},t) \ .
 \label{equation::C_760}
\end{eqnarray}
Für das zweite Gleichheitszeichen haben wir die Bewegungsgleichungen in der GENERIC-Form 
\eqref{equation::C_350} verwendet. Die Nebenbedingung für die Entropie \eqref{equation::C_430} 
in der adjungierten Form bewirkt, dass der erste Term mit den direkten Kopplungen an 
die relevanten Variablen weg fällt. Die Entropie-Gleichung vereinfacht sich also auf
\begin{equation}
 \frac{ d }{ dt } \, S[x(t)] = \sum_i \int d^d r \ \sum_k \int d^d r^\prime \int_{t_0}^t dt^\prime 
 \ \frac{ \delta S[x(t)] }{ \delta x_i(\mathbf{r},t) }
 \, M_{ik}(\mathbf{r},t;\mathbf{r}^\prime,t^\prime)
 \, \frac{ \delta S[x(t^\prime)] }{ \delta x_k(\mathbf{r}^\prime,t^\prime) }
 + \, \sum_i \int d^d r \ \frac{ \delta S[x(t)] }{ \delta x_i(\mathbf{r},t) }
 \, f_i(\mathbf{r},t) \ . 
 \label{equation::C_770}
\end{equation}
Der quadratische Term mit der Gedächtnismatrix ist ein Hinweis auf den zweiten Hauptsatz 
der Thermodynamik, welcher besagt, dass die Entropie mit der Zeit immer anwächst oder 
zumindest konstant bleibt. Wenn die Gedächtnismatrix \eqref{equation::C_320} symmetrisch 
und positiv definit wäre, dann wäre dieser Term immer positiv. Da wir jedoch bisher 
keine Näherung durchgeführt haben, ist die Zeitumkehrinvarianz der zugrunde liegenden 
mikroskopischen Quantentheorie nicht gebrochen, so dass der zweite Hauptsatz der 
Thermodynamik hier nicht gelten kann.

Wir fahren fort mit der Energie und erhalten
\begin{eqnarray}
 \frac{ d }{ dt } \, E[x(t)] &=& \sum_i \int d^d r \ \frac{ \delta E[x(t)] }{ \delta x_i(\mathbf{r},t) }
 \ \frac{ \partial x_i(\mathbf{r},t) }{ \partial t } \nonumber \\
 &=& \sum_i \int d^d r \ \sum_k \int d^d r^\prime \ \frac{ \delta E[x(t)] }{ \delta x_i(\mathbf{r},t) }
 \, L_{ik}(\mathbf{r},\mathbf{r}^\prime;t) \, \frac{ \delta E[x(t)] }{ \delta x_k(\mathbf{r}^\prime,t) } 
 \nonumber \\
 &&+ \, \sum_i \int d^d r \ \sum_k \int d^d r^\prime \int_{t_0}^t dt^\prime 
 \ \frac{ \delta E[x(t)] }{ \delta x_i(\mathbf{r},t) }
 \, M_{ik}(\mathbf{r},t;\mathbf{r}^\prime,t^\prime)
 \, \frac{ \delta S[x(t^\prime)] }{ \delta x_k(\mathbf{r}^\prime,t^\prime) } \nonumber \\
 &&+ \, \sum_i \int d^d r \ \frac{ \delta E[x(t)] }{ \delta x_i(\mathbf{r},t) }
 \, f_i(\mathbf{r},t) \ .
 \label{equation::C_780}
\end{eqnarray}
Der erste Term auf der rechten Seite ist null, weil die Possion-Matrix antisymmetrisch ist 
gemäß \eqref{equation::C_400}. Der zweite Term ist null wegen der Nebenbedingung für die 
Energie \eqref{equation::C_460} in der adjungierten Form. Der dritte Term ist null wegen der 
Nebenbedingung \eqref{equation::C_620}. Alle Terme auf der rechten Seite sind also null. 
Folglich ist die Energie $E[x(t)]$ zeitlich konstant, wie es für eine Erhaltungsgröße zu 
erwarten ist.

Für die weiteren Erhaltungsgrößen Impuls $\mathbf{P}[x(t)]$ und Teilchenzahl $N[x(t)]$ 
liefern die Nebenbedingungen analoge Ergebnisse. Eine Ausnahme ist der erste Term in der 
Gleichung für den Impuls. Wegen der Nebenbedingung \eqref{equation::C_550} ist nicht 
unmittelbar klar, dass dieser Term null ist. Wir finden
\begin{equation}
 \frac{ d }{ dt } \, \mathbf{P}[x(t)] = \sum_i \int d^d r \ \sum_k \int d^d r^\prime 
 \ \frac{ \delta \mathbf{P}[x(t)] }{ \delta x_i(\mathbf{r},t) }
 \, L_{ik}(\mathbf{r},\mathbf{r}^\prime;t) \, \frac{ \delta E[x(t)] }{ \delta x_k(\mathbf{r}^\prime,t) } 
 = \sum_i \int d^d r \ ( \nabla x_i(\mathbf{r},t) ) \, \frac{ \delta E[x(t)] }{ \delta x_i(\mathbf{r},t) }
 = \mathbf{0} \ .
 \label{equation::C_790}
\end{equation}
Der Term ist dennoch null, weil die Energie symmetrisch unter räumlichen Translationen ist. 
Dieses Argument gilt ganz allgemein, weil Erhaltungsgrößen immer mit Symmetrie-Eigenschaften 
verbunden sind. Man kann also allgemein zeigen, dass der erste Term immer null ergibt, 
auch wenn die zugehörige Nebenbedingung zunächst etwas von null Verschiedenes liefert.
Fassen wir nochmals zusammen, wo finden wir die Bewegungsgleichungen für die Energie, 
den Impuls und die Teilchenzahl
\begin{equation}
 \frac{ d }{ dt } \, E[x(t)] = 0 \ , \qquad
 \frac{ d }{ dt } \, \mathbf{P}[x(t)] = \mathbf{0} \ , \qquad
 \frac{ d }{ dt } \, N[x(t)] = 0 \ ,
 \label{equation::C_800}
\end{equation}
wie man sie für Erhaltungsgrößen erwartet.

\subsection{Thermisches Gleichgewicht}
\label{section::3H}
Die Bewegungsgleichungen für die relevanten Veriablen \eqref{equation::C_350} kann man 
umschreiben in die Form
\begin{eqnarray}
 \partial_t x_i(\mathbf{r},t) &=& - \, \mathbf{v} \cdot \nabla x_i(\mathbf{r},t)
 + \sum_k \int d^d r^\prime \ L_{ik}(\mathbf{r},\mathbf{r}^\prime;t)
 \, \frac{ \delta \Omega[x(t)] }{ \delta x_k(\mathbf{r}^\prime,t) } \nonumber \\
 &&- \, \frac{ 1 }{ T } \sum_k \int d^d r^\prime \int_{t_0}^t dt^\prime 
 \ M_{ik}(\mathbf{r},t;\mathbf{r}^\prime,t^\prime)
 \, \frac{ \delta \Omega[x(t^\prime)] }{ \delta x_k(\mathbf{r}^\prime,t^\prime) } \nonumber \\
 &&+ \, f_i(\mathbf{r},t)
 \label{equation::C_810}
\end{eqnarray}
\end{widetext}
mit dem großkanonischen thermodynamischen Potential
\begin{equation}
 \Omega[x(t)] = E[x(t)] - T \, S[x(t)] - \mathbf{v} \cdot \mathbf{P}[x(t)] - \mu \, N[x(t)] \ .
 \label{equation::C_820}
\end{equation}
Die Äquivalenz zwischen \eqref{equation::C_810} und der GENERIC-Form \eqref{equation::C_350} 
folgt aus den Nebenbedingungen. Vom ersten Term spaltet sich ein Term mit der Geschwindigkeit 
$\mathbf{v}$ ab, weil die rechte Seite der Nebenbedingung \eqref{equation::C_550} nicht null 
ist.

Im thermischen Gleichgewicht wird die Entropie $S[x(t)]$ maximal unter den Nebenbedingungen, 
dass die Energie $E[x(t)]$, der Impuls $\mathbf{P}[x(t)]$ und die Teilchenzahl $N[x(t)]$ fest 
Werte annehmen. Unter der Verwendung der geeigneten Lagrange-Parameter Temperatur $T$, 
Geschwindigkeit $\mathbf{v}$ und chemisches Potential $\mu$ führt das auf die notwendige 
Bedingung für das großkanonische thermodynamische Potential 
\begin{equation}
 \frac{ \delta \Omega[x(t)] }{ \delta x_k(\mathbf{r}^\prime,t) } = 0 \ .
 \label{equation::C_830}
\end{equation}
Zwei der Terme in \eqref{equation::C_810} sind somit null. Nehmen wir als Anfangsbedingung 
eine relevante Dichtematrix $\varrho(t_0)=\tilde{\varrho}(t_0)$, welche definiert ist in 
\eqref{equation::B_260}, so sind auch die fluktuierenden Kräfte null. Es verbleibt dann 
nur noch der allererste Term, und die Bewegungsgleichungen vereinfachen sich auf
\begin{equation}
 \partial_t x_i(\mathbf{r},t) = - \, \mathbf{v} \cdot \nabla x_i(\mathbf{r},t) \ .
 \label{equation::C_840}
\end{equation}
Die allgemeine Lösung dieser Gleichung lautet
\begin{equation}
 x_i(\mathbf{r},t) = \xi_i( \mathbf{r} - \mathbf{v} t ) \ ,
 \label{equation::C_850}
\end{equation}
wobei $\xi_i(\mathbf{r})$ zunächst eine beliebige Funktion ist. Letztere wird jedoch festgelegt 
durch die notwendige Bedingung \eqref{equation::C_830} für die Maximierung der Entropie 
unter den Nebenbedingungen. Im einfachsten Fall ist $\xi_i(\mathbf{r})=\xi_i$ eine Konstante. 
Dann ist das physikalische System räumlich homogen, wie man es von einer normalen Flüssigkeit 
im thermischen Gleichgewicht erwartet.

Falls dennoch räumliche Inhomogenitäten vorhanden sind, so zeigt die Lösung 
\eqref{equation::C_850}, dass diese sich mit einer konstanten Geschwindigkeit $\mathbf{v}$ 
im Raum bewegen. Wir kommen so dem tieferen Sinn der rechten Seite der Nebenbedingungen wie 
\eqref{equation::C_550} näher. Die Erhaltungsgröße hängt mit einer Symmetrietransformation 
zusammen und erzeugt diese über den Kommutator \eqref{equation::C_540}. Die Lösung bewegt 
sich dann entlang der Symmetrietransformation gleichförmig mit einer konstanten Geschwindigkeit.

\subsection{Poisson-Klammern}
\label{section::3I}
Dzyaloshinskii und Volovik \cite{DV80} haben einen eleganten Weg vorgestellt, wie man die 
reversiblen Anteile der hydrodynamischen Gleichungen mit Poisson-Klammern herleiten kann. 
Wir wollen hier zeigen, dass im GENERIC-Formalismus der erste Term auf der rechten Seite 
der Bewegungsgleichung \eqref{equation::C_350} genau diese Form hat. Mit der 
antisymmetrischen Poisson-Matrix \eqref{equation::C_390} können wir für zwei beliebige 
Funktionale $F[x(t)]$ und $G[x(t)]$ eine Poisson-Klammer definieren durch
\begin{widetext}
\begin{equation}
 \{ F[x(t)], G[x(t)] \} = \sum_i \int d^d r \ \sum_k \int d^d r^\prime 
 \ \frac{ \delta F[x(t)] }{ \delta x_i(\mathbf{r},t) }
 \, L_{ik}(\mathbf{r},\mathbf{r}^\prime;t) 
 \, \frac{ \delta G[x(t)] }{ \delta x_k(\mathbf{r}^\prime,t) } \ .
 \label{equation::C_860}
\end{equation}
\end{widetext}
Setzen wir hier $F[x(t)]=x_i(\mathbf{r},t)$ und $G[x(t)]=E[x(t)]$ ein, so bekommen wir mit 
\begin{equation}
 \{ x_i(\mathbf{r},t), E[x(t)] \} = \sum_k \int d^d r^\prime 
 \ L_{ik}(\mathbf{r},\mathbf{r}^\prime;t) \, \frac{ \delta E[x(t)] }{ \delta x_k(\mathbf{r}^\prime,t) }
 \label{equation::C_870}
\end{equation}
exakt den ersten Term auf der rechten Seite von \eqref{equation::C_350}. Somit lässt sich der 
reversible Anteil der hydrodynamischen Gleichungen in der Form
\begin{equation}
 \partial_t \, x_i(\mathbf{r},t) = \{ x_i(\mathbf{r},t), E[x(t)] \} + \cdots
 \label{equation::C_880}
\end{equation}
schreiben. Dieses Ergebnis stimmt genau mit der Formulierung von Dzyaloshinskii und Volovik 
\cite{DV80} überein.

Die Nebenbedingungen, welche mit der Poisson-Matrix \eqref{equation::C_390} formuliert werden, 
lassen sich ebenfalls durch Poisson-Klammern darstellen. Setzen wir $G[x(t)]=S[x(t)]$ ein, so 
erhalten wir aus \eqref{equation::C_420} die Nebenbedingung für die Entropie
\begin{equation}
 \{ x_i(\mathbf{r},t), S[x(t)] \} = 0 \ .
 \label{equation::C_890}
\end{equation}
Setzen wir weiterhin $\mathbf{P}[x(t)]$ und $N[x(t)]$ ein, so bekommen wir aus 
\eqref{equation::C_550} und \eqref{equation::C_560} die Nebenbedingungen für die 
Erhaltungsgrößen 
\begin{eqnarray}
 \{ x_i(\mathbf{r},t), \mathbf{P}[x(t)] \} &=& - \nabla x_i(\mathbf{r},t) \ ,
 \label{equation::C_900} \\
 \{ x_i(\mathbf{r},t), N[x(t)] \} &=& 0 \ .
 \label{equation::C_910}
\end{eqnarray}
Diese Poisson-Klammern korrespondieren zu den quantenmechanischen Kommutatoren 
\eqref{equation::C_540}. Die rechten Seiten sind im allgemeinen nicht null, weil die 
Erhaltungsgrößen mit Symmetrie-Transformationen zusammenhängen und diese im Sinne 
der Lie-Gruppen erzeugen.

Setzen wir nun für $F[x(t)]=E[x(t)]$ die Energie ein, so erhalten wir die Poisson-Klammern
\begin{eqnarray}
 \{ E[x(t)], S[x(t)] \} &=& 0 \ ,
 \label{equation::C_920} \\
 \{ E[x(t)], \mathbf{P}[x(t)] \} &=& \mathbf{0} \ ,
 \label{equation::C_930} \\
 \{ E[x(t)], N[x(t)] \} &=& 0 \ .
 \label{equation::C_940}
\end{eqnarray}
Die rechten Seiten sind hier immer null, weil die Energie $E[x(t)]$ symmetrisch unter den 
Transformationen ist. Für \eqref{equation::C_930} können wir das explizit nachprüfen analog 
zu \eqref{equation::C_790}. Die korrespondierenden Kommutatoren legen andererseits nahe, 
dass die rechten Seiten von \eqref{equation::C_930} und \eqref{equation::C_940} null sein 
müssen, weil $\mathbf{P}[x(t)]$ und $N[x(t)]$ Erhaltungsgrößen sind. Das großkanonische 
thermodynamische Potential $\Omega[x(t)]$ ist in \eqref{equation::C_820} als Linearkombination 
der Energie $E[x(t)]$, der Entropie $S[x(t)]$ und der Erhaltungsgrößen $\mathbf{P}[x(t)]$ und 
$N[x(t)]$ dargestellt. Folglich gilt ebenso 
\begin{eqnarray}
 \{ E[x(t)], \Omega[x(t)] \} = 0 \ .
 \label{equation::C_950}
\end{eqnarray}
Diese Gleichung werden wir später in Kapitel \ref{section::6} verwenden.

Die elementaren Poisson-Klammern bekommen wir, in dem wir $F[x(t)]=x_i(\mathbf{r},t)$ und 
$G[x(t)]=x_k(\mathbf{r}^\prime,t)$ in die allgemeine Poisson-Klammer \eqref{equation::C_860}
einsetzen. Verwenden wir weiterhin die Poisson-Matrix \eqref{equation::C_390}, so bekommen wir
\begin{eqnarray}
 \hspace{-10mm} L_{ik}(\mathbf{r},\mathbf{r}^\prime;t) &=& 
 \{ x_i(\mathbf{r},t), x_k(\mathbf{r}^\prime,t) \} \nonumber \\
 &=& ( i \hbar )^{-1} \, \mathrm{Sp}\{ \tilde{\varrho}(t) 
 \, [ a_i(\mathbf{r}), a_k(\mathbf{r}^\prime) ] \} \ .
 \label{equation::C_960}
\end{eqnarray}
Die elementaren Poisson-Klammern der relevanten Erwartungswerte lassen sich folglich 
berechnen aus den Erwartungswerten der Kommutatoren der relevanten Variablen, wobei wie 
üblich in der Quantentheorie ein Faktor $( i \hbar )^{-1}$ hinzugefügt wird. Wir 
bemerken jedoch, dass eine Jacobi-Identitiät mit den Poisson-Klammern für die relevanten 
Erwartungswerte im allgemeinen nicht gilt. Wegen der Erwartungswert-Bildung mit der 
Spur und der relevanten Dichtematrix in \eqref{equation::C_960} überträgt sich die 
Jacobi-Identität mit den Kommutatoren im allgemeinen nicht auf eine Jacobi-Identität 
mit den entsprechenden Poisson-Klammern.

\section{Hydrodynamik}
\label{section::4}
Bisher sind keine Näherungen durchgeführt worden. Alle Gleichungen von Kapitel \ref{section::3} 
gelten exakt. Es wurde lediglich mit \eqref{equation::B_260} eine relevante Dichtematrix 
$\tilde{\varrho}(t)$ definiert, welche die Entropie unter gewissen Nebenbedingungen maximiert. 
Diese kann man natürlich als Näherung für die exakte Dichtematrix $\varrho(t)$ betrachten. 
Die relevante Dichtematrix wird jedoch nicht als Näherung verwendet, sondern zur Definition 
der Entropie \eqref{equation::B_210} und zur Definition von zwei Projektionsoperatoren 
\eqref{equation::B_440} und \eqref{equation::B_500}, welche eine Aufteilung der physikalischen 
Variablen in relevante und restliche irrelevante erlaubt. Das exakte Ergebnis waren die 
Bewegungsgleichungen für die Erwartungswerte der relevanten Variablen \eqref{equation::C_350} 
zusammen mit einigen Nebenbedingungen. Die Aufteilung der Variablen liefert auf der rechten 
Seite drei Terme: den ersten Term für die direkten Kopplungen der relevanten Variablen, den 
zweiten Term für die indirekte Kopplung über die restlichen irrelevanten Variablen mit 
Gedächtniseffekten und den dritten Term für die verbleibenden fluktuierenden Kräfte durch 
die restlichen irrelevanten Variablen. Im Folgenden wollen wir zeigen, dass man aus den 
Bewegungsgleichungen \eqref{equation::C_350} mit ein paar wenigen Annahmen und Näherungen 
die hydrodynamischen Gleichungen für eine normale Flüssigkeit erhält.

\subsection{Hydrodynamische Näherung}
\label{section::4A}
In der Hydrodynamik werden Eigenschaften von normalen Flüssigkeiten untersucht, die auf großen 
Längenskalen und großen Zeitskalen stattfinden. Die mikroskopische Struktur der Flüssigkeit 
im einzelnen hat keinen Einfluss. Aus welchen Atomen oder Molekülen die Flüssigkeit besteht 
und wie sich diese bewegen spielt im Detail keine Rolle. Folglich werden sich die Variablen 
der Theorie, die Erwartungswerte $x_i(\mathbf{r},t)$ und die Lagrange-Parameter 
$\lambda_i(\mathbf{r},t)$, nur langsam mit der Ortsvariable $\mathbf{r}$ und langsam mit der 
Zeit $t$ verändern. Gradienten dieser Variablen werden zunächst vernachlässigt. Man nimmt 
daher näherungsweise an, dass sich die Flüssigkeit lokal in einem thermischen Gleichgewicht 
befindet \cite{GM62}. Das bedeutet, dass sich die relevante Dichtematrix \eqref{equation::B_260} 
nicht so stark von der großkanonischen Boltzmann-Verteilung \eqref{equation::B_300} 
unterscheiden sollte.

Wir haben mit \eqref{equation::C_270} gezeigt, dass im thermischen Gleichgewicht der 
Liouville-Operator $\mathsf{L}$ selbstadjungiert oder hermitesch ist, weil der Hamilton-Operator 
mit der Boltzmann-Verteilung vertauscht gemäß $[\tilde{\varrho}_\mathrm{eq},H]=0$. In der 
Hydrodynamik befindet sich die Flüssigkeit in einem Nichtgleichgewichtszustand. Dieser Zustand 
ist jedoch lokal nicht so weit vom Gleichgewicht entfernt. Man spricht hier von einem lokalen 
thermischen Gleichgewicht. Wir können daher \emph{näherungsweise} annehmen, dass die relevante 
Dichtematrix mit dem Hamilton-Operator vertauscht gemäß $[\tilde{\varrho}(t),H]\approx 0$. Dann 
ist nach \eqref{equation::C_270} auch der Liouville-Operator $\mathsf{L}$ \emph{näherungsweise} 
selbstadjungiert oder hermitesch, und es gilt
\begin{equation}
 ( Y_1 | \mathsf{L} Y_2 )_t \approx ( \mathsf{L} Y_1 | Y_2 )_t 
 \quad \mbox{wegen} \quad [\tilde{\varrho}(t), H] \approx 0 \ .
 \label{equation::D_010}
\end{equation}
In Folge lassen sich die Frequenzmatrix \eqref{equation::C_280} und die Gedächtnismatrix
\eqref{equation::C_320} mit dem Mori-Skalarprodukt auch für das Nichtgleichgewicht 
\emph{näherungsweise} in symmetrischer Form schreiben, wie die Formeln \eqref{equation::C_330} 
und \eqref{equation::C_340} für das thermische Gleichgewicht. Wir erhalten also näherungsweise 
die Frequenzmatrix
\begin{equation}
 \Omega_{ik}(\mathbf{r},\mathbf{r}^\prime;t) \approx -i \, k_B^{-1}
 ( a_k(\mathbf{r}^\prime) | \, \mathsf{L} \, | a_i(\mathbf{r}) )_t
 \label{equation::D_020}
\end{equation}
und die Gedächtnismatrix
\begin{equation}
 M_{ik}(\mathbf{r},t;\mathbf{r}^\prime,t^\prime) \approx k_B^{-1} 
 ( \mathsf{Q}(t^\prime) \, \mathsf{L} \, a_k(\mathbf{r}^\prime) | 
 U(t^\prime,t) | \mathsf{Q}(t) \, \mathsf{L} \, a_i(\mathbf{r}) )_t \ .
 \label{equation::D_030}
\end{equation}
Wir haben hier wieder das Mori-Skalarprodukt mit zwei senkrechten Strichen in der Mitte 
geschrieben, um jeweils den Operator hervorzuheben, den man im Mori-Skalarprodukt durch 
Adjungieren entweder nach vorne oder nach hinten stellen kann. Die Frequenzmatrix 
ist somit näherungsweise antisymmetrisch
\begin{equation}
 \Omega_{ik}(\mathbf{r},\mathbf{r}^\prime;t) \approx - \Omega_{ki}(\mathbf{r}^\prime,\mathbf{r};t) \ ,
 \label{equation::D_040}
\end{equation}
und die Gedächtnismatrix ist näherungsweise symmetrisch
\begin{equation}
 M_{ik}(\mathbf{r},t;\mathbf{r}^\prime,t^\prime) \approx M_{ki}(\mathbf{r}^\prime,t^\prime;\mathbf{r},t) \ .
 \label{equation::D_050}
\end{equation}
Die Poisson-Matrix vom GENERIC-Formalismus \eqref{equation::C_390} ist bereits im 
Nichtgleichgewicht antisymmetrisch gemäß \eqref{equation::C_400}. Hier brauchen wir 
keine Näherung durchführen.

Entscheidend für den Erfolg der hydrodynamischen Näherung ist die richtige Auswahl der 
relevanten Variablen. Man muss hier eine Trennung der räumlichen und zeitlichen Skalen erreichen. 
Das bedeutet, mit der Auswahl muss man alle räumlich und zeitlich langsamen Variablen erwischen, 
so dass die restlichen irrelevanten Variablen alle räumlich und zeitlich schnell variieren. 
Für die Bewegungsgleichung \eqref{equation::C_350} hat dies zur Folge, dass Gedächtniseffekte 
im zweiten Term vernachlässigbar werden und die fluktuierenden Kräfte im dritten Term auf 
kurzen Längenskalen und Zeitskalen variieren.

Konstante Variablen einer normalen Flüssigkeit sind die Erhaltungsgrößen Energie $E[x(t)]$, 
Impuls $\mathbf{P}[x(t)]$ und Teilchenzahl $N[x(t)]$. Die Operatoren hierfür sind in 
\eqref{equation::C_360}, \eqref{equation::C_490} und \eqref{equation::C_500} als 
Linearkombinationen der relevanten Variablen $a_i(\mathbf{r})$ dargestellt. Als langsame 
relevante Variablen sind daher die Dichten dieser Erhaltungsgrößen geeignet. Wir wählen 
die Massendichte, die Impulsdichte und die Energiedichte definiert durch die Linearkombinationen
\begin{eqnarray}
 \rho(\mathbf{r}) &=& \sum_i m \, n_i \, a_i(\mathbf{r}) \ ,
 \label{equation::D_060} \\
 \mathbf{j}(\mathbf{r}) &=& \sum_i \mathbf{p}_i \, a_i(\mathbf{r}) \ ,
 \label{equation::D_070} \\
 \varepsilon(\mathbf{r}) &=& \sum_i \varepsilon_i \, a_i(\mathbf{r}) \ ,
 \label{equation::D_080}
\end{eqnarray}
wobei $n_i$, $\mathbf{p}_i$ und $\varepsilon_i$ die entsprechenden Koeffizienten sind. 
Anstelle der Teilchendichte $n(\mathbf{r})$ verwendet man in der Hydrodynmik üblicherweise 
die Massendichte $\rho(\mathbf{r})$. Der Unterschied ist ein Faktor $m$ für die Masse eines 
Teilchens. Beachten wir, dass die Impulsdichte drei räumliche Komponenten hat, so stellen 
wir fest, dass die relevanten Variablen einer normalen Flüssigkeit aus genau fünf verschiedenen 
Dichten bestehen. Da die lineare Abbildung in \eqref{equation::D_060}-\eqref{equation::D_080} 
umkehrbar eindeutig sein muss, kann der Index $i=1,\ldots,5$ genau fünf Werte annehmen.

Wir vernachlässigen Gedächtniseffekte, indem wir die Gedächtnismatrix mit einer Delta-Funktion 
in der Zeit schreiben gemäß
\begin{equation}
 M_{ik}(\mathbf{r},t;\mathbf{r}^\prime,t^\prime) \approx 
 2 \, M_{ik}(\mathbf{r},\mathbf{r}^\prime;t) \, \delta( t - t^\prime ) \ .
 \label{equation::D_090}
\end{equation}
Der Faktor $2$ ist erforderlich, weil die Zeit-Integration in den Gedächtnistermen der 
Bewegungsgleichungen immer nur über das halbe Intervall der Delta-Funktion reicht. Die 
Matrix $M_{ik}(\mathbf{r},\mathbf{r}^\prime;t)$, welche auf der rechten Seite von 
\eqref{equation::D_090} definiert wird, bezeichnet man als \emph{Onsager-Matrix}. Wegen 
\eqref{equation::D_050} ist sie symmetrisch. Die \emph{Poisson-Matrix} 
$L_{ik}(\mathbf{r},\mathbf{r}^\prime;t)$, welche in \eqref{equation::C_390} definiert ist, 
ist dagegen antisymmetrisch. Es gelten also die zwei Symmetriebedingungen
\begin{eqnarray}
 L_{ik}(\mathbf{r},\mathbf{r}^\prime;t) &=& -L_{ki}(\mathbf{r}^\prime,\mathbf{r};t) \ ,
 \label{equation::D_100} \\
 M_{ik}(\mathbf{r},\mathbf{r}^\prime;t) &=& +M_{ki}(\mathbf{r}^\prime,\mathbf{r};t) \ .
 \label{equation::D_110}
\end{eqnarray}
Setzen wir nun die Gedächtnismatrix \eqref{equation::D_090} in die Bewegungsgleichung des 
GENERIC-Formalismus \eqref{equation::C_350} ein und führen die Integration über die Zeit aus, 
so vereinfacht sich diese Gleichung auf
\begin{eqnarray}
 \partial_t x_i(\mathbf{r},t) &=& \sum_k \int d^d r^\prime \ L_{ik}(\mathbf{r},\mathbf{r}^\prime;t)
 \, \frac{ \delta E[x(t)] }{ \delta x_k(\mathbf{r}^\prime,t) } \nonumber \\
 &&+ \, \sum_k \int d^d r^\prime \ M_{ik}(\mathbf{r},\mathbf{r}^\prime;t)
 \, \frac{ \delta S[x(t^\prime)] }{ \delta x_k(\mathbf{r}^\prime,t^\prime) } \nonumber \\
 &&+ \, f_i(\mathbf{r},t) \ .
 \label{equation::D_120}
\end{eqnarray}
Ebenso vereinfacht sich die Entropie-Gleichung \eqref{equation::C_770}, und wir erhalten
\begin{widetext}
\begin{equation}
 \hspace{-10mm} \frac{ d }{ dt } \, S[x(t)] = \sum_i \int d^d r \ \sum_k \int d^d r^\prime
 \ \frac{ \delta S[x(t)] }{ \delta x_i(\mathbf{r},t) }
 \, M_{ik}(\mathbf{r},\mathbf{r}^\prime,t)
 \, \frac{ \delta S[x(t)] }{ \delta x_k(\mathbf{r}^\prime,t) }
 + \, \sum_i \int d^d r \ \frac{ \delta S[x(t)] }{ \delta x_i(\mathbf{r},t) }
 \, f_i(\mathbf{r},t) \ . 
 \label{equation::D_130}
\end{equation}
\end{widetext}
Die wichtigsten Nebenbedingungen des GENERIC-For\-malismus vereinfachen sich auf
\begin{eqnarray}
 \hspace{-10mm} \sum_k \int d^d r^\prime \ L_{ik}(\mathbf{r},\mathbf{r}^\prime;t)
 \, \frac{ \delta S[x(t)] }{ \delta x_k(\mathbf{r}^\prime,t) } &=& 0 \ ,
 \label{equation::D_140} \\
 \hspace{-10mm} \sum_k \int d^d r^\prime \ M_{ik}(\mathbf{r},\mathbf{r}^\prime;t)
 \, \frac{ \delta E[x(t)] }{ \delta x_k(\mathbf{r}^\prime,t) } &=& 0 \ .
 \label{equation::D_150}
\end{eqnarray}
Die weiteren Nebenbedingungen für die Erhaltungsgrößen Impuls $\mathbf{P}[x(t)]$ und Teilchenzahl 
$N[x(t)]$ vereinfachen sich entsprechend. Für den Impuls erhalten wir
\begin{eqnarray}
 \hspace{-10mm} \sum_k \int d^d r^\prime \ L_{ik}(\mathbf{r},\mathbf{r}^\prime;t)
 \, \frac{ \delta \mathbf{P}[x(t)] }{ \delta x_k(\mathbf{r}^\prime,t) } &=& 
 - \nabla \, x_i(\mathbf{r},t) \ ,
 \label{equation::D_160} \\
 \hspace{-10mm} \sum_k \int d^d r^\prime \ M_{ik}(\mathbf{r},\mathbf{r}^\prime;t)
 \, \frac{ \delta \mathbf{P}[x(t)] }{ \delta x_k(\mathbf{r}^\prime,t) } &=& \mathbf{0} \ ,
 \label{equation::D_170}
\end{eqnarray}
und für die Teilchenzahl
\begin{eqnarray}
 \hspace{-10mm} \sum_k \int d^d r^\prime \ L_{ik}(\mathbf{r},\mathbf{r}^\prime;t)
 \, \frac{ \delta N[x(t)] }{ \delta x_k(\mathbf{r}^\prime,t) } &=& 0 \ ,
 \label{equation::D_180} \\
 \hspace{-10mm} \sum_k \int d^d r^\prime \ M_{ik}(\mathbf{r},\mathbf{r}^\prime;t)
 \, \frac{ \delta N[x(t)] }{ \delta x_k(\mathbf{r}^\prime,t) } &=& 0 \ .
 \label{equation::D_190}
\end{eqnarray}
Entsprechende Nebenbedingungen gelten auch für die fluktuierenden Kräfte $f_i(\mathbf{r},t)$. 

Die Gleichungen \eqref{equation::D_120}-\eqref{equation::D_190} zusammen mit den 
Symmetriebedingungen \eqref{equation::D_100} und \eqref{equation::D_110} wurden ursprünglich 
von Grmela und Öttinger \cite{GO97A,GO97B,Ot05} aufgestellt und aus den mikroskopischen Theorien 
für klassische Flüssigkeiten und Quantenflüssigkeiten hergeleitet. In ihrer Gesamtheit 
definieren sie den GENERIC-Formalismus. In ihrer ursprünglichen Form enthalten sie keine 
Gedächtniseffekte. Unsere Gleichungen in Kapitel \ref{section::3} dagegen wurden ohne 
irgendeine Näherung hergeleitet und beinhalten somit alle Gedächtniseffekte. Folglich sind 
die Bewegungsgleichungen und Nebenbedingungen in Kapitel \ref{section::3} eine Erweiterung 
des GENERIC-Formalismus, welche Gedächtniseffekte mit einschließt.

Die Onsager-Matrix $M_{ik}(\mathbf{r},\mathbf{r}^\prime;t)$ ist im allgemeinen 
\emph{positiv semidefinit}. Das bedeutet, ihre Eigenwerte sind entweder positiv oder null.
Für die Entropiegleichung \eqref{equation::D_130} hat dies die Folge, dass der quadratische 
Term immer einen Beitrag größer oder gleich null liefert. Wenn wir die Fluktuationen im 
zweiten Term weglassen, dann wächst die Entropie monoton mit der Zeit an. Folglich gilt 
der zweite Hauptsatz der Thermodynamik und die Invarianz unter Zeitumkehr wird gebrochen. 
In Kapitel \ref{section::5} stellen wir jedoch fest, dass der zweite fluktuierende Term 
auch negative Beiträge zur Entropie liefert und somit die Zeitumkehrinvarianz wieder herstellt.

\subsection{Reaktive Beiträge}
\label{section::4B}
In der exakten Theorie hat die Gedächtnismatrix $M_{ik}(\mathbf{r},t;\mathbf{r}^\prime,t^\prime)$ 
ursprünglich zwei Arten von Beiträgen, nämlich \emph{reaktive} und \emph{dissipative}. 
Die erste Art von Beiträgen ist reversibel in der Zeit. Jedoch, wenn wir mit der Formel 
\eqref{equation::D_090} die Näherung durchführen und Gedächtniseffekte vernachlässigen, 
dann fallen die reaktiven Terme weg, und nur die dissipativen Terme verbleiben. Folglich 
ist der zweite Term in der GENERIC-Bewegungsgleichung \eqref{equation::D_120} rein 
dissipativ, denn die Onsager-Matrix $M_{ik}(\mathbf{r},\mathbf{r}^\prime;t)$ ist 
symmetrisch gemäß \eqref{equation::D_110} und positiv definit. Andererseits gibt es 
komplexere Flüssigkeiten, in denen reaktive Terme vorhanden sind und für die Eigenschaften 
auf großen Zeitskalen und großen Wellenlängen eine wichtige Rolle spielen. Solche 
reaktiven Terme wurden zum ersten Mal von Forster \cite{Fo74a,Fo74b,Fo75} für nematische 
Flüssigkristalle hergeleitet und untersucht.

Wir können reaktive Terme in unserer Theorie berücksichtigen, indem wir für die 
Gedächtnismatrix die Näherungsformel
\begin{eqnarray}
 M_{ik}(\mathbf{r},t;\mathbf{r}^\prime,t^\prime) &\approx& 
 2 \, K_{ik}(\mathbf{r},\mathbf{r}^\prime;t) \, \varepsilon( t - t^\prime ) 
 \, \delta( t - t^\prime ) \nonumber\\ 
 &&+ \ 2 \, M_{ik}(\mathbf{r},\mathbf{r}^\prime;t) \, \delta( t - t^\prime )
 \label{equation::D_200}
\end{eqnarray}
verwenden. Der erste Term ist hier neu und enthält die reaktiven Beiträge. Die 
Vorzeichenfunktion $\varepsilon( t - t^\prime )$ wird definiert durch 
$\varepsilon( t - t^\prime ) = \pm 1$ für $t-t^\prime \,^>_<\, 0$ und bewirkt, dass 
der erste Term unterschiedliche Vorzeichen für $t>t^\prime$ und $t<t^\prime$ hat. 
Die Stärke des reaktiven Terms wird durch die \emph{reaktive Matrix} 
$K_{ik}(\mathbf{r},\mathbf{r}^\prime;t)$ beschrieben. Damit die Gedächtnismatrix die 
Symmetrie-Bedingung \eqref{equation::D_050} erfüllt, muss die reaktive Matrix 
antisymmetrisch sein gemäß 
\begin{equation}
 K_{ik}(\mathbf{r},\mathbf{r}^\prime;t) \ = \ -K_{ki}(\mathbf{r}^\prime,\mathbf{r};t) \ .
 \label{equation::D_210}
\end{equation}
Andererseits beschreibt der zweite Term von \eqref{equation::D_200} die dissipativen 
Beiträge und hat dieselben Eigenschaften wie zuvor. Setzen wir nun die Gedächtnismatrix 
\eqref{equation::D_200} in die Bewegungsgleichung des GENERIC-Formalismus 
\eqref{equation::C_350} ein und führen die Integration über die Zeit aus, so erhalten 
wir eine Gleichung ähnlich wie \eqref{equation::D_120}. Jedoch wird im zweiten Term 
die Onsager-Matrix $M_{ik}(\mathbf{r},\mathbf{r}^\prime;t)$ ersetzt durch die Summe der 
Matrizen $K_{ik}(\mathbf{r},\mathbf{r}^\prime;t) + M_{ik}(\mathbf{r},\mathbf{r}^\prime;t)$. 
Auf diese Weise wird die ursprüngliche Bewegungsgleichung des GENERIC-Formalismus 
\cite{GO97A,GO97B} erweitert um einen zusätzlichen reaktiven Term.

Später in Abschnitt \ref{section::4E} werden wir zeigen, dass die fluktuierenden Kräfte 
$f_i(\mathbf{r},t)$ in der Bewegungsgleichung stochastisch und gaußisch sind, wobei die 
Korrelationen durch die Gedächtnismatrix $M_{ik}(\mathbf{r},t;\mathbf{r}^\prime,t^\prime)$ 
beschrieben werden. Wenn wir die einfache Näherungsformel $\eqref{equation::D_090}$ 
verwenden, dann stellt es sich heraus, dass die GENERIC-Bewegungsgleichung 
\eqref{equation::D_120} als Langevin-Gleichung einer stochastischen Theorie 
interpretiert werden kann. Das werden wir in Kapitel \ref{section::6} zeigen. 
Andererseits, wenn wir die komplexere Näherungsformel $\eqref{equation::D_200}$ 
verwenden, dann stellt es sich heraus, dass die reaktiven Terme mit der Vorzeichenfunktion 
$\varepsilon( t - t^\prime )$ \emph{nicht kompatibel} zu einer gewöhnlichen stochastischen 
Theorie sind. Aus diesem Grunde betrachten wie in dieser Arbeit nur gewöhnliche einfache 
Flüssigkeiten, welche durch die Gedächtnismatrix \eqref{equation::D_090} beschrieben 
werden, in der die reaktiven Terme näherungsweise null sind und nur die dissipativen 
Terme berücksichtigt werden. Dennoch ist es für zukünftige Arbeiten interessant, 
den GENERIC-Formalismus zu erweitern, um auch reaktive Terme zu berücksichtigen. 
Damit lassen sich dann komplexerer Flüssigkeiten beschreiben wie die nematischen 
Flüssigkristalle, welche von Forster untersucht wurden \cite{Fo74a,Fo74b,Fo75}.

\subsection{Hydrodynamische Gleichungen für eine \break normale Flüssigkeit ohne Dissipation}
\label{section::4C}
Nachdem wir die hydrodynamischen Gleichungen mit \eqref{equation::D_120} in ihrer allgemeinen 
Form hergeleitet haben, wollen wir nun ihre spezielle Form für eine normale Flüssigkeit finden. 
Die relevanten Variablen $\rho(\mathbf{r})$, $\mathbf{j}(\mathbf{r})$ und $\varepsilon(\mathbf{r})$ 
haben wir bereits mit \eqref{equation::D_060}-\eqref{equation::D_080} ausgewählt. Zunächst 
vernachlässigen wir die Effekte der Dissipation und der Fluktutionen und betrachten nur den 
ersten Term auf der rechten Seite von \eqref{equation::D_120}. Daher berechnen wir im Folgenden 
die elementaren Poisson-Klammern und die Poisson-Matrix \eqref{equation::C_960}. Dazu müssen 
wir die Kommutatoren der relevanten Variablen berechnen und dann die Erwartungswerte bilden. 
Für eine erste einfache Berechnung betrachten wir zunächst ein nichtrelativistisches 
Vielteilchensystem mit nur einer Teilchensorte und ohne Wechselwirkung in zweiter Quantisierung 
und erhalten damit Ergebnisse für die Poisson-Klammern, welche schon die richtige Form haben. 
Später erweitern wir auf mehrere Teilchensorten, fügen die Wechselwirkung mit einem lokalen 
Feld wie z.B.\ dem elektromagnetischen Feld hinzu und argumentieren, dass sich dadurch an 
den Ergebnissen nichts mehr ändert. Es seien also $\psi(\mathbf{r})$ und $\psi^+(\mathbf{r})$ 
die Feldoperatoren von Bosonen oder Fermionen, und die zugehörigen Kommutatoren $(-)$ oder 
Antikommutatoren $(+)$ seien definiert durch
\begin{eqnarray}
 { [ \psi(\mathbf{r}), \psi(\mathbf{r}) ]_\mp } &=& 0 \ ,
 \label{equation::D_220} \\
 { [ \psi(\mathbf{r}), \psi^+(\mathbf{r}) ]_\mp } &=& i \hbar 
 \, \delta( \mathbf{r} - \mathbf{r}^\prime ) \ ,
 \label{equation::D_230} \\
 { [ \psi(\mathbf{r}), \psi(\mathbf{r}) ]_\mp } &=& 0 \ .
 \label{equation::D_240}
\end{eqnarray}
Die relevanten Variablen schreiben wir dann in der Form
\begin{eqnarray}
 \rho(\mathbf{r}) &=& m \, \psi^+(\mathbf{r}) \, \psi(\mathbf{r}) \ ,
 \label{equation::D_250} \\
 \mathbf{j}(\mathbf{r}) &=& \frac{ \hbar }{ 2i } \, \left\{ \psi^+(\mathbf{r}) 
 \, [ \nabla \psi(\mathbf{r}) ] - [ \nabla \psi^+(\mathbf{r}) ] \, \psi(\mathbf{r}) \right\} 
 \ , \hspace{10mm} 
 \label{equation::D_260} \\
 \varepsilon(\mathbf{r}) &=& \frac{ \hbar^2 }{ 2 m } \, [ \nabla \psi^+(\mathbf{r}) ] 
 \cdot [ \nabla \psi(\mathbf{r}) ] \ .
 \label{equation::D_270}
\end{eqnarray}
Weiterhin benötigen wir den Spannungstensor $\Pi_{ik}(\mathbf{r})$ und die Energiestromdichte 
$\mathbf{j}_E(\mathbf{r})$, definiert durch
\begin{widetext}
\begin{eqnarray}
 \Pi_{ik}({r}) &=& \frac{ \hbar^2 }{ 4 m } \left\{ 
 [ \partial_i \psi^+(\mathbf{r}) ] \, [ \partial_k \psi(\mathbf{r}) ]
 + [ \partial_k \psi^+(\mathbf{r}) ] \, [ \partial_i \psi(\mathbf{r}) ]
 - \psi^+(\mathbf{r}) \, [ \partial_i \partial_k \psi(\mathbf{r}) ]
 - [ \partial_i \partial_k \psi^+(\mathbf{r}) ] \, \psi(\mathbf{r}) \right\} \ ,
 \label{equation::D_280} \\
 j_{E,i}(\mathbf{r}) &=& \frac{ \hbar^3 }{ 8 m^2i } \, \left\{ 
 [ \partial_k \psi^+(\mathbf{r}) ] \, [ \partial_i \partial_k \psi(\mathbf{r}) ]
 + [ \partial_i \psi^+(\mathbf{r}) ] \, [ \partial_k \partial_k \psi(\mathbf{r}) ]
 - [ \partial_i \partial_k \psi^+(\mathbf{r}) ] \, [ \partial_k \psi(\mathbf{r}) ]
 - [ \partial_k \partial_k \psi^+(\mathbf{r}) ] \, [ \partial_i \psi(\mathbf{r}) ] \right\} 
 \ . \hspace{10mm}
 \label{equation::D_290}
\end{eqnarray}
Wir berechnen zuerst die Kommutatoren der relevanten Varablen 
\eqref{equation::D_250}-\eqref{equation::D_270} in allen Kombinationen mit Hilfe der elementaren 
Kommutatoren oder Antikommutatoren \eqref{equation::D_220}-\eqref{equation::D_240}. Es ist 
hierbei unerheblich, ob die Teilchen Bosonen oder Fermionen sind. Die Ergebnisse sind die 
gleichen. Wir berechnen danach die Erwartungswerte gemäß \eqref{equation::C_960}, multiplizieren 
mit einem Faktor $(i\hbar)^{-1}$ und erhalten dann die elementaren Poisson-Klammern
\begin{eqnarray}
 \{ \rho(\mathbf{r},t), \rho(\mathbf{r}^\prime,t) \} &=& 0 \ ,
 \label{equation::D_300} \\
 \{ \mathbf{j}(\mathbf{r},t), \rho(\mathbf{r}^\prime,t) \} &=& - \, \rho(\mathbf{r},t) \, \nabla 
 \, \delta( \mathbf{r} - \mathbf{r}^\prime ) \ ,
 \label{equation::D_310} \\
 \{ j_i(\mathbf{r},t), j_k(\mathbf{r}^\prime,t) \} &=& - \, \left[ j_k(\mathbf{r},t) \, \partial_i
 + \partial_k \, j_i(\mathbf{r},t) \right] \, \delta( \mathbf{r} - \mathbf{r}^\prime ) \ ,
 \label{equation::D_320} \\
 \{ \rho(\mathbf{r},t), \varepsilon(\mathbf{r}^\prime,t) \} &=&  - \, \nabla \cdot \mathbf{j}(\mathbf{r},t)
 \, \delta( \mathbf{r} - \mathbf{r}^\prime ) \ ,
 \label{equation::D_330} \\
 \{ j_i(\mathbf{r},t), \varepsilon(\mathbf{r}^\prime,t) \} &=& - \, \left[ 
 \varepsilon(\mathbf{r},t) \, \partial_i + \partial_k \, \Pi_{ik}(\mathbf{r},t) 
 - \frac{ \hbar^2 }{ 4 m^2 } \, \partial_k \, ( \partial_k \rho (\mathbf{r},t) ) \, \partial_i 
 \right] \, \delta( \mathbf{r} - \mathbf{r}^\prime ) \ ,
 \label{equation::D_340} \\
 \{ \varepsilon(\mathbf{r},t), \varepsilon(\mathbf{r}^\prime,t) \} &=&  - \, 
 \left[ \nabla \cdot \mathbf{j}_E(\mathbf{r},t) + \mathbf{j}_E(\mathbf{r},t) \cdot \nabla \right]
 \, \delta( \mathbf{r} - \mathbf{r}^\prime ) \ .
 \label{equation::D_350}
\end{eqnarray}
\end{widetext}
Die physikalischen Größen auf den rechten Seiten sind alle Erwartungswerte der Operatoren. 
Wir verwenden hier die gleiche Notation und nehmen an, dass aus dem Kontext klar wird, was 
gemeint ist. Die räumlichen Differentialoperatoren $\partial_i$ und $\nabla$ wirken auf alle 
Ortsvariablen $\mathbf{r}$, die rechts von ihnen stehen, auch auf die Ortsvariable $\mathbf{r}$ 
in der Delta-Funktion. Eine Ausnahme ist der letzte Term in \eqref{equation::D_340}. Hier wird 
ein Differentialoperator explizit durch runde Klammern auf die nachfolgende Dichte beschränkt. 
Dieser Term hat einen expliziten Faktor $\hbar^2$ und ist somit eine Quantenkorrektur. Folglich 
dürfen wir für eine klassische Flüssigkeit den letzten Term in \eqref{equation::D_340} 
weglassen.

Nachdem wir die Poisson-Klammern \eqref{equation::D_300}-\eqref{equation::D_350} für ein 
nichtrelativistisches Vielteilchensystem mit nur einer Teilchensorte und ohne Wechselwirkung 
berechnet haben, stellen wir fest, dass sie sehr robust sind und sich nicht ändern, wenn 
allgemeinere und komplexere mikroskopische Theorien mit mehreren Teilchensorten und mit 
Wechselwirkungen betrachtet werden. Mehrere Teilchensorten werden berücksichtigt, in dem 
man den Feldoperatoren $\psi^{\ }_a(\mathbf{r})$ und $\psi^+_a(\mathbf{r})$ einen Index 
$a$ hinzufügt, welcher die Teilchensorte abzählt. Die elementaren Vertauschungsregeln 
für diese Feldoperatoren \eqref{equation::D_220}-\eqref{equation::D_240} werden 
entsprechend erweitert. Nachdem man analoge Berechnungen durchgeführt hat, erhält man 
unverändert dieselben Poisson-Klammern \eqref{equation::D_300}-\eqref{equation::D_350}. 

In einem nächsten Schritt betrachten wir die Beiträge für das elektromagnetische Feld, 
zunächst ohne eine Wechselwirkung mit den Teilchen. Wir fügen zu der Energiedichte des 
Vielteilchensystems \eqref{equation::D_270} den Standard-Ausdruck für die Energiedichte 
des elektromagnetischen Feldes hinzu. Analog fügen wir zu der Energiestromdichte des 
Vielteilchensystems \eqref{equation::D_290} die Energiestromdichte des elektromagnetischen 
Feldes, also den Poynting-Vektor, hinzu. Weiterhin wird der Poynting-Vektor geteilt durch 
$c^2$ zu der Impulsdichte \eqref{equation::D_260} und der Maxwellsche Spannungstensor zu 
der Impulsstromdichte \eqref{equation::D_280} hinzu addiert. Anschließend werden die 
Poisson-Klammern erneut berechnet, in dem zusätzlich die Quanten-Kommutatoren für das 
elektrische und das magnetische Feld verwendet werden, welche wohl bekannt aus der 
Quantenelektrodynamik sind. Als Ergebnis erhalten wir wieder die Poisson-Klammern 
\eqref{equation::D_300}-\eqref{equation::D_350} ohne irgend eine Änderung. 

In einem weiteren Schritt geben wir den Teilchen eine Ladung $e$ und ein magnetisches 
Moment $\boldsymbol{\mu}$, um Wechselwirkungen zwischen den Teilchen und dem 
elektromagnetischen Feld zu berücksichtigen. Entsprechend fügen wir Wechselwirkungs-Beiträge 
zu den Dichten und den Stromdichten \eqref{equation::D_250}-\eqref{equation::D_290} hinzu. 
Als Ergebnis finden wir, dass die Poisson-Klammern auch durch diese zusätzlichen 
Beiträgen nicht verändert werden. Wir stellen fest: Mit den bisher betrachteten 
Beiträgen lässt sich die Materie durch ein ziemlich fundamentales mikroskopisches 
Modell beschreiben, in dem die Teilchen mit den Atomkernen und den umgebenden Elektronen 
identifiziert werden, welche alle mit dem elektromagnetische Feld wechselwirken.

Als mikroskopisches Modell für die Teilchen und ihre Wechselwirkungen kann allgemeiner 
eine relativistische Quantenfeldtheorie betrachtet werden, welche auf Klein-Gordon-Feldern 
und Dirac-Feldern basiert. Die elektromagnetischen Felder für die Wechselwirkungen 
können allgemeiner ersetzt werden durch die nichtabelschen Eichfelder für die 
elektroschwache und die starke Wechselwirkung. Auf diese Weise ist es möglich, die 
Poisson-Klammern \eqref{equation::D_300}-\eqref{equation::D_350} sogar für die 
fundamentalste Theorie der Materie, das Standard-Modell der Elementarteilchen und 
ihrer Wechselwirkungen \cite{S13} zu beweisen.

Wir schließen aus unseren Überlegungen, dass die Poisson-Klammern 
\eqref{equation::D_300}-\eqref{equation::D_350} sehr allgemein sind und für alle 
mikroskopischen Modelle der Materie gelten sollten, die eine normal Flüssigkeit 
beschreiben. Sie sind gültig für relativistische und nichtrelativistische Modelle, 
für klassische Theorien und Quantentheorien. Allerdings sind die Berechnungen, um 
die Poisson-Klammern für ein spezielles mikroskopisches Modell zu beweisen, 
meistens ziemlich kompliziert und aufwendig. 

Die Poisson-Klammern wurde ursprünglich von Dzyaloshinskii und Volovik \cite{DV80} 
verwendet, um die reversiblen Terme in den hydrodynamischen Gleichungen herzuleiten. 
Diese Autoren verwendeten Symmetrieargumente um die Poisson-Klammern zu bestimmen. 
Allerdings betrachteten Sie die Entropiedichte $\sigma(\mathbf{r},t)$ anstelle der 
Energiedichte $\varepsilon(\mathbf{r},t)$. Dennoch stimmen ihre Poisson-Klammern 
mit unseren überein, wie wir weiter unten zeigen werden.

Alternativ können die hydrodynamischen Gleichungen durch ein Variationsprinzip mit 
einer Lagrange-Funktion und phänomenologischen Variablen formuliert werden. Eine 
zugehörige Hamilton-Funktion kann man herleiten, welche die Poisson-Klammern für 
die hydrodynamischen Variablen liefert. Auf diese Wiese haben Enz und Turski \cite{ET79} 
zuerst die drei Poisson-Klammern \eqref{equation::D_300}-\eqref{equation::D_320} für 
die Massendichte $\rho(\mathbf{r},t)$ und die Impulsdichte $\mathbf{j}(\mathbf{r},t)$ 
erhalten. Van Saarloos \emph{et al.}\ \cite{SBM81} fanden die letzteren drei 
Poisson-Klammern \eqref{equation::D_330}-\eqref{equation::D_350} mit der Energiedichte 
$\varepsilon(\mathbf{r},t)$. Im Ergebnis stellen wir fest: Die Poisson-Klammern für 
die hydrodynamischen Variablen einer normalen Flüssigkeit sind seit langer Zeit 
bekannt. Es gibt eine allgemeine Übereinstimmung darüber, wie sie aussehen sollen. 

Wir setzen nun unsere Berechnungen fort. Die Erwartungswerte der Operatoren 
\eqref{equation::D_260}-\eqref{equation::D_290} kann man für eine klassischen 
Flüssigkeit einfach ausrechnen. Wir erhalten
\begin{eqnarray}
 \mathbf{j}(\mathbf{r},t) &=& \rho(\mathbf{r},t) \, \mathbf{v}(\mathbf{r},t) \ ,
 \label{equation::D_360} \\
 \varepsilon(\mathbf{r},t) &=& u(\mathbf{r},t) 
 + \frac{ \mathbf{j}(\mathbf{r},t)^2 }{ 2\, \rho(\mathbf{r},t) } \ ,
 \label{equation::D_370} \\
 \Pi_{ik}(\mathbf{r},t) &=& p(\mathbf{r},t) \, \delta_{ik} 
 + \rho(\mathbf{r},t) \, v_i(\mathbf{r},t) \, v_k(\mathbf{r},t) \ , \hspace{10mm}
 \label{equation::D_380} \\
 \mathbf{j}_E(\mathbf{r},t) &=& [ \varepsilon(\mathbf{r},t) 
 + p(\mathbf{r},t) ] \, \mathbf{v}(\mathbf{r},t) \ .
 \label{equation::D_390}
\end{eqnarray}
Hierbei ist $u(\mathbf{r},t)$ die aus der Thermodynamik bekannte \emph{innere Energiedichte},
$p(\mathbf{r},t)$ der \emph{Druck} und $\mathbf{v}(\mathbf{r},t)$ das Geschwindigkeitsfeld.
Die spezielle Form dieser physikalischen Größen mit dem Geschwindigkeitsfeld gibt die 
\emph{Galilei-Invarianz} einer normalen Flüssigkeit wieder.

Die Formel \eqref{equation::D_370} zeigt, dass sich die Energiedichte $\varepsilon(\mathbf{r},t)$ 
in zwei Anteile aufspaltet, die innere Energiedichte $u(\mathbf{r},t)$ und einen kinetischen 
Anteil. Dieser Zusammenhang bietet nun die Möglichkeit, aus 
\eqref{equation::D_330}-\eqref{equation::D_350} auch die entsprechenden elementaren 
Poisson-Klammern für die innere Energiedichte $u(\mathbf{r},t)$ herzuleiten. Wir finden
\begin{eqnarray}
 \{ \rho(\mathbf{r},t), u(\mathbf{r}^\prime,t) \} &=& 0 \ ,
 \label{equation::D_400} \\
 \{ \mathbf{j}(\mathbf{r},t), u(\mathbf{r}^\prime,t) \} &=& - \, \left[ 
 u(\mathbf{r},t) \, \nabla + \nabla \, p(\mathbf{r},t) \right] \nonumber \\
 &&\times \, \delta( \mathbf{r} - \mathbf{r}^\prime ) \ ,
 \label{equation::D_410} \\
 \{ u(\mathbf{r},t), u(\mathbf{r}^\prime,t) \} &=& 0 \ .
 \label{equation::D_420}
\end{eqnarray}
Die neuen Poisson-Klammern für die innere Energiedichte $u(\mathbf{r},t)$ haben 
offensichtlich eine deutlich einfachere Struktur als jene für die gesamte Energiedichte 
$\varepsilon(\mathbf{r},t)$, weil sie nicht vom Geschwindigkeitsfeld 
$\mathbf{v}(\mathbf{r},t)$ abhängen dürfen.

Die Entropie $S(t)$ ist nach der Definition \eqref{equation::B_180} oder \eqref{equation::B_210} 
eine globale Größe. In der Hydrodynamik wird jedoch auf lokaler Ebene ein thermisches 
Gleichgewicht angenommen. Daher sollte die Entropie auch lokal als Entropiedichte 
$\sigma(\mathbf{r},t)$ existieren, so dass die globale Entropie gegeben ist durch das Integral 
$S(t) = \int d^dr \, \sigma(\mathbf{r},t)$. Es wird eine lokale Differentialgleichung für 
die Entropiedichte geben, die einer Kontinuitätsgleichung ähnlich ist, jedoch mit einer 
Erweiterung. Auf der rechten Seite wird ein Quellterm stehen mit Beiträgen von der 
Dissipation und den Fluktuationen, der eine ähnliche Struktur hat wie die Terme auf der 
rechten Seite der globalen Entropiegleichung \eqref{equation::D_130}. Um die reversiblen 
Anteile der lokalen Entropiegleichung zu finden, benötigen wir die Poisson-Klammern mit 
der Entropiedicht $\sigma(\mathbf{r},t)$. Diese werden wir im folgenden herleiten.

Aus der Thermodynamik ist bekannt, dass die innere Energiedichte $u(\mathbf{r},t)$ von der 
Entropiedichte $\sigma(\mathbf{r},t)$ und der Massendichte $\rho(\mathbf{r},t)$ abhängt. 
Es gilt der lokale Zusammenhang
\begin{equation}
 u(\mathbf{r},t) = u( \sigma(\mathbf{r},t), \rho(\mathbf{r},t) )
 \label{equation::D_430}
\end{equation}
und die thermodynamische Relation
\begin{equation}
 du(\mathbf{r},t) = T(\mathbf{r},t) \, d\sigma(\mathbf{r},t) 
 + m^{-1} \mu(\mathbf{r},t) \, d\rho(\mathbf{r},t) \ .
 \label{equation::D_440}
\end{equation}
Hierbei sind $T(\mathbf{r},t)$ die Temperatur und $\mu(\mathbf{r},t)$ das chemische Potential. 
Aus letzterer Gleichung lässt sich ein Zusammenhang für die Poisson-Klammern herleiten, und zwar
\begin{eqnarray}
 \{ x_i(\mathbf{r},t), u(\mathbf{r}^\prime,t) \} &=& T(\mathbf{r}^\prime,t) 
 \, \{ x_i(\mathbf{r},t), \sigma(\mathbf{r}^\prime,t) \} \nonumber \\
 &&+ m^{-1} \mu(\mathbf{r}^\prime,t) \, \{ x_i(\mathbf{r},t), \rho(\mathbf{r}^\prime,t) \} \ .
 \nonumber \\
 \label{equation::D_450}
\end{eqnarray}
Für $x_i(\mathbf{r},t)$ kann jede beliebige unserer relevanten Variablen eingesetzt 
werden. Wir finden nun, dass die elementaren Poisson-Klammern mit der Entropie-Dichte 
$\sigma(\mathbf{r},t)$ definiert werden durch
\begin{eqnarray}
 \{ \rho(\mathbf{r},t), \sigma(\mathbf{r}^\prime,t) \} &=& 0 \ ,
 \label{equation::D_460} \\
 \{ \mathbf{j}(\mathbf{r},t), \sigma(\mathbf{r}^\prime,t) \} &=& - \,
 \sigma(\mathbf{r},t) \, \nabla \, \delta( \mathbf{r} - \mathbf{r}^\prime ) \ ,
 \label{equation::D_470} \\
 \{ \sigma(\mathbf{r},t), \sigma(\mathbf{r}^\prime,t) \} &=& 0 \ .
 \label{equation::D_480}
\end{eqnarray}
Mit Hilfe von \eqref{equation::D_450} kann man zeigen, dass diese Poisson-Klammern für die 
Entropiedichte $\sigma(\mathbf{r},t)$ kompatibel sind mit jenen für die innere Energiedichte 
$u(\mathbf{r},t)$ und die Massendichte $\rho(\mathbf{r},t)$.

Die Poisson-Klammern für die verschiedenen relevanten Variablen sind nun komplett. Wir 
stellen fest, dass sie mit jenen von Dzyaloshinskii und Volovik \cite{DV80} übereinstimmen. 
Die reversiblen Anteile der hydrodynamischen Gleichungen \eqref{equation::C_880} lassen 
sich daher schreiben in der Form
\begin{equation}
 \partial_t x_i(\mathbf{r},t) = \{ x_i(\mathbf{r},t) , E[\sigma(t),\mathbf{j}(t),\rho(t)] \} \ .
 \label{equation::D_490}
\end{equation}
Da wir alle erforderlichen elementaren Poisson-Klammern gefunden haben, können wir die Energie 
$E[\sigma(t),\mathbf{j}(t),\rho(t)]$ als Funktional in den natürlichen Variablen Entropiedichte 
$\sigma(\mathbf{r},t)$, Impulsdichte $\mathbf{j}(\mathbf{r},t)$ und Massendichte 
$\rho(\mathbf{r},t)$ verwenden. Die zugehörigen Funktional-Ableitungen sind 
\begin{eqnarray}
 \frac{ \delta E[\sigma(t),\mathbf{j}(t),\rho(t)] }{ \delta \sigma(\mathbf{r},t) }
 &=& T(\mathbf{r},t) \ ,
 \label{equation::D_500} \\
 \frac{ \delta E[\sigma(t),\mathbf{j}(t),\rho(t)] }{ \delta \mathbf{j}(\mathbf{r},t) }
 &=& \mathbf{v}(\mathbf{r},t) \ ,
 \label{equation::D_510} \\
 \frac{ \delta E[\sigma(t),\mathbf{j}(t),\rho(t)] }{ \delta \rho(\mathbf{r},t) }
 &=& m^{-1} \, \mu(\mathbf{r},t) \ .
 \label{equation::D_520}
\end{eqnarray}
Setzten wir nun in \eqref{equation::D_490} für $x_i(\mathbf{r},t)$ alle unsere relevanten 
Variablen ein, so erhalten wir die hydrodynamischen Gleichungen
\begin{eqnarray}
 \partial_t \, \rho(\mathbf{r},t) &=& - \, \nabla \cdot \mathbf{j}(\mathbf{r},t) \ ,
 \label{equation::D_530} \\
 \partial_t \, j_i(\mathbf{r},t) &=& - \, \partial_k \, \Pi_{ik}(\mathbf{r},t) \ ,
 \label{equation::D_540} \\
 \partial_t \, \sigma(\mathbf{r},t) &=& - \, \nabla \cdot \mathbf{q}(\mathbf{r},t) 
 + \frac{ R(\mathbf{r},t) }{ T(\mathbf{r},t) } \ ,
 \label{equation::D_550} \\
 \partial_t \, \varepsilon(\mathbf{r},t) &=& - \, \nabla \cdot \mathbf{j}_E(\mathbf{r},t) \ .
 \label{equation::D_560}
\end{eqnarray}
Die ersten drei Gleichungen bekommt man durch Verwendung der Poisson-Klammern 
\eqref{equation::D_300}-\eqref{equation::D_320} und \eqref{equation::D_460}-\eqref{equation::D_480} 
zusammen mit den Funktional-Ableitungen \eqref{equation::D_500}-\eqref{equation::D_520}. 
Die vierte Gleichung für die Energiedichte folgt direkt aus der Poisson-Klammer \eqref{equation::D_350}.
Die Stromdichten auf der rechten Seite sind gegeben durch die Formeln \eqref{equation::D_360},
\eqref{equation::D_380} und \eqref{equation::D_390}. Zusätzlich finden wir die 
Entropiestromdichte
\begin{equation}
 \mathbf{q}(\mathbf{r},t) = \sigma(\mathbf{r},t) \, \mathbf{v}(\mathbf{r},t) \ .
 \label{equation::D_570}
\end{equation}
Die hydrodynamischen Gleichungen \eqref{equation::D_530}-\eqref{equation::D_560} 
wurden hier für eine ideale Flüssigkeit hergeleitet. Sie sind jedoch in dieser 
Form unverändert gültig auch für eine reale Flüssigkeit, wenn die Effekte von 
Dissipation und Fluktuationen hinzugefügt werden. Die erforderlichen Änderungen 
treten in den Stromdichten in Form von zusätzlichen Beiträgen auf, wie wir in 
den nachfolgenden Abschnitten sehen werden. 

Auf der rechten Seite der Entropie-Gleichung \eqref{equation::D_550} haben wir für spätere 
Betrachtungen einen Quellterm hinzugefügt. Dieser ist hier jedoch null, und wir haben  
\begin{equation}
 R(\mathbf{r},t) = 0 \ .
 \label{equation::D_580}
\end{equation}
Die hydrodynamischen Gleichungen \eqref{equation::D_530}-\eqref{equation::D_560} 
für unsere normale Flüssigkeit haben die aus elementaren Lehrbüchern bekannte Form 
\cite{LL06}. Sie enthalten bisher jedoch nur die reversiblen Terme. Effekte von 
Dissipation und Fluktuation wurden bisher nicht berücksichtigt.

\subsection{Hydrodynamische Gleichungen für eine \break normale Flüssigkeit mit Dissipation}
\label{section::4D}
Nun fügen wir die Terme der Dissipation hinzu, lassen die Fluktuationen jedoch noch weg.
Die Effekte der Dissipation sind im zweiten Term der hydrodynamischen Gleichungen 
\eqref{equation::D_120} enthalten. Ihre Stärke wird beschrieben durch die Onsager-Matrix 
$M_{ik}(\mathbf{r},\mathbf{r}^\prime;t)$. In der Hydrodynamik werden nur Phänomene auf 
großen Längenskalen betrachtet. Daher approximieren wir die räumliche Struktur durch eine 
Delta-Funktion. Die relevanten Variablen, die wir betrachten, sind alles Dichten von 
Erhaltungsgrößen. Aus diesem Grunde fügen wir noch zwei räumliche Differentialoperatoren 
ein. Wir machen also für die Onsager-Matrix den Ansatz
\begin{equation}
 M_{ik}(\mathbf{r},\mathbf{r}^\prime;t) = - \partial_m \, N_{im,kn}(x(\mathbf{r},t)) \, \partial_n 
 \, \delta( \mathbf{r} - \mathbf{r}^\prime ) \ .
 \label{equation::D_590}
\end{equation}
Wir motivieren die zwei räumlichen Differentialoperatoren mit der Darstellung der 
Gedächtnismatrix in der Form \eqref{equation::C_720} zusammen mit \eqref{equation::C_730}.
Der Ansatz ist lokal. Folglich hängt die Marix $N_{im,kn}=N_{im,kn}(x(\mathbf{r},t))$ lokal 
von den relevanten Variablen $x_i(\mathbf{r},t)$ ab. Setzen wir dies in \eqref{equation::D_120} 
ein, so bekommen wir die hydrodynamischen Gleichungen in der Form
\begin{eqnarray}
 \partial_t \, x_i(\mathbf{r},t) &=& \{ x_i(\mathbf{r},t), E[x(t)] \} \nonumber \\
 &&- \sum_{kn} \partial_m \, N_{im,kn} \, \partial_n 
 \, \frac{ \delta S[x(t)] }{ \delta x_k(\mathbf{r},t) } \nonumber \\
 &&+ \, f_i(\mathbf{r},t) \ .
 \label{equation::D_600}
\end{eqnarray}
Wir benötigen als nächstes die Funktional-Ableitungen der Entropie $S[x(t)]$. Nach der 
Thermodynamik sind die natürlichen Variablen der Entropie die Energiedichte 
$\varepsilon(\mathbf{r},t)$, die Impulsdichte $\mathbf{j}(\mathbf{r},t)$ und die 
Massendichte $\rho(\mathbf{r},t)$. Wir identifizieren daher
\begin{equation}
 S[x(t)] = S[\varepsilon(t), \mathbf{j}(t), \rho(t)]
 \label{equation::D_610}
\end{equation}
und finden die Funktional-Ableitungen
\begin{eqnarray}
 \frac{ \delta S[\varepsilon(t),\mathbf{j}(t),\rho(t)] }{ \delta \varepsilon(\mathbf{r},t) }
 &=& \frac{ 1 }{ T(\mathbf{r},t) } \ ,
 \label{equation::D_620} \\
 \frac{ \delta S[\varepsilon(t),\mathbf{j}(t),\rho(t)] }{ \delta \mathbf{j}(\mathbf{r},t) }
 &=& - \, \frac{ \mathbf{v}(\mathbf{r},t) }{ T(\mathbf{r},t) }\ ,
 \label{equation::D_630} \\
 \frac{ \delta S[\varepsilon(t),\mathbf{j}(t),\rho(t)] }{ \delta \rho(\mathbf{r},t) }
 &=& -\, \frac{ 1 }{ m } \, \frac{ \mu(\mathbf{r},t) }{ T(\mathbf{r},t) }\ .
 \label{equation::D_640}
\end{eqnarray}
Da die Funktional-Ableitungen in \eqref{equation::D_600} nur in Form von Gradienten vorkommen, 
können wir sie umformen gemäß
\begin{eqnarray}
 \partial_n \frac{ \delta S }{ \delta \varepsilon } &=& \partial_n \frac{ 1 }{ T } 
 = - \frac{ 1 }{ T^2 } \partial_n T 
 = - \frac{ 1 }{ T^2 } \partial_n \frac{ \delta E }{ \delta \sigma } \ ,
 \label{equation::D_650} \\
 \partial_n \frac{ \delta S }{ \delta j_i } &=& - \partial_n \frac{ v_i }{ T }
 = - \frac{ 1 }{ T } \partial_n v_i + \frac{ v_i }{ T^2 } \partial_n T \nonumber \\
 &=& - \frac{ 1 }{ T } \partial_n \frac{ \delta E }{ \delta j_i } 
 + \frac{ v_i }{ T^2 } \partial_n \frac{ \delta E }{ \delta \sigma } \ ,
 \label{equation::D_660} \\
 \partial_n \frac{ \delta S }{ \delta \rho } &=& - \partial_n \frac{ \mu }{ m T }
 = - \frac{ 1 }{ m T } \partial_n \mu + \frac{ \mu }{ m T^2 } \partial_n T \nonumber \\
 &=& - \frac{ 1 }{ T } \partial_n \frac{ \delta E }{ \delta \rho } 
 + \frac{ \mu }{ m T^2 } \partial_n \frac{ \delta E }{ \delta \sigma } \ .
 \label{equation::D_670}
\end{eqnarray}
Folglich lassen sich auch die hydrodynamischen Gleichungen \eqref{equation::D_600} umformen. 
Mit einer neuen Matrix $\Lambda_{im,kn}$ bekommen wir dann 
\begin{eqnarray}
 \partial_t \, x_i(\mathbf{r},t) &=& \{ x_i(\mathbf{r},t), E[x(t)] \} \nonumber \\
 &&+ \, \sum_{kn} \partial_m \, \Lambda_{im,kn} \, \partial_n 
 \, \frac{ \delta E[x(t)] }{ \delta x_k(\mathbf{r},t) } \nonumber \\
 &&+ \, \delta_{i\sigma} \frac{ R(\mathbf{r},t) }{ T(\mathbf{r},t) } \, + \, f_i(\mathbf{r},t) \ .
 \label{equation::D_680}
\end{eqnarray}
Die relevanten Variablen $x_i(\mathbf{r},t)$ werden hier identifiziert durch die natürlichen 
Variablen der Energie $E[x(t)]$. Dies sind die Entropiedichte $\sigma(\mathbf{r},t)$, die 
Impulsdichte $\mathbf{j}(\mathbf{r},t)$ und die Massendichte $\rho(\mathbf{r},t)$. Da hier 
die Entropiegleichung mit enthalten ist, wurde der dritten Zeile ein Term mit einer 
Entropie-Produktionsrate hinzugefügt. Diesen Term berechnet man aus dem quadratischen 
Term der Entropiegleichung \eqref{equation::D_130}. Wir finden
\begin{equation}
 R(\mathbf{r},t) = \sum_{im,kn} \left( \partial_m 
 \frac{ \delta E[x(t)] }{ \delta x_i(\mathbf{r},t) } \right)
 \Lambda_{im,kn} \left( \partial_n 
 \frac{ \delta E[x(t)] }{ \delta x_k(\mathbf{r},t) } \right) \ .
 \label{equation::D_690}
\end{equation}
Weil $R(\mathbf{r},t)$ eine Energiedichte pro Zeiteinheit ist, handelt es sich hier um 
die durch Reibung in der Flüssigkeit erzeugte Wärme. Wir stellen fest, dass die hydrodynamischen 
Gleichungen \eqref{equation::D_680} zusammen mit der Formel für die erzeugte Reibungswärme 
\eqref{equation::D_690} mit den hydrodynamischen Gleichungen in der allgemeine Form von 
Dzyaloshinskii und Volovik \cite{DV80} übereinstimmen.

In den allgemeinen hydrodynamischen Gleichungen \eqref{equation::D_680} zusammen mit der 
Formel für die erzeugte Wärme \eqref{equation::D_690} sind die spezifischen Eigenschaften der 
Flüssigkeit, welche Effekte der Dissipation und Dämpfungen betreffen, zusammengefasst in 
Elementen der Matrix $\Lambda_{im,kn}$. Im Prinzip kann man diese Matrix aus der mikroskopischen 
Theorie der Flüssigkeit berechnen. Dazu berechnet man zunächst die Gedächtnismatrix 
$M_{ik}(\mathbf{r},t;\mathbf{r}^\prime,t^\prime)$ mit der Formel \eqref{equation::D_030}. 
Im zweiten Schritt bestimmt man die Onsager-Matrix $M_{ik}(\mathbf{r},\mathbf{r}^\prime;t)$, 
indem man Gedächtniseffekte vernachlässigt mit dem Ansatz \eqref{equation::D_090} oder mit 
der Umkehrformel
\begin{equation}
 M_{ik}(\mathbf{r},\mathbf{r}^\prime;t) = \int_{t_0}^t dt^\prime 
 \ M_{ik}(\mathbf{r},t;\mathbf{r}^\prime,t^\prime) \ .
 \label{equation::D_700}
\end{equation}
Im dritten Schritt berechnet man die Matrix $N_{im,kn}$, indem man die räumlich nichtlokalen 
Effekte vernachlässigt und die Formel \eqref{equation::D_590} invertiert. Schließlich bekommt 
man die Matrix $\Lambda_{im,kn}$, indem man die Energiedichte $\varepsilon(\mathbf{r},t)$ 
in den ursprünglichen relevanten Variablen austauscht durch die Entropiedichte 
$\sigma(\mathbf{r},t)$.

Die Berechnung ist in der Theorie möglich, in der Praxis jedoch viel zu kompliziert und daher 
nicht durchführbar. In der Praxis wird ein anderer Weg beschritten. Mit Argumenten zur 
Symmetrie des Systems wird die Anzahl der von Null verschiedenen Matrixelemente $\Lambda_{im,kn}$ 
stark reduziert, so dass nur ganz wenige Parameter übrigbleiben. Diese Parameter werden 
am Ende durch Vergleich mit dem Experiment bestimmt. In diesem Sinne wird die Hydrodynamik zu 
einer phänomenologischen Theorie.

Für die Massendichte $\rho(\mathbf{r},t)$ gilt die Kontinuitätsgleichung \eqref{equation::D_530} 
bereits exakt auf mikroskopischer Ebene. Man kann sie für die Operatorausdrücke 
\eqref{equation::D_250} und \eqref{equation::D_260} in der zweiten Quantisierung leicht 
beweisen. Daher wird die Massendichte $\rho(\mathbf{r},t)$ von dissipativen Effekten nicht 
betroffen. Es verbleiben die Impulsdichte $\mathbf{j}(\mathbf{r},t)$ und die Entropiedichte 
$\sigma(\mathbf{r},t)$. Wir nehmen an, dass die der Flüssigkeit zugrunde liegende mikroskopische 
Theorie invariant unter Raumspiegelungen ist. Unter dieser Transformation ändert die 
Impulsdichte das Vorzeichen, die Entropiedichte jedoch nicht. Folglich gibt es in der 
Onsager-Matrix keine Kopplung zwischen der Impulsdichte $\mathbf{j}(\mathbf{r},t)$ und der 
Entropiedichte $\sigma(\mathbf{r},t)$. Die Matrix $\Lambda_{im,kn}$ ist in den einzelnen 
relevanten Variablen also weitgehend diagonal.

Der zweite Term in der allgemeinen hydrodynamischen Gleichung \eqref{equation::D_680} hat 
offensichtlich die Form einer Divergenz von Stromdichten. Wir entnehmen daraus die dissipativen 
Beiträge zu den allgemeinen Stromdichten
\begin{equation}
 \Delta j_{im}(\mathbf{r},t) = - \sum_{kn} \Lambda_{im,kn} \, \partial_n 
 \, \frac{ \delta E[x(t)] }{ \delta x_k(\mathbf{r},t) } \ .
 \label{equation::D_710}
\end{equation}
Folglich wird sich die Struktur der speziellen hydrodynamischen Gleichungen 
\eqref{equation::D_530}-\eqref{equation::D_560} durch die Effekte der Dissipation 
nicht ändern. Nur die Entropie-Gleichung bekommt wegen der Erzeugung von Wärme und 
Entropie durch Reibung einen Zusatzterm. Berücksichtigen wir, dass die Matrix 
$\Lambda_{im,kn}$ weitgehend diagonal ist, so finden wir die dissipativen Stromdichten
\begin{eqnarray}
 \Delta j_k &=& 0 \ ,
 \label{equation::D_720} \\
 \Delta \Pi_{ik} &=& - \, \Lambda^{jj}_{ik,mn} \, \partial_n v_m \ ,
 \label{equation::D_730} \\
 \Delta q_k &=& - \, \Lambda^{\sigma\sigma}_{kn} \, \partial_n T \ ,
 \label{equation::D_740} \\
 \Delta j_{E,k} &=& T \Delta q_k + v_i \Delta \Pi_{ik} + m^{-1} \mu \Delta j_k 
 \label{equation::D_750}
\end{eqnarray}
und die erzeugte Wärme
\begin{equation}
 R = ( \partial_k v_i ) \, \Lambda^{jj}_{ik,mn} \, ( \partial_n v_m ) 
 + ( \partial_k T ) \, \Lambda^{\sigma\sigma}_{kn} \, ( \partial_n T ) \ .
 \label{equation::D_760}
\end{equation}
Zur Vereinfachung der Schreibweise lassen wir ab jetzt die Argumente $\mathbf{r}$ und $t$ 
weg. Nach der Summenkonvention wird automatisch über doppelt auftretende Indizes summiert.

Eine normale Flüssigkeit ist isotrop. Das bedeutet, sie ist invariant unter Rotationen und 
Raumspiegelungen. Das schränkt die Matrizen für die Dissipation weiter ein. Wir finden
\begin{eqnarray}
 \Lambda^{jj}_{ik,mn} &=& \eta \, \left( \delta_{im} \delta_{kn} + \delta_{in} \delta_{km} 
 - \frac{ 2 }{ d } \, \delta_{ik} \delta_{mn} \right) \nonumber \\
 &&+ \, \zeta \, \delta_{ik} \delta_{mn} \ ,
 \label{equation::D_770} \\
 \Lambda^{\sigma\sigma}_{kn} &=& \frac{ \varkappa }{ T } \, \delta_{kn} \ .
 \label{equation::D_780}
\end{eqnarray}
Von der gesamten dissipativen Matrix $\Lambda_{im,kn}$ verbleiben also gerade mal drei 
Parameter, die \emph{Scherviskosität} $\eta$, die \emph{Volumenviskosität} $\zeta$ und die 
\emph{Wärmeleitfähigkeit} $\varkappa$. Diese drei Parameter beschreiben die dissipativen 
Effekte in einer normalen Flüssigkeit vollständig. Sie werden üblicherweise phänomenologisch 
bestimmt durch Vergleiche mit Experimenten.

Wir setzen nun die Matrizen \eqref{equation::D_770} und \eqref{equation::D_780} ein. Damit  
finden wir den Zusatzbeitrag für den Spannungstensor
\begin{equation}
 \Delta \Pi_{ik} = - \eta \left( \partial_i v_k + \partial_k v_i - \frac{ 2 }{ d } 
 \, \delta_{ik} \, \partial_n v_n \right) - \zeta \, \delta_{ik} \, \partial_n v_n
 \label{equation::D_790}
\end{equation}
und für die Entropiestromdichte
\begin{equation}
 \Delta q_k = - \frac{ \varkappa }{ T } \, \partial_k T \ .
 \label{equation::D_800}
\end{equation}
Addieren wir die Zusatzbeiträge \eqref{equation::D_720}-\eqref{equation::D_750} zu den 
reversiblen Beiträgen \eqref{equation::D_360}, \eqref{equation::D_380}, \eqref{equation::D_570} 
und \eqref{equation::D_390}, so erhalten wir die gesamten Stromdichten
\begin{eqnarray}
 j_k &=& \rho \, v_k \ ,
 \label{equation::D_810} \\
 \Pi_{ik} &=& p \, \delta_{ik} + \rho \, v_i \, v_k + \Delta \Pi_{ik} \ ,
 \label{equation::D_820} \\
 q_k &=& \sigma \, v_k - \frac{ \varkappa }{ T } \, \partial_k T \ ,
 \label{equation::D_830} \\
 j_{E,k} &=& ( \varepsilon + p ) v_k - \varkappa \, \partial_k T + v_i \Delta \Pi_{ik} \ .
 \hspace{5mm}
 \label{equation::D_840}
\end{eqnarray}
Der zweite Term in der Energiestromdichte \eqref{equation::D_840} stellt die Wärmestromdichte 
dar. Dies erklärt die Bezeichnung \emph{Wärmeleitfähigkeit} für den Parameter $\varkappa$. 
Aus \eqref{equation::D_760} finden wir schließlich die erzeugte Wärmedichte
\begin{eqnarray}
 R &=& \eta \left[ ( \partial_i v_k ) ( \partial_i v_k ) + ( \partial_i v_k ) ( \partial_k v_i ) 
 - \frac{ 2 }{ d } \, ( \partial_i v_i ) ( \partial_k v_k ) \right] \nonumber \\
 &&+ \, \zeta \, ( \nabla \cdot \mathbf{v} )^2 + \frac{ \varkappa }{ T } \, ( \nabla T )^2 \ .
 \label{equation::D_850}
\end{eqnarray}

Wir fassen zusammen. Die hydrodynamischen Gleichungen für eine normale Flüssigkeit mit 
Dissipation sind gegeben durch \eqref{equation::D_530}-\eqref{equation::D_560}, zusammen 
mit den Stromdichten \eqref{equation::D_810}-\eqref{equation::D_840} und der erzeugten 
Wärmedichte \eqref{equation::D_850}. Sie stimmen überein mit den hydrodynamischen 
Gleichungen, die in den bekannten Lehrbüchern zu finden sind \cite{LL06}. Drei von 
diesen sind reine Kontinuitätsgleichungen. Das bedeutet, die zugehörigen physikalischen 
Größen Energie, Impuls und Massendichte sind Erhaltungsgrößen. Eine Ausnahme bildet die 
Gleichung für die Entropiedichte \eqref{equation::D_550}. Hier steht mit der erzeugten 
Wärmedichte $R$ ein Quellterm auf der rechten Seite. Da die erzeugte Wärmedichte 
\eqref{equation::D_850} eine quadratische Form hat und die Parameter $\eta$, $\zeta$ 
und $\varkappa$ positiv sind, ist sie immer größer gleich null. Folglich ist der 
Quellterm der Entropie immer größer gleich null. Diese Eigenschaft liefert den 
zweiten Hauptsatz der Thermodynamik, so dass die gesamte Entropie des Systems immer 
anwachsen muss.

\subsection{Hydrodynamische Fluktuationen}
\label{section::4E}
Die fluktuierenden Kräfte stellen den dritten Term in der allgemeinen hydrodynamischen 
Gleichung \eqref{equation::D_120} dar. Sie werden definiert durch die Formel 
\eqref{equation::C_610}. Definieren wir die quantenmechanischen Operatoren der 
fluktuierenden Kräfte im Heisenberg Bild durch
\begin{equation}
 \hat{f}_i(\mathbf{r},t) = i \, \mathsf{Q}(t_0) \, U(t_0,t) \, \mathsf{Q}(t) 
 \, \mathsf{L} \, \hat{a}_i(\mathbf{r}) \ ,
 \label{equation::D_860}
\end{equation}
so berechnen sich ihre Erwartungswerte mit
\begin{equation}
 f_i(\mathbf{r},t) = \mathrm{Sp}\{ \hat{\varrho}(t_0) \, \hat{f}_i(\mathbf{r},t) \}
 \label{equation::D_870} \ .
\end{equation}
Zur Unterscheidung zwischen Operatoren und Erwartungswerten kennzeichnen wir hier wieder 
die quantenmechanischen Operatoren mit einem Dach.

Wir nehmen an, dass durch den orthogonalen Projektionsoperator $\mathsf{Q}(t)$ alle physikalischen 
Freiheitsgrade entfernt werden, welche sich auf großen Längenskalen und großen Zeitskalen 
bewegen. Folglich werden sich die fluktuierenden Kräfte $f_i(\mathbf{r},t)$ auf kurze 
Längenskalen und kurze Zeitskalen beschränken und nahezu stochastisch verhalten. Wir 
nehmen weiterhin an, dass physikalische Zustände immer durch reine quantenmechanische 
Zustände beschrieben werden, so dass zur Anfangszeit $t_0$ die Dichtematrix gegeben 
ist durch $\hat{\varrho}(t_0)= \vert\Psi_0\rangle \langle\Psi_0\vert$. Da eine normal 
Flüssigkeit ein klassisches System ist, erwarten wir, dass die fluktuierenden Kräfte 
$\hat{f}_i(\mathbf{r},t)$ näherungsweise scharf definierte klassische Variablen sind. 
Das bedeutet, dass Erwartungswerte von Produkten von fluktuierenden Kräften näherungsweise 
faktorisieren. Für einen Erwartungswert mit zwei fluktuierenden Kräften erhalten wir also 
näherungsweise
\begin{eqnarray}
 &&\mathrm{Sp}\{ \hat{\varrho}(t_0) \, \hat{f}_i(\mathbf{r},t)
 \, \hat{f}_k(\mathbf{r}^\prime,t^\prime) \} \, \approx \nonumber\\
 &&\approx \, \mathrm{Sp}\{ \hat{\varrho}(t_0) \, \hat{f}_i(\mathbf{r},t) \}
 \, \mathrm{Sp}\{ \hat{\varrho}(t_0) \, \hat{f}_k(\mathbf{r}^\prime,t^\prime) \} \nonumber \\
 &&= \, f_i(\mathbf{r},t) \, f_k(\mathbf{r}^\prime,t^\prime) \ .
 \label{equation::D_880}
\end{eqnarray}
Wir stellen uns nun vor, dass das Experiment mit der normalen Flüssigkeit mehrfach 
durchgeführt wird. Die Dichtematrix $\hat{\varrho}^{(n)}(t_0)$ ist dann für jedes Experiment mit 
Nummer $n=1,\ldots,N$ unterschiedlich. Wir wollen jedoch annehmen, dass sie im statistischen 
Mittel näherungsweise mit einer relevanten Dichtematrix \eqref{equation::B_260} übereinstimmt,
so dass gilt
\begin{equation}
 \langle \hat{\varrho}(t_0) \rangle = \frac{ 1 }{ N } \sum_{n=1}^N \hat{\varrho}^{(n)}(t_0)
 \approx \hat{\tilde{\varrho}}(t_0) \ .
 \label{equation::D_890}
\end{equation}
Es folgt dann das statistische Mittel der fluktuierenden Kräfte
\begin{equation}
 \langle f_i(\mathbf{r},t) \rangle \approx \mathrm{Sp}\{ \hat{\tilde{\varrho}}(t_0) 
 \, \hat{f}_i(\mathbf{r},t) \} = 0
 \label{equation::D_900}
\end{equation}
und die Korrelationsfunktion
\begin{eqnarray}
 &&\langle f_i(\mathbf{r},t) \, f_k(\mathbf{r}^\prime,t^\prime) \rangle \approx \nonumber \\
 &&\approx \, \mathrm{Sp}\{ \hat{\tilde{\varrho}}(t_0) \, \hat{f}_i(\mathbf{r},t) 
 \, \hat{f}_k(\mathbf{r}^\prime,t^\prime) \} \nonumber \\
 &&\approx \, \int_0^1 d\alpha \ \mathrm{Sp}\{ \hat{\tilde{\varrho}}(t_0)^\alpha 
 \, \hat{f}_i(\mathbf{r},t) \, \hat{\tilde{\varrho}}(t_0)^{1-\alpha}
 \, \hat{f}_k(\mathbf{r}^\prime,t^\prime) \} \nonumber \\
 &&= ( f_i(\mathbf{r},t) | f_k(\mathbf{r}^\prime,t^\prime) )_{t_0} \ .
 \label{equation::D_910}
\end{eqnarray}
Nach einigen Umformungen lässt sich die Korrelationsfunktion der fluktuierenden Kräfte also 
in Form eines Mori-Skalarprodukts schreiben. Setzen wir hier die Operatoren der fluktuierenden 
Kräfte \eqref{equation::D_860} ein und beachten, dass der Zeitentwicklungsoperators $U(t_0,t)$ 
im Mori-Skalarprodukt näherungsweise adjungiert werden kann, so finden wir
\begin{widetext}
\begin{eqnarray}
 \langle f_i(\mathbf{r},t) \, f_k(\mathbf{r}^\prime,t^\prime) \rangle &\approx& 
 ( f_i(\mathbf{r},t) | f_k(\mathbf{r}^\prime,t^\prime) )_{t_0} \nonumber \\
 &=& ( \mathsf{Q}(t_0) \, U(t_0,t) \, \mathsf{Q}(t) \, \mathsf{L} \, \hat{a}_i(\mathbf{r}) |
 \mathsf{Q}(t_0) \, U(t_0,t^\prime) \, \mathsf{Q}(t^\prime) \, \mathsf{L} 
 \, \hat{a}_k(\mathbf{r}^\prime) )_{t_0} \nonumber \\
 &\approx& ( \mathsf{Q}(t) \, \mathsf{L} \, \hat{a}_i(\mathbf{r}) |
 [U(t_0,t)]^+ \, \mathsf{Q}(t_0) \, \mathsf{Q}(t_0) \, U(t_0,t^\prime) \, \mathsf{Q}(t^\prime) \, \mathsf{L} 
 \, \hat{a}_k(\mathbf{r}^\prime) )_t \nonumber \\
 &=& ( \mathsf{Q}(t) \, \mathsf{L} \, \hat{a}_i(\mathbf{r}) |
 [U(t_0,t)]^+ \, U(t_0,t^\prime) \, \mathsf{Q}(t^\prime) \, \mathsf{L} 
 \, \hat{a}_k(\mathbf{r}^\prime) )_t \nonumber \\
 &=& ( \mathsf{Q}(t) \, \mathsf{L} \, \hat{a}_i(\mathbf{r}) |
 U(t,t^\prime) \, \mathsf{Q}(t^\prime) \, \mathsf{L} \, \hat{a}_k(\mathbf{r}^\prime) )_t
 \, = \, k_B \, M_{ki}(\mathbf{r}^\prime,t^\prime;\mathbf{r},t)
 \, \approx \, k_B \, M_{ik}(\mathbf{r},t;\mathbf{r}^\prime,t^\prime) \ .
 \label{equation::D_920}
\end{eqnarray}
\end{widetext}
Das letzte Gleichheitszeichen folgt aus der Definition der Gedächtnismatrix 
\eqref{equation::C_320}. Wir stellen also fest, dass sich die Korrelationsfunktion der 
fluktuierenden Kräfte mit der Gedächtnismatrix $M_{ik}(\mathbf{r},t;\mathbf{r}^\prime,t^\prime)$ 
in Verbindung bringen lässt. 

Die Bezeichnung \emph{näherungsweise} betrifft immer die hydrodynamische Näherung, die wir 
hier verwenden. In diesem Sinne werden wir ab jetzt zur Vereinfachung in den Gleichungen die 
Näherungszeichen durch Gleichheitszeichen ersetzen. Da wir Gedächtniseffekte vernachlässigen, 
dürfen wir die Gedächtnismatrix durch die Onsager-Matrix darstellen gemäß \eqref{equation::D_090}. 
So erhalten wir für die Korrelationsfunktion der fluktuierenden Kräfte das Ergebnis
\begin{equation}
 \langle f_i(\mathbf{r},t) \, f_k(\mathbf{r}^\prime,t^\prime) \rangle =
 2 \, k_B \, M_{ik}(\mathbf{r},\mathbf{r}^\prime;t) \, \delta( t - t^\prime ) \ .
 \label{equation::D_930}
\end{equation}

Die wesentlichen Eigenschaften der fluktuierenden Kräfte $f_i(\mathbf{r},t)$ werden durch den 
Erwartungswert \eqref{equation::D_900} und die Korrelationsfunktion \eqref{equation::D_930} 
beschrieben. Betrachten wir diese beiden Gleichungen genauer, so stellen wir fest, dass sie 
genau den Definitionsgleichungen für \emph{gaußische stochastische Kräfte} entsprechen. Wir 
kommen also zu dem Ergebnis, dass man die fluktuierenden Kräfte $f_i(\mathbf{r},t)$ als 
gaußische stochastische Kräfte interpretieren kann. Folglich ist die allgemeine hydrodynamische 
Gleichung \eqref{equation::D_120} eine stochastische Differentialgleichung. Es wird somit 
für die Differentialrechnung ein stochastischer Formalismus benötigt. Da wir die allgmeinen 
hydrodynamischen Gleichungen unter Annahme der Zeitumkehrinvarianz in der zugrunde liegenden 
mikroskopischen Theorie hergeleitet haben, ist der Differentialoperator für die Zeitableitung 
$\partial_t$ symmetrisch definiert. In diesem Sinne ist \eqref{equation::D_120} eine 
stochastische Differentialgleichung im \emph{Stratonovich-Formalismus}.

Die Onsager-Matrix $M_{ik}(\mathbf{r},\mathbf{r}^\prime;t)$ tritt an zwei Stellen auf. 
Zum einen legt sie im zweiten Term der allgemeinen hydrodynamischen Gleichungen 
\eqref{equation::D_120} die Stärke der \emph{Dissipationen} fest. Zum anderen legt sie in 
\eqref{equation::D_930} die Stärke der \emph{Fluktuationen} der stochastischen Kräfte fest. 
Dieser Zusammenhang zwischen Fluktuation und Dissipation ist bekannt unter dem Namen 
\emph{Fluktuations-Dissipations-Theorem der zweiten Art}.

Im Folgenden wollen wir die stochastischen Kräfte speziell für unsere normale Flüssigkeit 
untersuchen. Für die relevanten Variablen Energiedichte $\varepsilon(\mathbf{r},t)$, 
Impulsdichte $\mathbf{j}(\mathbf{r},t)$ und Massendichte $\rho(\mathbf{r},t)$ dürfen wir 
die Onsager-Matrix mit dem Ansatz \eqref{equation::D_590} verwenden. Es folgt dann die 
Korrelationsfunktion
\begin{equation}
 \langle f_i(\mathbf{r},t) \, f_k(\mathbf{r}^\prime,t^\prime) \rangle =
 - \, \partial_m \, 2 \, k_B \, N_{im,kn} \, \partial_n 
 \, \delta( \mathbf{r} - \mathbf{r}^\prime ) \, \delta( t - t^\prime ) \ .
 \label{equation::D_940}
\end{equation}
Die stochastischen Kräfte sind also sowohl auf der räumlichen Längenskala als auch auf 
der zeitlichen Skala extrem kurzreichweitig.

Wegen den räumlichen Differentialoperatoren in \eqref{equation::D_940} ist es zweckmäßig, die 
stochastischen Kräfte $f_i(\mathbf{r},t)$ der Dichten als Ableitungen von neuen stochastischen 
Kräften $g_{ik}(\mathbf{r},t)$ der zugehörigen Stromdichten zu schreiben gemäß
\begin{equation}
 f_i(\mathbf{r},t) = - \, \partial_k \, g_{ik}(\mathbf{r},t) \ .
 \label{equation::D_950}
\end{equation}
Diese Darstellung der fluktuierenden Kräfte haben wir schon früher in 
\eqref{equation::C_740} zusammen mit \eqref{equation::C_750} hergeleitet. Sie gilt 
allgemein, wenn die hydrodynamischen Variablen Dichten von Erhaltungsgrößen sind. 
Die Erwartungswerte der neuen stochastischen Kräfte sind dann ebenfalls null, und 
für die Korrelationsfunktion erhalten wir
\begin{equation}
 \langle g_{ik}(\mathbf{r},t) \, g_{mn}(\mathbf{r}^\prime,t^\prime) \rangle =
 2 \, k_B \, N_{ik,mn} \, \delta( \mathbf{r} - \mathbf{r}^\prime ) \, \delta( t - t^\prime ) \ .
 \label{equation::D_960}
\end{equation}

Der Übergang zu den relevanten Variablen Entropiedichte $\sigma(\mathbf{r},t)$, 
Impulsdichte $\mathbf{j}(\mathbf{r},t)$ und Massendichte $\rho(\mathbf{r},t)$ erfolgt, 
indem wir die Matrix $N_{ik,mn}$ durch $\Lambda_{ik,mn}$ ersetzen. Da die 
Funktional-Ableitungen der Entropie \eqref{equation::D_620}-\eqref{equation::D_640} einen 
zusätzlichen Faktor $1/T$ haben, die Funktional-Ableitungen der Energie 
\eqref{equation::D_500}-\eqref{equation::D_520} jedoch nicht, müssen wir in der 
Korrelationsfunktion noch einen zusätzlichen Faktor $T$ hinzufügen. Wir erhalten somit 
in diesem Fall
\begin{equation}
 \langle g_{ik}(\mathbf{r},t) \, g_{mn}(\mathbf{r}^\prime,t^\prime) \rangle =
 2 \, k_B T \, \Lambda_{ik,mn} \, \delta( \mathbf{r} - \mathbf{r}^\prime ) \, \delta( t - t^\prime ) \ .
 \label{equation::D_970}
\end{equation}
Wir weisen darauf hin, dass die Temperatur $T = T(\mathbf{r},t)$ nicht konstant ist, 
sondern fluktuiert. Sie wird in \eqref{equation::D_500} als Funktionalableitung der 
Energie $E[x]$ definiert und hängt somit im allgemeinen von den hydrodynamischen 
Variablen $x_i(\mathbf{r},t)$ ab.

Im vorherigen Abschnitt haben wir festgestellt, dass aus Symmetriegründen die meisten 
Elemente der Matrix $\Lambda_{ik,mn}$ null sind. Es gibt keine von Null verschiedene 
Matrixelemente für die Massendichte. Da es für die Massendichte keine Dissipation gibt, 
fallen auch die stochastischen Kräfte für die Massenstromdichte weg. Wir haben also 
\begin{equation}
 g^j_k(\mathbf{r},t) = 0 \ .
 \label{equation::D_980}
\end{equation}
Es verbleiben die stochastischen Kräfte für den Spannungstensor $g^\Pi_{ik}(\mathbf{r},t)$ 
und für die Entropiestromdichte $g^q_k(\mathbf{r},t)$. Nach \eqref{equation::D_900} 
sind die Erwartungswerte null. Wir haben also 
\begin{eqnarray}
 \langle g^\Pi_{ik}(\mathbf{r},t) \rangle &=& 0 \ ,
 \label{equation::D_990} \\
 \langle g^q_k(\mathbf{r},t) \rangle &=& 0 \ .
 \label{equation::D_1000}
\end{eqnarray}
Wir bemerken zur Schreibweise: In \eqref{equation::D_980}-\eqref{equation::D_1000} 
und in den nachfolgenden Formeln bezeichnet der obere Index jeweils die Stromdichte, 
zu welcher die jeweilige stochastische Kraft gehört. Wegen der Zeitumkehrinvarianz 
gibt es keine Korrelationen zwischen den beiden stochastischen Kräften 
$g^\Pi_{ik}(\mathbf{r},t)$ und $g^q_k(\mathbf{r},t)$, so dass
\begin{equation}
 \langle g^\Pi_{ik}(\mathbf{r},t) \, g^q_n(\mathbf{r}^\prime,t^\prime) \rangle = 0 \ .
 \label{equation::D_1010}
\end{equation}
Die verbleibenden Korrelationen werden durch die Matrizen \eqref{equation::D_770} und 
\eqref{equation::D_780} beschrieben. Wir erhalten damit 
\begin{widetext}
\begin{eqnarray}
 \langle g^\Pi_{ik}(\mathbf{r},t) \, g^\Pi_{mn}(\mathbf{r}^\prime,t^\prime) \rangle &=&
 2 \, k_B T \, \Lambda^{jj}_{ik,mn}
 \, \delta( \mathbf{r} - \mathbf{r}^\prime ) \, \delta( t - t^\prime ) \nonumber \\
 &=& 2 \, k_B T \, \left[ \eta \left( \delta_{im} \delta_{kn} + \delta_{in} \delta_{km} 
 - \frac{ 2 }{ d } \delta_{ik} \delta_{mn} \right)
 + \zeta \, \delta_{ik} \delta_{mn} \right]
 \, \delta( \mathbf{r} - \mathbf{r}^\prime ) \, \delta( t - t^\prime ) \hspace{6mm}
 \label{equation::D_1020}
\end{eqnarray}
und
\begin{eqnarray}
 \langle g^q_k(\mathbf{r},t) \, g^q_n(\mathbf{r}^\prime,t^\prime) \rangle &=&
 2 \, k_B T \, \Lambda^{\sigma\sigma}_{kn}
 \, \delta( \mathbf{r} - \mathbf{r}^\prime ) \, \delta( t - t^\prime ) \nonumber \\
 &=& 2 \, k_B \, \varkappa \, \delta_{kn}
 \, \delta( \mathbf{r} - \mathbf{r}^\prime ) \, \delta( t - t^\prime ) \ .
 \label{equation::D_1030}
\end{eqnarray}
\end{widetext}
Die gaußischen stochastischen Kräfte für die Massenstromdichte $g^j_k(\mathbf{r},t)$, für den 
Spannungstensor $g^\Pi_{ik}(\mathbf{r},t)$ und für die Entropiestromdichte $g^q_k(\mathbf{r},t)$ 
sind mit den Gleichungen \eqref{equation::D_980}-\eqref{equation::D_1030} eindeutig definiert. 
Es verbleibt die stochastische Kraft für die Energiestromdichte. Diese bekommt man aus der 
thermodynamischen Relation
\begin{equation}
 g^{j_E}_k(\mathbf{r},t) = T(\mathbf{r},t) \, g^q_k(\mathbf{r},t) 
 + v_i(\mathbf{r},t) \, g^\Pi_{ik}(\mathbf{r},t) \ .
 \label{equation::D_1040}
\end{equation}
Am Ende müssen die stochastischen Kräfte in die hydrodynamischen Gleichungen eingefügt werden. 
Man erreicht dies, indem man die stochastischen Kräfte zu den Stromdichten 
\eqref{equation::D_810}-\eqref{equation::D_840} hinzu addiert. Wir finden also
\begin{eqnarray}
 j_k &=& \rho \, v_k \ ,
 \label{equation::D_1050} \\
 \Pi_{ik} &=& p \, \delta_{ik} + \rho \, v_i \, v_k + \Delta \Pi_{ik} 
 + g^\Pi_{ik}(\mathbf{r},t) \ , \hspace{5mm}
 \label{equation::D_1060} \\
 q_k &=& \sigma \, v_k 
 - \frac{ \varkappa }{ T } \, \partial_k T + g^q_k(\mathbf{r},t) \ , \hspace{5mm}
 \label{equation::D_1070} \\
 j_{E,k} &=& ( \varepsilon + p ) v_k - \varkappa \, \partial_k T + v_i \Delta \Pi_{ik} 
 + g^{j_E}_k(\mathbf{r},t) \ . \qquad
 \hspace{4mm}
 \label{equation::D_1080}
\end{eqnarray}
Für die Entropiedichte $\sigma(\mathbf{r},t)$ sind die stochastischen Kräfte noch nicht 
vollständig berücksichtigt. Die Darstellung durch die Divergenz der Entropiestromdichte 
gemäß \eqref{equation::D_950} ist hier nicht ausreichend. Es kommt auch hier ein Quellterm 
hinzu. Wir berücksichtigen ihn, in dem wir zu der erzeugten Wärme \eqref{equation::D_760} 
auch einen stochastischen Term hinzufügen. Wir bekommen somit
\begin{eqnarray}
 R &=& ( \partial_k v_i ) \, \Lambda^{jj}_{ik,mn} \, ( \partial_n v_m ) 
 + ( \partial_k T ) \, \Lambda^{\sigma\sigma}_{kn} \, ( \partial_n T ) \nonumber \\
 &&- \, ( \partial_k v_i ) \, g^\Pi_{ik}(\mathbf{r},t) - ( \partial_k T ) \, g^q_k(\mathbf{r},t) \ .
 \label{equation::D_1090}
\end{eqnarray}
Die beiden negativen Vorzeichen hängen mit dem Minuszeichen in \eqref{equation::D_950} zusammen.

Im Ergebnis stellen wir fest: Die hydrodynamischen Gleichungen 
\eqref{equation::D_530}-\eqref{equation::D_560} bleiben unverändert. Änderungen, um die 
Fluktuationen zu berücksichtigen, erfolgen allein in den Stromdichten 
\eqref{equation::D_1050}-\eqref{equation::D_1080} und in der erzeugten Wärme 
\eqref{equation::D_1090} durch die zusätzlichen stochastischen Terme. Untersuchungen zu 
hydrodynamische Fluktuationen sind auch in elementaren Lehrbüchern \cite{LL09} zu finden. 
Unsere Ergebnisse stimmen mit den darin beschriebenen überein, insbesondere die 
Korrelationsfunktionen \eqref{equation::D_1020} und \eqref{equation::D_1030}.

\section{Zeitumkehrinvarianz und der zweite Hauptsatz der Thermodynamik}
\label{section::5}
Während die zugrunde liegende mikroskopische Theorie die Invarianz unter Zeitumkehr erfüllt, 
ist bekannt, dass die statistische Mechanik des Nichtgleichgewichts und die daraus folgenden 
hydrodynamischen Gleichungen diese Invarianz brechen. Denn nach dem zweiten Hauptsatz der 
Thermodynamik sollte die Entropie mit der Zeit immer anwachsen oder zumindest gleich bleiben 
gemäß $dS/dt \geq 0$. Diese Aussagen stellen die allgemeine Lehrmeinung in der statistischen 
Physik für das Nichtgleichgewicht dar. Wir untersuchen nun in wie weit diese Aussagen 
vereinbar mit unserer Theorie sind und ob sie modifiziert oder neu interpretiert werden müssen.

In Kapitel \ref{section::3} haben wir die verallgemeinerten hydrodynamischen Gleichungen 
\eqref{equation::C_350} zusammen mit der Entropiegleichung \eqref{equation::C_770} aus der 
mikroskopischen Theorie hergeleitet. Da keine Näherungen gemacht wurden und die Gleichungen 
exakt sind, müssen diese invariant unter der Zeitumkehr sein. Daher kann für die Entropie 
$S(t)$ in \eqref{equation::C_770} nicht der zweite Hauptsatz der Thermodynamik gelten.

Die Invarianz unter Zeitumkehr kann jedoch gebrochen werden, wenn man die Terme auf den 
rechten Seiten der Gleichungen einzeln betrachtet. Der erste Term auf der rechten Seite der 
Entropiegleichung \eqref{equation::C_770} ist ein dissipativer Term und hat eine 
\emph{quadratische Form}. Es ist daher naheliegend, dass dieser Term positiv definit ist 
und ein Anwachsen der Entropie entsprechend dem zweiten Hauptsatz der Thermodynamik 
bewirkt. Es gibt hier noch eine Unsicherheit. Im Nichtgleichgewicht ist die Onsager-Matrix 
\eqref{equation::C_290} nicht symmetrisch, und es kann nicht vollständig garantiert werden, 
dass sie positiv definit ist. Betrachtet man jedoch eine Flüssigkeit im Rahmen der 
Hydrodynamik, so kann man annehmen, dass zumindest lokal ein thermisches Gleichgewicht 
gilt. Folglich sollte die Onsager-Matrix \eqref{equation::C_290} auf lokaler Ebene 
symmetrisch und positiv definit sein.

Wir kommen also zu dem Schluss: Da nach der Zeitumkehrinvarianz eine Möglichkeit bestehen 
muss, dass die Entropie $S(t)$ mit der Zeit auch wieder abnimmt, so kann dies nur mit 
dem zweiten Term der Entropiegleichung \eqref{equation::C_770} geschehen, dem 
\emph{fluktuierenden Term}. 

Betrachten wir die drei Terme auf der rechten Seite der verallgemeinerten hydrodynamischen 
Gleichungen \eqref{equation::C_350}, so kommen wir weiterhin zu den drei Ergebnissen: 
\begin{itemize}
\item[(a)] Die reversiblen Terme erfüllen die Zeitumkehrinvarianz.
\item[(b)] Für sich alleine genommen brechen die dissipativen Terme und die fluktuierenden 
Terme die Zeitumkehrinvarianz. 
\item[(c)] Die Summe der dissipativen und der fluktuierenden Terme erfüllt jedoch die Invarianz 
unter Zeitumkehr.
\end{itemize}

Man kann nun die Frage stellen, welchen Einfluss die Näherungen, die zu den hydrodynamischen 
Gleichung der normalen Flüssigkeit in der bekannten Form führen, auf die Struktur der Terme 
und auf die drei Ergebnisse haben. Zunächst stellen wir fest, dass die Struktur der Terme 
unverändert bleibt. Auf allen Stufen der Näherungen gibt es reversible, dissipative und 
fluktuierende Terme. Selbst die erzeugte Wärme pro Volumen und Zeit $R(\mathbf{r},t)$, 
welche ganz am Ende unserer Überlegungen auftaucht und in \eqref{equation::D_1090} definiert 
wird, hat dieselbe Struktur wie die rechte Seite der Entropie-Gleichung \eqref{equation::C_770} 
am Anfang und ohne Näherung: Sie enthält einen dissipativen und einen fluktuierenden Term. 
Der dissipative Term führt zum Anwachsen der Entropie. Ein Rückgang der Entropie ist nur 
durch den fluktuierenden Term möglich.

Die reversiblen Terme werden durch die Näherungen nicht verändert. Daher bleibt Ergebnis (a) 
unverändert bestehen. Wenn die dissipativen Terme die Zeitumkehrinvarianz brechen, so werden 
die Näherungen daran qualitativ nichts ändern. Folglich bleibt auch Ergebnis (b) bestehen. 
Die Summe der dissipativen und fluktuierenden Terme wird jedoch durch die Näherungen so 
beeinflusst, dass die Zeitumkehrinvarianz nicht mehr exakt sondern nur noch näherungsweise 
gilt. Folglich wird Ergebnis (c) nur noch näherungsweise gelten. Das muss einem insbesondere 
dann klar sein, wenn man in den hydrodynamischen Gleichungen die fluktuierenden Terme durch 
gaußische stochastische Kräfte modelliert. Qualitativ ändert das jedoch nichts.

Wir fassen nochmals zusammen. Um die Zeitumkehrinvarianz in den hydrodynamischen Gleichungen 
im Zusammenhang mit dem zweiten Hauptsatz der Thermodynamik zu verstehen, ist es wichtig,
die \emph{fluktuierenden Terme} zu betrachten. Nur die fluktuierenden Terme können eine 
Abnahme der Entropie $S(t)$ bewirken. In diesem Sinne muss die allgemeine Lehrmeinung 
über den zweiten Hauptsatz der Thermodynamik erweitert und neu interpretiert werden. 
Wir mögen sagen dass die fluktuierenden Terme in den hydrodynamischen Gleichungen die 
Zeitumkehrinvarianz \emph{wiederherstellen}.

\section{Langevin- und Fokker-Planck-Gleichung}
\label{section::6}
Nachdem in Kapitel \ref{section::4} die Gedächtniseffekte vernachlässigt wurden, haben wir 
hydrodynamische Gleichungen erhalten, welche die Form von stochastischen Gleichungen besitzen. 
Die fluktuierenden Terme wurden als gaußische stochastische Kräfte modelliert. Diese 
Eigenschaften haben wir sowohl für die allgemeinen hydrodynamischen Gleichungen in der 
GENERIC-Form erhalten, als auch für die speziellen hydrodynamischen Gleichungen einer 
normalen Flüssigkeit. Wir wollen daher untersuchen, ob diese Gleichungen kompatibel mit 
der allgemeinen Theorie der stochastischen Prozesse sind.

\subsection{Stochastische Prozesse}
\label{section::6A}
Zur Beschreibung eines stochastischen Prozesses betrachten wir die Variablen $x_i(t)$. 
Wir lassen die Ortsvariable $\mathbf{r}$ weg und berücksichtigen nur den Index $i$, um 
einfachere Formeln zu schreiben. Die Ortsvariable kann man jederzeit wieder hinzufügen, 
um allgemeinere Formeln zu erhalten. Wir folgen der Darstellung der stochastischen Theorie 
für Nichtgleichgewichtssysteme von Graham und Haken \cite{GH71A,GH71B,Gr73}. Wenn keine 
Gedächtniseffekte vorhanden sind, dann handelt es sich um einen Markov-Prozess. Die 
Zeitentwicklung der Wahrscheinlichkeitsverteilung $P(x,t)$ der stochastischen Variablen 
wird beschrieben durch eine Master-Gleichung. Führt man eine Kramers-Moyal-Entwicklung 
durch und bricht nach der zweiten Ordnung ab, so erhält man die Fokker-Planck-Gleichung
\begin{equation}
 \frac{ \partial }{\partial t } P(x,t) = \left[ - \frac{ \partial }{ \partial x_i } K^{(1)}_i(x) 
 + \frac{ \partial^2 }{ \partial x_i \partial x_j } K^{(2)}_{ij}(x) \right] P(x,t) \ .
 \label{equation::F_010}
\end{equation}
Zur Vereinfachung der Schreibweise verwenden wir hier die Summenkonvention und nehmen 
an, dass über alle zweimal auftretende Indizes automatisch summiert wird.
Die Differentialoperatoren auf der rechten Seite wirken über die eckige Klammer hinaus 
auch auf die Verteilungsfunktion $P(x,t)$. Die ersten beiden Kramers-Moyal-Koeffizienten 
sind $K^{(1)}_i(x)$ und $K^{(2)}_{ij}(x)$. Sie hängen im allgemeinen wiederum von den 
stochastischen Variablen $x_i$ ab. Der zweite Koeffizient ist eine symmetrische Matrix 
$K^{(2)}_{ij}(x)=+K^{(2)}_{ji}(x)$.

Wir definieren nun die fluktuierenden Kräfte
\begin{equation}
 f_i(t) = B_{im}(x(t)) \, \varepsilon_m(t)
 \label{equation::F_020}
\end{equation}
mit einer Matrix $B_{im}(x(t))$ und gaußisch verteilten elementaren stochastischen 
Kräften $\varepsilon_m(t)$ mit den Eigenschaften 
\begin{equation}
 \langle \varepsilon_m(t) \rangle = 0 \ , \qquad 
 \langle \varepsilon_m(t) \varepsilon_n(t^\prime) \rangle = \delta_{mn} \, \delta(t-t^\prime) \ .
 \label{equation::F_030}
\end{equation}
Die Anzahl der elementaren stochastischen Kräfte $\varepsilon_m(t)$ ist größer oder gleich 
der Anzahl der fluktuierenden Kräfte $f_i(t)$. Folglich muss die Matrix $B_{im}(x)$ nicht 
unbedingt quadratisch sein. Wir verlangen jedoch die Bedingung
\begin{equation}
 B_{im}(x) \,  B_{jm}(x) = 2 \, K^{(2)}_{ij}(x) \ .
 \label{equation::F_040}
\end{equation}
Nach Graham und Haken \cite{GH71B,Gr73} findet man dann für die stochastischen 
Variablen $x_i(t)$ die Langevin-Gleichung 
\begin{equation}
 \partial_t x_i(t) = K^{(1)}_i(x(t)) - \frac{ 1 }{ 2 } 
 \frac{ \partial B_{im}(x(t)) }{ \partial x_j(t) } B_{jm}(x(t)) + f_i(t) \ ,
 \label{equation::F_050}
\end{equation}
welche zur Fokker-Planck-Gleichung \eqref{equation::F_010} äquivalent ist. Die 
zeitliche Ableitung auf der linken Seite der Langevin-Gleichung ist symmetrisch 
definiert. Folglich ist \eqref{equation::F_050} eine stochastische 
Differentialgleichung im \emph{Stratonovich-Formalismus} \cite{St63,vK07}. Der 
zweite Term mit der Matrix $B_{im}(x(t))$ und deren Ableitung ist hierbei ein wohl 
bekannter Term.

Wir stellen fest: Wenn der zweite Kramers-Moyal-Koeffizient $K^{(2)}_{ij}(x(t))$ und 
die Matrix $B_{im}(x(t))$ selbst von den stochastischen Variablen $x_i(t)$ abhängen, 
dann enthalten die Definitionen der fluktuierenden Kräfte \eqref{equation::F_020} und 
die Langevin-Gleichung \eqref{equation::F_050} zusätzliche Terme mit partiellen 
Ableitungen dieser Matrizen nach den stochastischen Variablen. Das macht hier die 
Theorie subtil und kompliziert, muss jedoch sorgfältig berücksichtigt werden.

Wir untersuchen nun, unter welchen Bedingungen die Boltzmann-Verteilung
\begin{equation}
 P_\mathrm{eq}(x) = Z^{-1} \, e^{-F(x)}
 \label{equation::F_060}
\end{equation}
mit einer vorgegebenen freien Energie $F(x)$ eine stationäre Lösung der 
Fokker-Planck-Gleichung \eqref{equation::F_010} ist. Diese Boltzmann-Verteilung 
soll das thermische Gleichgewicht beschreiben. Wir setzen ein und formen 
um gemäß
\begin{eqnarray}
 0 &=& \frac{ \partial }{\partial t } P_\mathrm{eq}(x) 
 = \frac{ \partial }{ \partial x_i } \left[ - K^{(1)}_i 
 + \frac{ \partial }{ \partial x_j } K^{(2)}_{ij} \right] P_\mathrm{eq}(x) \nonumber \\
 &=& \frac{ \partial }{ \partial x_i } \left[ - K^{(1)}_i 
 + \frac{ \partial K^{(2)}_{ij} }{ \partial x_j } + K^{(2)}_{ij} 
 \frac{ \partial }{ \partial x_j } \right] P_\mathrm{eq}(x) \nonumber \\
 &=& \frac{ \partial }{ \partial x_i } \left[ - K^{(1)}_i 
 + \frac{ \partial K^{(2)}_{ij} }{ \partial x_j } - K^{(2)}_{ij} 
 \frac{ \partial F }{ \partial x_j } \right] P_\mathrm{eq}(x) \ . \qquad
 \label{equation::F_070}
\end{eqnarray}
Wir schreiben nun den ersten Kramers-Moyal-Koeffi\-zienten in der Form 
\begin{equation}
 K^{(1)}_i(x) = V_i(x) - K^{(2)}_{ij}(x) \frac{ \partial F(x) }{ \partial x_j }
 + \frac{ \partial K^{(2)}_{ij}(x) }{ \partial x_j } 
 \label{equation::F_080}
\end{equation}
mit einem reversiblen Term $V_i(x)$ und zwei dissipativen Termen, welche durch den zweiten 
Kramers-Moyal-Koeffizienten parametrisiert werden. Aus \eqref{equation::F_070} folgt dann 
für den reversiblen Term die Bedingung
\begin{equation}
 \frac{ \partial }{\partial t } P_\mathrm{eq}(x) = 
 \frac{ \partial }{ \partial x_i } \bigl[ - V_i(x) P_\mathrm{eq}(x) \bigr] = 0 \ .
 \label{equation::F_090}
\end{equation}
Diese Bedingung lässt sich einfach lösen, wenn wir den reversiblen Term in 
der Form
\begin{equation}
 V_i(x) = - A^{(2)}_{ij}(x) \frac{ \partial F(x) }{ \partial x_j }
 + \frac{ \partial A^{(2)}_{ij}(x) }{ \partial x_j } 
 \label{equation::F_100}
\end{equation}
mit einer antisymmetrischen Matrix $A^{(2)}_{ij}(x)=-A^{(2)}_{ji}(x)$ schreiben. Es folgt 
dann nämlich
\begin{equation}
 \frac{ \partial }{ \partial x_i } \bigl[ V_i(x) P_\mathrm{eq}(x) \bigr] 
 = \frac{ \partial^2 }{ \partial x_i \partial x_j } \bigl[ A^{(2)}_{ij}(x) P_\mathrm{eq}(x) \bigr]
 = 0
 \label{equation::F_110}
\end{equation}
aus der Antisymmetrie dieser Matrix. Die reversiblen Terme \eqref{equation::F_100} haben fast 
dieselbe Form wie die dissipativen Terme in \eqref{equation::F_080}. Der Unterschied besteht 
in dem Symmetrieverhalten der Matrizen $A^{(2)}_{ij}(x)=-A^{(2)}_{ji}(x)$ und 
$K^{(2)}_{ij}(x)=+K^{(2)}_{ji}(x)$.

Setzen wir nun den ersten Kramers-Moyal-Koeffizien\-ten \eqref{equation::F_080} in die 
Langevin-Gleichung \eqref{equation::F_050} ein so finden wir
\begin{eqnarray}
 \partial_t x_i(t) &=& V_i(x(t)) - K^{(2)}_{ij}(x(t)) \frac{ \partial F(x(t)) }{ \partial x_j }
 \nonumber\\
 &&+ \frac{ 1 }{ 2 } B_{im}(x(t)) \frac{ \partial B_{jm}(x(t)) }{ \partial x_j(t) } 
 + f_i(t) \ . \qquad
 \label{equation::F_120}
\end{eqnarray}
Der dritte Term in dieser Gleichung ergibt sich aus der Summe des zweiten Terms von 
\eqref{equation::F_050} und des dritten Terms von \eqref{equation::F_080}, wobei der 
zweite Kramers-Moyal-Koeffizient in der Form \eqref{equation::F_040} eingesetzt 
wurde. Diese Langevin-Gleichung in der Stratonovich-Form ist kompatibel mit der 
Boltzmann-Verteilung \eqref{equation::F_060} für das thermische Gleichgewicht, wenn 
der reversible Term $V_i(x(t))$ entweder die Form \eqref{equation::F_100} hat oder 
zumindest die Bedingung \eqref{equation::F_090} erfüllt.

\subsection{GENERIC-Formalismus}
\label{section::6B}
In Kapitel \ref{section::4} haben wir die allgemeinen hydrodynamischen Gleichungen des 
GENERIC-Formalismus hergeleitet. Um einfachere Formeln zu schreiben, lassen wir wieder 
die Ortsvariable $\mathbf{r}$ weg und verwenden die Summenkonvention für die Indizes. 
Aus \eqref{equation::D_120} finden wir dann die Langevin-Gleichung
\begin{eqnarray}
 \partial_t x_i(t) &=& L_{ik}(x(t)) \, \frac{ \partial E(x(t)) }{ \partial x_k(t) } \nonumber \\
 &&+ \, M_{ik}(x(t)) \, \frac{ \partial S(x(t)) }{ \partial x_k(t) } 
 + \, f_i(t) \ ,
 \label{equation::F_130}
\end{eqnarray}
und aus \eqref{equation::D_140} und \eqref{equation::D_150} folgen die zugehörigen 
Nebenbedingungen
\begin{eqnarray}
 L_{ik}(x(t)) \, \frac{ \partial S(x(t)) }{ \partial x_k(t) } &=& 0 \ ,
 \label{equation::F_140} \\
 M_{ik}(x(t)) \, \frac{ \partial E(x(t)) }{ \partial x_k(t) } &=& 0 \ .
 \label{equation::F_150}
\end{eqnarray}
Weitere Nebenbedingungen gibt es für Erhaltungsgrößen wie Impuls $\mathbf{P}(x(t))$ und 
Teilchenzahl $N(x(t))$ wie üblich im GENERIC-Formalismus.

Wir zeigen nun, dass die Langevin-Gleichung \eqref{equation::F_130} eine stochastische 
Differentialgleichung im \emph{Stratonovich-Formalismus} ist \cite{vK07}. Dazu gehen wir 
zur ursprünglichen exakten Bewegungsgleichung \eqref{equation::C_350} mit Gedächtniseffekten 
zurück. Wir integrieren diese Gleichung über ein zeitliches Intervall mit der Ausdehnung 
$\Delta t$. Weiterhin nehmen wir an, die Gedächtnis-Effekte haben eine zeitliche 
Ausdehnung von der Größenordnung $\Delta t_M$. Im Ergebnis bekommen wir somit eine 
Integralgleichung mit zwei Zeitskalen $\Delta t$ und $\Delta t_M$, wobei auf der 
linken Seite eine Differenz von zwei hydrodynamischen Variablen zu unterschiedlichen 
Zeiten steht. Die Langevin-Gleichung \eqref{equation::F_130} erhalten wir daraus in 
dem Grenzfall, dass beide Zeitskalen $\Delta t$ und $\Delta t_M$ infinitesimal klein 
werden. Es kommt jedoch auf die Reihenfolge der zwei Grenzwertprozesse an. \emph{Zuerst} 
vernachlässigen wir die Gedächtniseffekte (Markov-Näherung) und führen den 
Grenzwertprozess $\Delta t_M \to 0 $ durch. Wir dürfen folglich die Gedächtnismatrix 
ersetzen durch die Formel \eqref{equation::D_090} mit einer Delta-Funktion in der Zeit. 
Das Integral über die Zeit im zweiten Term auf der rechten Seite von der ursprünglichen 
Bewegungsgleichung \eqref{equation::C_350} lässt sich nun explizit ausführen, und wir 
erhalten daraus den zweiten Term von \eqref{equation::F_130}. Im Ergebnis finden wir 
eine Integralgleichung, welche der integrierten Form von \eqref{equation::F_130} 
entspricht. \emph{Danach} werten wir das verbleibende Integral mit dem Mittelwertsatz 
aus und führen den Grenzwertprozess $\Delta t \to 0 $ durch. Im Ergebnis erhalten wir 
dann die Langevin-Gleichung \eqref{equation::F_130}, wobei die zeitlich Ableitung 
auf der linken Seite \emph{symmetrisch} definiert wird. Diese ist somit eine 
stochastische Differentialgleichung im Stratonovich Formalismus.

In Analogie zu \eqref{equation::C_820} definieren wir das großkanonische thermodynamische 
Potential
\begin{equation}
 \Omega(x(t)) = E(x(t)) - T \, S(x(t)) - \mathbf{v} \cdot \mathbf{P}(x(t)) - \mu \, N(x(t))
 \label{equation::F_160}
\end{equation}
mit drei konstanten Lagrange-Parametern, Temperatur $T$, Geschwindigkeit $\mathbf{v}$ und 
chemisches Potential $\mu$. Unter Verwendung der Nebenbedingungen formen wir damit den 
dissipativen Term der Langevin-Gleichung \eqref{equation::F_130} um. Wir erhalten dann
\begin{eqnarray}
 \partial_t x_i(t) &=& L_{ik}(x(t)) \, \frac{ \partial E(x(t)) }{ \partial x_k(t) } \nonumber \\
 &&- \frac{ 1 }{ T } \, M_{ik}(x(t)) \, \frac{ \partial \Omega(x(t)) }{ \partial x_k(t) } 
 + \, f_i(t) \ . \qquad
 \label{equation::F_170}
\end{eqnarray}
Wir wollen nun zeigen, dass diese Langevin-Gleichung mit der großkanonischen 
Boltzmann-Verteilung
\begin{equation}
 P_\mathrm{eq}(x) = Z^{-1} \, \exp \left( - \frac{ \Omega(x) }{ k_B T } \right)
 \label{equation::F_180}
\end{equation}
kompatibel ist. Dazu vergleichen wir im Detail die Formeln des GENERIC-Formalismus 
\eqref{equation::F_170} und \eqref{equation::F_180} mit den entsprechenden Formeln 
der allgemeinen stochastischen Theorie \eqref{equation::F_120} und \eqref{equation::F_060}. 
Zunächst identifizieren wir die freie Energie $F(x)=\Omega(x)/k_B T$ und den zweiten 
Kramers-Moyal-Koeffizienten $K^{(2)}_{ij}(x(t)) = k_B M_{ij}(x(t))$. Damit ist 
gewährleistet, dass die Verteilungsfunktionen \eqref{equation::F_180} und 
\eqref{equation::F_060} formal übereinstimmen. Weiterhin stellen wir fest, dass der 
zweite Term unserer Langevin-Gleichung \eqref{equation::F_170} mit dem zweiten Term 
der entsprechenden Langevin-Gleichung der stochastischen Theorie \eqref{equation::F_120} 
übereinstimmt.

In Abschnitt \ref{section::4E} haben wir aus der mikroskopischen Theorie hergeleitet, 
dass $f_i(t)$ gaußische stochastische Kräfte sind, welche durch die Erwartungswerte 
\eqref{equation::D_900} und durch die Korrelationsfunktionen \eqref{equation::D_930} 
eindeutig festgelegt werden. Diese müssen mit den fluktuierenden Kräften der 
stochastischen Theorie in der Darstellung \eqref{equation::F_020} vereinbar sein. 
Letzteres ist offensichtlich erfüllt, weil wir den zweiten Kramers-Moyal-Koeffizienten 
bereits mit $K^{(2)}_{ij}(x(t)) = k_B M_{ij}(x(t))$ identifiziert haben.

Wir stellen jedoch fest, dass dem dritten Term in \eqref{equation::F_120} kein 
entsprechender Term in \eqref{equation::F_170} zugeordnet werden kann. Aus diesem 
Grunde müssen wir fordern, dass dieser dritte Term null ist, so dass gilt 
\begin{equation}
 \frac{ 1 }{ 2 } \, B_{im}(x(t)) \, \frac{ \partial B_{jm}(x(t)) }{ \partial x_j(t) } = 0 \ .
 \label{equation::F_190}
\end{equation}
Da die Matrix $B_{im}(x(t))$ im allgemeinen von null verschieden und nicht singulär 
ist, müssen wir also die Bedingung
\begin{equation}
 \frac{ \partial B_{im}(x(t)) }{ \partial x_i(t) } = 0
 \label{equation::F_200}
\end{equation}
verlangen. Wir wollen nun zeigen, dass diese Bedingung für eine normale Flüssigkeit 
allgemein erfüllt wird. Die Matrix $B_{im}(x(t))$ hängt über \eqref{equation::F_040} 
mit dem zweiten Kramers-Moyal-Koeffizienten $K^{(2)}_{ij}(x(t)) = k_B M_{ij}(x(t))$ 
und folglich mit der Onsager-Matrix $M_{ij}(x(t))$ zusammen. In der lokalen Näherung 
wird die Onsager-Matrix durch den Ansatz \eqref{equation::D_590} mit zwei räumlichen 
Differentialoperatoren dargestellt. Für die Matrix $B_{im}(x(t))$ finden wir daher 
eine analoge Darstellung mit einem räumlichen Differentialoperator. Fügen wir 
vorübergehend die Ortsvariable $\mathbf{r}$ wieder hinzu, so erhalten wir 
\begin{equation}
 B_{im}(\mathbf{r},\mathbf{r}_1;t) = \partial_k \, C_{ik,m}(x(\mathbf{r},t))
 \, \delta( \mathbf{r} - \mathbf{r}_1 )
 \label{equation::F_210}
\end{equation}
mit einer Matrix $C_{ik,m}(x(\mathbf{r},t))$, für die in Analogie zu \eqref{equation::F_040} 
die Gleichung 
\begin{equation}
 \sum_m C_{ik,m}(x(\mathbf{r},t)) \, C_{jl,m}(x(\mathbf{r},t)) = 
 2 \, k_B \, N_{ik,jl}(x(\mathbf{r},t))
 \label{equation::F_220}
\end{equation}
gilt. Wir setzen nun \eqref{equation::F_210} in die Bedingung \eqref{equation::F_200} 
ein, berechnen sorgfältig die Funktional-Ableitungen und finden
\begin{widetext}
\begin{equation}
 \sum_i \int d^d r \ \frac{ \delta B_{im}(\mathbf{r},\mathbf{r}_1;t) }{ \delta x_i(\mathbf{r},t) } \ = 
 \ \sum_{ik} \int d^d r \ \partial_k \, \frac{ \partial C_{ik,m}(x(\mathbf{r},t)) }{ \partial x_i(\mathbf{r},t) } 
 \, \delta( \mathbf{0} ) \ = \ 0 \ .
 \label{equation::F_230}
\end{equation}
\end{widetext}
Durch die Funktional-Ableitung kommt ein unendlicher Faktor $\delta( \mathbf{0} )$ hinzu, 
der einer räumlichen Delta-Funktion bei $\mathbf{r} = \mathbf{0}$ entspricht. Der Integrand 
des räumlichen Integrals ist offensichtlich eine räumliche Divergenz. Daher lässt sich mit 
dem Satz von Gauß das räumliche Integral in ein Oberflächenintegral umwandeln. Weil die 
Oberfläche im Unendlichen liegt, ist das Integral null. Somit ist gezeigt, dass für eine 
normale Flüssigkeit die Bedingung \eqref{equation::F_200} erfüllt ist.

Als nächstes betrachten wir den reversiblen Term von der Langevin-Gleichung 
\eqref{equation::F_120}. Wir identifizieren diesen mit dem ersten Term in unserer Gleichung 
\eqref{equation::F_170} und erhalten
\begin{equation}
 V_i(x(t)) = L_{ik}(x(t)) \, \frac{ \partial E(x(t)) }{ \partial x_k(t) } \ .
 \label{equation::F_240}
\end{equation}
Wir müssen jetzt noch nachprüfen, dass die Bedingung \eqref{equation::F_090} erfüllt ist. 
Leider hat unser reversibler Term \eqref{equation::F_240} nicht die Form \eqref{equation::F_100}. 
Zwar kann man die antisymmetrische Matrix $A^{(2)}_{ij}(x(t))=k_B T \, L_{ik}(x(t))$ 
identifizieren. Es fehlt jedoch der zweite Term in \eqref{equation::F_100} mit der Ableitung 
dieser Matrix. Weiterhin identifizieren wir die freie Energie $F(x(t))=E(x(t))/k_B T$, benötigen 
würden wir jedoch $F(x(t))=\Omega(x(t))/k_B T$. Eine Umformung mit der Formel 
\eqref{equation::F_160} und den Nebenbedingungen des GENERIC-Formalismus ist nicht 
selbstverständlich, weil die rechten Seiten dieser Nebenbedingungen im allgemeinen nicht null sind. 

Wir müssen also \eqref{equation::F_090} explizit nachprüfen und zum einen Symmetrieargumente 
und zum anderen die Eigenschaften von Erhaltungsgrößen anwenden. Dazu setzen wir den reversiblen 
Term \eqref{equation::F_240} zusammen mit der Boltzmann-Verteilung \eqref{equation::F_180} 
in die Bedingung \eqref{equation::F_090} ein. Nach einigen Umformungen erhalten wir
\begin{eqnarray}
 0 &=& \frac{ \partial }{ \partial x_i } \bigl[ V_i(x) P_\mathrm{eq}(x) \bigr] \nonumber \\
 &=& \left[ \frac{ \partial V_i(x) }{ \partial x_i } 
 - \frac{ 1 }{ k_B T } \frac{ \partial \Omega(x) }{ \partial x_i } V_i(x) \right] 
 P_\mathrm{eq}(x) \nonumber \\
 &=& \left[ \frac{ \partial V_i(x) }{ \partial x_i } 
 - \frac{ 1 }{ k_B T } \frac{ \partial \Omega(x) }{ \partial x_i } L_{ik}(x) 
 \frac{ \partial E(x) }{ \partial x_k } \right]  P_\mathrm{eq}(x) \nonumber \\ 
 &=& \left[ \frac{ \partial V_i(x) }{ \partial x_i } 
 - \frac{ 1 }{ k_B T } \{ \Omega(x), E(x) \} \right]  P_\mathrm{eq}(x) \ . 
 \label{equation::F_250}
\end{eqnarray}
Der zweite Term ist null. Das haben wir bereits in Kapitel \ref{section::3} mit der 
Poisson-Klammer \eqref{equation::C_950} gezeigt. Wir finden also für den reversiblen 
Term die notwendige Bedingung 
\begin{equation}
 \frac{ \partial V_i(x(t)) }{ \partial x_i(t) } = 0 \ .
 \label{equation::F_260}
\end{equation}
Wir fügen jetzt zwischendurch die Ortsvariable $\mathbf{r}$ wieder ein. Für eine normale 
Flüssigkeit sind die stochastischen Variablen $x_i(\mathbf{r},t)$ die Dichten von 
Erhaltungsgrößen. In Kapitel \ref{section::4C} haben wir gezeigt, dass die reversiblen 
Terme sich als Divergenzen von Stromdichten darstellen lassen. Daher finden wir
\begin{equation}
 V_i(x(\mathbf{r},t)) = - \nabla \cdot \mathbf{J}_i(x(\mathbf{r},t)) \ . 
 \label{equation::F_270}
\end{equation}
Wir setzen dies in die notwendige Bedingung \eqref{equation::F_260} ein und erhalten
\begin{eqnarray}
 \sum_i \int d^d r \, \frac{ \delta V_i(x(\mathbf{r},t)) }{ \delta x_i(\mathbf{r},t) }
 &=& - \sum_i \int d^d r \ \nabla \cdot
 \frac{ \partial \mathbf{J}_i(\mathbf{r},t) }{ \partial x_i(\mathbf{r},t) } 
 \, \delta( \mathbf{0} ) \nonumber \\
 &=& 0 \ .
 \label{equation::F_280}
\end{eqnarray}
Durch die Funktional-Ableitung kommt wiederum ein unendlicher Faktor $\delta( \mathbf{0} )$ 
hinzu, der einer räumlichen Delta-Funktion bei $\mathbf{r} = \mathbf{0}$ entspricht. 
Nach dem Satz von Gauß formen wir das Integral über den Raum in ein Oberflächenintegral 
um. Weil die Oberfläche im Unendlichen liegt, ist das Integral null. Somit ist die 
notwendige Bedingung für die reversiblen Terme erfüllt. 

Wir fassen zusammen: Mit den Gleichungen \eqref{equation::F_200} und \eqref{equation::F_260} 
gibt es zwei Bedingungen, welche erfüllt werden müssen, damit unsere hydrodynamischen 
Gleichungen für eine normale Flüssigkeit mit gaußischen Fluktuationen kompatibel sind 
mit der großkanonischen Boltzmann-Verteilung \eqref{equation::F_180} im thermischen 
Gleichgewicht. Die Bedingungen \eqref{equation::F_200} und \eqref{equation::F_260} 
werden hier besonders einfach erfüllt, weil sie sich auf Integrale über 
\emph{Divergenzen von Vektorfunktionen} zurückführen lassen, die mit dem Satz von Gauß 
in Oberflächenintegrale umgewandelt werden können und somit null ergeben. Man sieht 
das deutlich an den Ausdrücken \eqref{equation::F_210} und \eqref{equation::F_270} und 
den Gleichungen \eqref{equation::F_230} und \eqref{equation::F_280}. Diese Tatsache 
hängt damit zusammen, dass wir für die relevanten hydrodynamischen Variablen die 
\emph{Dichten von Erhaltungsgrößen} gewählt haben. Als Folge davon sind die 
hydrodynamischen Gleichungen Kontinuitätsgleichungen, so dass auf den rechten Seiten 
immer \emph{Divergenzen von Stromdichten} stehen.

Die hydrodynamischen Gleichungen können erweitert werden durch zusätzliche Variablen 
$x_i(\mathbf{r},t)$, welche die Ordnungsparameter für die Brechung von bestimmten 
Symmetrien darstellen und Phasenübergänge zweiter Ordnung beschreiben \cite{CL95}. 
Beispiele wären die Magnetisierung $\mathbf{m}(\mathbf{r},t)$ eines ferromagnetischen 
Systems oder die Kondensatwellenfunktion $\Psi(\mathbf{r},t)$ von superfluidem 
$^4\mathrm{He}$. In diesen Fällen lässt sich die Bedingung \eqref{equation::F_200} 
im allgemeinen nicht erfüllen, wenn die entsprechenden Komponenten der Onsager-Matrix 
$M_{ij}(x(t))$ und folglich die entsprechenden Komponenten der Matrix $B_{im}(x(t))$ 
nicht konstant sind sondern von den hydrodynamischen Variablen $x_i(t)$ abhängen. Man 
kann jedoch andererseits durch Symmetrieargumente zeigen, dass die reversiblen Terme 
\eqref{equation::F_240} die notwendige Bedingung \eqref{equation::F_260} immer erfüllen. 
Das reicht allerdings nicht aus. Wir können im allgemeinen nicht garantieren, dass die 
erweiterten hydrodynamischen Gleichungen mit Fluktuationen kompatibel sind mit der 
großkanonischen Boltzmann-Verteilung \eqref{equation::F_180} im thermischen Gleichgewicht.

Die Langevin-Gleichung des GENERIC-Formalismus \eqref{equation::F_130} wurde unter 
der Annahme hergeleitet, dass die relevante Dichtematrix durch eine lokale großkanonische  
Verteilung der physikalischen Variablen in der Form \eqref{equation::B_260} gegeben 
ist \cite{Ot00}. Hierbei kommt es nicht darauf an, ob das zugrunde liegende mikroskopische 
System quantenmechanisch oder klassisch ist. Wir hatten jedoch einige Schwierigkeiten 
zu zeigen, dass das thermische Gleichgewicht zumindest für eine normale Flüssigkeit durch 
die großkanonische Boltzmann-Verteilung \eqref{equation::F_180} mit dem großkanonischen 
thermodynamischen Potential \eqref{equation::F_160} beschrieben wird. 

Alternativ ist im GENRIC-Formalismus eine direkte Herleitung der Fokker-Planck-Gleichung 
\eqref{equation::F_010} möglich \cite{Ot98}. Man betrachtet hierzu die 
\emph{Funktionswerte} der Verteilungsfunktion $f(x,t)$ der physikalischen 
Variablen $x_i(\mathbf{r},t)$ als \emph{relevante Variablen} und baut damit den 
Projektionsoperator-Formalismus auf. Das zugrunde liegende Ensemble ist in diesem 
Fall \emph{mikrokanonisch}. Folglich ist die Herleitung hier nur klassisch möglich, 
nicht quantenmechanisch. Man findet als Ergebnis, dass die großkanonische 
Boltzmann-Verteilung \eqref{equation::F_180} automatisch eine stationäre Lösung der 
Fokker-Planck-Gleichung für das thermische Gleichgewicht ist. Sowohl der dissipative 
als auch der reversible Term haben mit \eqref{equation::F_080} und \eqref{equation::F_100} 
die erforderlichen Formen. Andererseits sind hier die Formeln für die Poisson-Matrix 
$L_{ik}(x)$ und die Formeln für die Onsager-Matrix $M_{ik}(x)$ wesentlich komplizierter 
sind \cite{Ot98}. Eine ähnliche Herleitung der Fokker-Planck-Gleichung für die 
nichtlineare Hydrodynamik mit Fluktuationen wurde zuvor von Zubarev und Morozov 
gegeben \cite{ZM83}, und die zugehörigen Langevin-Gleichungen wurden von Morozov 
\cite{Mo84} und weiterhin von Kim und Mazenko \cite{KM91} gefunden. Der 
GENERIC-Formalismus ist jedoch wesentlich eleganter für die Herleitung der Gleichungen 
und für die Behandlung der Nichtlinearitäten und Fluktuationen.

Zum Schluss untersuchen wir noch einen Zusammenhang von besonderer Art. In der 
normalen Flüssigkeit sind die Temperatur $T(\mathbf{r},t)$, die Geschwindigkeit 
$\mathbf{v}(\mathbf{r},t)$ und das chemische Potential $\mu(\mathbf{r},t)$ lokale 
fluktuierende Größen, die durch die Funktional-Ableitungen der Entropie 
\eqref{equation::D_620}-\eqref{equation::D_640} definiert sind. Andererseits haben 
wir die konstanten Lagrange-Parameter $T$, $\mathbf{v}$ und $\mu$. Wir wollen 
herausfinden, wie diese miteinander zusammenhängen. Dazu berechnen wir das Integral 
von der Ableitung der Boltzmann-Verteilung und finden mit dem Satz von Gauß
\begin{equation}
 \int Dx \ \frac{ \partial }{ \partial x_i } \, P_\mathrm{eq}(x) = 0 \ .
 \label{equation::F_290}
\end{equation}
Das Maß $Dx$ bedeutet, dass das Integral über alle hydrodynamischen Variablen $x_i$ 
berechnet wird. Wenn die hydrodynamischen Variablen $x_i(\mathbf{r})$ auch noch von 
der Ortsvariablen $\mathbf{r}$ abhängen, dann ist das Integral ein 
\emph{Funktionalintegral}, und die Ableitungen sind \emph{Funktionalableitungen} 
\cite{Am84}. Berechnen wir nun für die normale Flüssigkeit das Funktionalintegral 
mit den Funktional-Ableitungen nach der Energiedichte $\varepsilon(\mathbf{r},t)$, 
nach der Impulsdichte $\mathbf{j}(\mathbf{r},t)$ und nach der Massendichte 
$\rho(\mathbf{r},t)$, so finden wir die Beziehungen
\begin{eqnarray}
 \left\langle \frac{ 1 }{ T(\mathbf{r},t) } \right\rangle_\mathrm{eq} &=& \frac{ 1 }{ T } \ ,
 \label{equation::F_300} \\
 \left\langle \frac{ \mathbf{v}(\mathbf{r},t) }{ T(\mathbf{r},t) } \right\rangle_\mathrm{eq} 
 &=& \frac{ \mathbf{v} }{ T } \ ,
 \label{equation::F_310} \\
 \left\langle \frac{ \mu(\mathbf{r},t) }{ T(\mathbf{r},t) } \right\rangle_\mathrm{eq} 
 &=& \frac{ \mu }{ T } \ ,
 \label{equation::F_320}
\end{eqnarray}
wobei die Erwartungswerte im thermischen Gleichgewicht berechnet werden. Die 
Lagrange-Parameter $T$, $\mathbf{v}$ und $\mu$ entsprechen also nicht direkt den 
physikalischen Größen Temperatur, Geschwindigkeit und chemisches Potential, sondern 
sie hängen über bestimmte Erwartungswerte mit diesen physikalischen Größen zusammen.

\subsection{Entropie im thermischen Gleichgewicht}
\label{section::6C}
Die Entropiegleichung \eqref{equation::D_130} lässt sich ebenfalls einfacher schreiben, 
wenn wir die Ortsvariable $\mathbf{r}$ weglassen und die Summenkonvention für die Indizes 
verwenden. Wir erhalten dann
\begin{eqnarray}
 \frac{ d }{ dt } \, S(x(t)) &=& \frac{ \partial S(x(t)) }{ \partial x_i(t) }
 \, M_{ik}(x(t)) \, \frac{ \partial S(x(t)) }{ \partial x_k(t) } \nonumber \\
 && + \, \frac{ \partial S(x(t)) }{ \partial x_i(t) } \, f_i(t) \ . 
 \label{equation::F_330}
\end{eqnarray}
Der erste Term auf der rechten Seite ist der dissipative Term. Da die Onsager-Matrix 
$ M_{ik}(x(t))$ positiv definit ist, liefert er immer einen Beitrag größer gleich null. 
Dieser Term bewirkt also das Anwachsen der Entropie und folglich den zweiten Hauptsatz 
der Thermodynamik. 

Im thermischen Gleichgewicht gibt es auch Fluktuationen. Folglich wird auch im thermischen 
Gleichgewicht der erste dissipative Term immer größer oder gleich null sein. Diesem Trend muss 
der zweite fluktuierende Term entgegenwirken. Denn im thermischen Gleichgewicht hat die Entropie 
bereits ihr Maximum erreicht und muss im Mittel konstant bleiben. Ein langfristiger Drift 
der Entropie nach oben oder unten wäre ein Widerspruch. Wir wollen das hier jetzt nachprüfen 
und damit zeigen, dass unsere stochastische Theorie für die Hydrodynamik einer normalen 
Flüssigkeit in sich konsistent ist.

Im Folgenden berechnen wir den ersten Kramers-Moyal-Koeffizienten für die Entropie 
$K^{(1)}_S(x(t))$. Wir tun dies so, wie man gewöhnlich aus stochastischen Gleichungen 
die Kramers-Moyal-Koeffizienten berechnet. Dazu integrieren wir die Gleichung 
\eqref{equation::F_330} und erhalten
\begin{eqnarray}
 S(x(t_2)) &=& S(x(t_1)) \nonumber \\
 && + \, \int_{t_1}^{t_2} dt \ \biggl[ \frac{ \partial S(x(t)) }{ \partial x_i(t) }
 \, M_{ik}(x(t)) \, \frac{ \partial S(x(t)) }{ \partial x_k(t) } \nonumber \\
 && \hspace{17mm} + \, \frac{ \partial S(x(t)) }{ \partial x_i(t) } \, f_i(t) \biggr] \ .
 \label{equation::F_340}
\end{eqnarray}
In derselben Weise integrieren wir auch die Langevin-Gleichung \eqref{equation::F_130} und 
erhalten eine Integralgleichung für die stochastischen Variablen $x_i(t)$. Nachfolgend 
ersetzen wir auf der rechten Seite von \eqref{equation::F_340} die stochastischen Variablen 
$x_i(t)$ für Zeiten $t>t_1$ wiederholt mit dieser Integralgleichung, so dass wir eine 
verschachtelte Störungsreihe bekommen, die nur von den stochastischen Variablen $x_i(t_1)$ 
zum Zeitpunkt $t_1$ abhängen, und von den elementaren gaußischen stochastischen Kräften 
$\varepsilon_m(t)$ mit Zeiten $t$ im Intervall $t_1<t<t_2$. Wir nehmen an, dass die Differenz 
$\Delta t= t_2 - t_1$ klein ist und entwickeln die rechte Seite nach Potenzen von $\Delta t$ 
bis zur ersten Ordnung.

Anschließend bilden wir den Mittelwert bezüglich der elementaren stochastischen Kräfte 
$\varepsilon_m(t)$ mit Zeiten $t$ im Intervall $t_1<t<t_2$. Dies entspricht einem Mittelwert 
mit einer bedingten zusammenhängenden Wahrscheinlichkeitsverteilung. Als Ergebnis erhalten 
wir dann
\begin{eqnarray}
 \langle S(x(t_2)) \rangle_\mathrm{bedingt} &=& S(x(t_1)) \, 
 + \, ( t_2 - t_1 ) \, K^{(1)}_S(x(t_1)) \nonumber \\ 
 && + \, O( ( t_2 - t_1 )^2 )
 \label{equation::F_350}
\end{eqnarray}
mit dem ersten Kramers-Moyal-Koeffizienten
\begin{eqnarray}
 K^{(1)}_S(x(t)) &=& \frac{ \partial S(x(t)) }{ \partial x_i(t) }
 \, M_{ik}(x(t)) \, \frac{ \partial S(x(t)) }{ \partial x_k(t) } \nonumber \\
 && + \, \frac{ \partial }{ \partial x_k(t) } \frac{ \partial S(x(t)) }{ \partial x_i(t) } 
 \, k_B \, M_{ik}(x(t)) \ . \qquad
 \label{equation::F_360}
\end{eqnarray}
Wir bilden weiterhin den Erwartungswert mit der Wahrscheinlichkeitsverteilung $P(x(t_1),t_1)$ 
für die stochastischen Variablen $x_i(t_1)$ zum Zeitpunkt $t_1$. Aus \eqref{equation::F_350} 
erhalten wir dann
\begin{eqnarray}
 \langle S(x(t_2)) \rangle &=& \langle S(x(t_1)) \rangle \, 
 + \, ( t_2 - t_1 ) \, \langle K^{(1)}_S(x(t_1)) \rangle \nonumber \\ 
 && + \, O( ( t_2 - t_1 )^2 ) \ ,
 \label{equation::F_370}
\end{eqnarray}
und im Limes $\Delta t = t_2 - t_1 \to 0$ folgt
\begin{equation}
 \frac{ d }{ d t } \langle S(x(t)) \rangle = \langle K^{(1)}_S(x(t)) \rangle \ .
 \label{equation::F_380}
\end{equation}

Wir berechnen nun den Erwartungswert des Kramers-Moyal-Koeffizienten mit der großkanonischen 
Boltzmann-Verteilung \eqref{equation::F_180} und erhalten
\begin{widetext}
\begin{equation}
 \langle K^{(1)}_S(x) \rangle_\mathrm{eq} = \int Dx \ \biggl[ \frac{ \partial S(x) }{ \partial x_i }
 \, M_{ik}(x) \, \frac{ \partial S(x) }{ \partial x_k } 
 + \, \left( \frac{ \partial }{ \partial x_k } \frac{ \partial S(x) }{ \partial x_i } 
 k_B \, M_{ik}(x) \right) \biggr] P_\mathrm{eq}(x) \ .
 \label{equation::F_390}
\end{equation}
Es gelten die Nebenbedingung \eqref{equation::F_150} für die Energie $E(x)$ und ebenso für 
die weiteren Erhaltungsgrößen $\mathbf{P}(x)$ und $N(x)$. Daher dürfen wir im ersten Term 
mit Hilfe der Formel \eqref{equation::F_160} die Entropie durch das großkanonische 
thermodynamische Potential ersetzen gemäß $S(x) \to - \Omega(x)/T$. Wir bekommen dann 
\begin{eqnarray}
 \langle K^{(1)}_S(x) \rangle_\mathrm{eq} &=& \int Dx \ \biggl[ - \frac{ \partial S(x) }{ \partial x_i }
 \, M_{ik}(x) \, \frac{ 1 }{ T } \, \frac{ \partial \Omega(x) }{ \partial x_k } \,
 + \, \left( \frac{ \partial }{ \partial x_k } \frac{ \partial S(x) }{ \partial x_i } 
 k_B \, M_{ik}(x) \right) \biggr] P_\mathrm{eq}(x) \nonumber \\
 &=& \int Dx \ \biggl[ \frac{ \partial S(x) }{ \partial x_i }
 \, k_B M_{ik}(x) \, \frac{ \partial }{ \partial x_k } \,
 + \, \left( \frac{ \partial }{ \partial x_k } \frac{ \partial S(x) }{ \partial x_i } 
 k_B \, M_{ik}(x) \right) \biggr] P_\mathrm{eq}(x) \nonumber \\
 &=& \int Dx \ \frac{ \partial }{ \partial x_k } \left[ \frac{ \partial S(x) }{ \partial x_i } 
 k_B \, M_{ik}(x) \, P_\mathrm{eq}(x) \right] \ .
 \label{equation::F_400}
\end{eqnarray}
\end{widetext}
Das letzte Integral über die stochastischen Variablen lässt sich mit dem Satz von Gauß 
in ein Oberflächenintegral umformen und ergibt somit null. Als Ergebnis finden wir folglich  
\begin{equation}
 \frac{ d }{ d t } \langle S(x(t)) \rangle_\mathrm{eq} 
 = \langle K^{(1)}_S(x(t)) \rangle_\mathrm{eq} = 0 \ .
 \label{equation::F_410}
\end{equation}
Im thermischen Gleichgewicht bleibt also die Entropie $S(x(t))$ im Mittel konstant, wie 
wir es erwartet haben. Unsere stochastischen Gleichungen mit den gaußisch fluktuierenden 
Kräften \eqref{equation::F_020} erfüllen diese Erwartung.

\section{Fluktuations-Theorem}
\label{section::7}
Nach unseren Untersuchungen gilt der zweite Hauptsatz der Thermodynamik nur im Mittel.
Das bedeutet, die mittlere Entropie $\langle S(x(t)) \rangle$ wächst mit der Zeit an 
oder bleibt zuminest konstant. Folglich gilt die Ungleichung
\begin{equation}
 \langle \Delta S \rangle = \langle S(x(t^\prime)) - S(x(t)) \rangle \geq 0
 \label{equation::G_010}
\end{equation}
für $\Delta t = t^\prime - t > 0 $. Fluktuationen bewirken jedoch, dass für kurze 
Zeitintervalle $\Delta t$ die Entropie $S(x(t))$ vorübergehend auch mal sinken 
kann, so dass auch mal $\Delta S = S(x(t^\prime)) - S(x(t)) < 0$ für bestimmte 
Zeitintervalle $\Delta t = t^\prime - t$ möglich ist.

Ähnliche Fragestellungen wurden vor zwanzig Jahren von Evans \emph{et al.}\ \cite{Ev93,Ev94} 
im Rahmen von Computersimulationen an mesoskopischen Vielteilchen-Systemen untersucht. Diese 
Untersuchungen führten zur Formulierung des sogenannten \emph{Fluktuations-Theorems}. 
Eine zusammenfassende Darstellung ist in dem Übersichtsartikel von Evans und Searles 
\cite{Ev02a} zu finden. Experimentell wurde das Fluktuations-Theorem in kleinen Systemen 
mit kolloidalen Teilchen nachgewiesen \cite{Ev02b}. Wir wollen hier untersuchen, in wie 
weit das Fluktuations-Theorem auf die Hydrodynamik einer normalen Flüssigkeit anwendbar ist 
und ob es überhaupt gilt.

Während Evans \emph{et al.}\ \cite{Ev93,Ev94} das Fluktuations-Theorem für 
Vielteilchen-Systeme beobachteten, formulierten und heuristisch erklärten, wurden 
mathematische Beweise später von Gallavotti und Cohen \cite{GC95a,GC95b}, von Kurchan 
\cite{Ku98} und von Lebowitz und Spohn \cite{LS99} geliefert. Eine einfache Herleitung 
des Fluktuations-Theorems wurde von Crooks \cite{Cr98,Cr99,Cr00} entwickelt, und zwar 
für physikalische Systeme, welche allgemein durch stochastische Prozesse ohne Gedächtnis, 
also Markov-Prozesse, beschrieben werden. Eine normale Flüssigkeit, welche 
durch die hydrodynamischen Gleichungen mit Fluktuationen von Kapitel \ref{section::4} 
beschrieben wird, gehört zu dieser Klasse von physikalischen Systemen. Daher werden 
wir im Folgenden die Herleitung des Fluktuations-Theorem nach Crooks \cite{Cr98,Cr99} 
im Rahmen des GENERIC-Formalismus darstellen.

\subsection{Mikroreversibilität und detailliertes Gleichgewicht}
\label{section::7A}
Das Fluktuations-Theorem ist eine Folge von Mikroreversibilität und der Existenz eines 
detailliertes Gleichgewichts. Aus diesem Grunde werden wir uns zunächst mit diesen 
Fragestellungen befassen. Wir nehmen an, es seien $t$ und $t^\prime$ zwei Zeitpunkte, 
wobei $t$ \emph{früher} und $t^\prime$ \emph{später} ist. Die Differenz 
$\Delta t = t^\prime - t$ ist also nach Annahme positiv. Die hydrodynamischen Variablen 
zu diesen Zeitpunkten bezeichnen wir kurz mit $x_i$ und $x^\prime_i$. 

Die Fokker-Planck-Gleichung \eqref{equation::F_010} und alternativ die 
Langevin-Gleichung \eqref{equation::F_050} zusammen mit den stochastischen Kräften 
\eqref{equation::F_020} beschreiben einen stochastischen Prozess, der zeitlich in 
\emph{Vorwärtsrichtung} fortschreitet. Als Lösung finden wir die bedingte 
Wahrscheinlichkeit $P_F( x^\prime | x )$, welche die Verteilung der hydrodynamischen 
Variablen $x^\prime_i$ zum späteren Zeitpunkt $t^\prime$ darstellt, wenn sich das 
System zum früheren Zeitpunkt $t$ in einem Zustand mit den hydrodynamischen 
Variablen $x_i$ befindet. Der Index $F$ soll andeuten, dass es sich um einen 
\emph{Vorwärtsprozess} handelt.

Alternativ kann man eine Fokker-Planck-Gleichung und eine Langevin-Gleichung angeben, 
welche den zeitlich umgekehrten stochastischen Prozess beschreibt, der zeitlich in 
\emph{Rückwärtsrichtung} verläuft. Die Gleichungen ändern sich dann in sofern, dass 
sich das Vorzeichen des dissipativen Terms im ersten Kramers-Moyal-Koeffizienten 
umdreht. Als Lösung finden wir in diesem Fall die bedingte Wahrscheinlichkeit 
$P_R( x | x^\prime )$, welche die Verteilung der hydrodynamischen Variablen $x_i$ 
zum früheren Zeitpunkt $t$ darstellt, wenn sich das System zum späteren Zeitpunkt 
$t^\prime$ in einem Zustand mit den hydrodynamischen Variablen $x^\prime_i$ befindet. 
Der Index $R$ soll andeuten, dass es sich um einen \emph{Rückwärtsprozess} handelt.

Das mikroskopische physikalische System, welches den hydrodynamischen Gleichungen 
mit Fluktuationen zugrunde liegt, ist symmetrisch unter Zeitumkehr. Das bedeutet, 
auf mikroskopischer Ebene müssen physikalische Prozesse in gleicher Weise zeitlich 
vorwärts und zeitlich rückwärts ablaufen können. In der statistischen Physik der 
stochastischen Prozesse führt diese \emph{Mikroreversibilität} auf die Existenz 
eines \emph{detaillierten Gleichgewichts} mit einer Verteilungsfunktion 
$P_\mathrm{eq}(x)$. Die Wahrscheinlichkeit für den Vorwärtsprozess muss gleich der 
Wahrscheinlichkeit für den Rückwärtsprozess sein. Es gilt folglich die Bedingung 
\begin{equation}
 P_F( x^\prime | x ) \ P_\mathrm{eq}(x) \ = \  P_R( x | x^\prime ) \ P_\mathrm{eq}(x^\prime) \ .
 \label{equation::G_020}
\end{equation}
Setzen wir hier für das thermische Gleichgewicht unsere großkanonische Boltzmann-Verteilung 
\eqref{equation::F_180} ein, und kürzen wir auf beiden Seiten den Normierungsfaktor $Z$ 
heraus, so erhalten wir 
\begin{equation}
 P_F( x^\prime | x ) \ \exp \left( - \frac{ \Omega(x) }{ k_B T } \right) 
 \ = \  P_R( x | x^\prime ) \ \exp \left( - \frac{ \Omega(x^\prime) }{ k_B T } \right)
 \label{equation::G_030}
\end{equation}
mit dem großkanonischen thermodynamischen Potential $\Omega(x)$, definiert in der 
Gleichung \eqref{equation::F_160}.

In einer normalen Flüssigkeit sind die Energie $E(x)$, der Impuls $\mathbf{P}(x)$ 
und die Teilchenzahl $N(x)$ Erhaltungsgrößen. Folglich enthalten die bedingten 
Wahrscheinlichkeiten $P_F( x^\prime | x )$ und $P_R( x | x^\prime )$ Faktoren mit 
Delta-Funktionen in diesen Erhaltungsgrößen, und zwar $\delta( E(x^\prime) - E(x) )$, 
$\delta( \mathbf{P}(x^\prime) - \mathbf{P}(x) )$ und $\delta( N(x^\prime) - N(x) )$. 
Setzen wir nun das großkanonische thermodynamische Potential \eqref{equation::F_160} 
ein, so stellen wir fest, dass sich in \eqref{equation::G_030} alle Terme mit der 
Energie $E(x)$, dem Impuls $\mathbf{P}(x)$ und der Teilchenzahl $N(x)$ auf beiden 
Seiten wegkürzen. Es bleiben nur die Terme mit der Entropie $S(x)$ übrig, und die 
Bedingung für das detaillierte Gleichgewicht vereinfacht sich auf
\begin{equation}
 P_F( x^\prime | x ) \ \exp\bigl( S(x) / k_B \bigr) 
 \ = \  P_R( x | x^\prime ) \ \exp\bigl( S(x^\prime) / k_B \bigr) \ .
 \label{equation::G_040}
\end{equation}
Die Bedingung für das detaillierte Gleichgewicht kann man auch direkt aus der 
Fokker-Planck-Gleichung \eqref{equation::F_010} beweisen, siehe dazu Graham und  
Haken \cite{GH71A,GH71B,Gr73}. Alternativ kann man den Beweis mit einer 
Funktional-Integral-Darstellung der dynamischen stochastischen Prozesse von Janssen 
\cite{Ja76} und de Dominicis \cite{DD76} durchführen. Im Ergebnis findet man eine 
Gleichung wie \eqref{equation::G_040}, wobei im Argument der Exponentialfunktionen 
auf beiden Seiten immer jenes Funktional steht, welches den dissipativen Term in 
der Langevin-Gleichung bestimmt. In der Langevin-Gleichung des GENERIC-Formalismus 
\eqref{equation::F_130} ist der zweite Term auf der rechten Seite der dissipative 
Term. Dieser setzt sich zusammen aus dem Produkt der Onsager Matrix $M_{ik}(x)$ 
und der Ableitung der Entropie $\partial S(x) / \partial x_k$. Folglich ist für 
den GENERIC-Formalismus die Bedingung für Mikroreversibilität und das detaillierte 
Gleichgewicht durch die Gleichung \eqref{equation::G_040} gegeben, wobei im Argument 
der Experimentalfunktion auf beiden Seiten die Entropie $S(x)$ steht.

Wir stellen fest, dass die Gleichung \eqref{equation::G_040} die allgemeinere und 
fundamentalere Form für die Bedingung des detaillierte Gleichgewichts ist. Sie gilt 
im GENERIC-Formalismus auch dann, wenn die Energie $E(x)$ und der Impuls $\mathbf{P}(x)$ 
nicht erhalten sind. Letzteres ist der Fall, wenn die Flüssigkeit in einem Volumen 
eingesperrt ist, welches sich zeitlich verändert. Man kann diese physikalische 
Situation erreichen, indem man zu der Energie $E(x)$ einen Term mit einem zeitlich und 
räumlich veränderlichen Potential hinzufügt. Die Gleichungen \eqref{equation::G_030} 
und \eqref{equation::G_020} sind dagegen speziellere Bedingungen, welche nur dann 
erfüllt sind, wenn die Erhaltungssätze gelten.

Wir weisen darauf hin, dass die Gleichung \eqref{equation::G_040} für den 
GENERIC-Formalismus unter der Annahme bewiesen wird, dass die Matrix $B_{im}(x)$ 
in der Definition der fluktuierenden Kräfte \eqref{equation::F_020} und der reversible 
Term \eqref{equation::F_240} die notwendigen Bedingungen \eqref{equation::F_200} und 
\eqref{equation::F_260} erfüllen. Diese zwei Bedingungen wurden in Abschnitt 
\ref{section::6B} verwendet, um zu zeigen, dass die großkanonische Boltzmann-Verteilung 
\eqref{equation::F_180} die Verteilung für das thermische Gleichgewicht ist.

\subsection{Das Fluktuations-Theorem in seiner ursprünglichen Form}
\label{section::7B}
Wir nehmen nun an, der Anfangszustand für den Vorwärtsprozess wird zur Zeit $t$ durch 
eine Wahrscheinlichkeitsverteilung $P_{F,0}(x)$ beschrieben. Entsprechend nehmen wir 
an, der Anfangszustand für den Rückwärtsprozess wird zur Zeit $t^\prime$ durch eine 
Wahrscheinlichkeitsverteilung $P_{R,0}(x^\prime)$ beschrieben. Damit definieren wir 
die Verbundwahrscheinlichkeiten 
\begin{eqnarray}
 P_F( x^\prime, x ) &=& P_F( x^\prime | x ) \ P_{F,0}(x) \ ,
 \label{equation::G_050} \\
 P_R( x, x^\prime ) &=& P_R( x | x^\prime ) \ P_{R,0}(x^\prime) \ ,
 \label{equation::G_060}
\end{eqnarray}
welche den Vorwärtsprozess und den Rückwärtsprozess zwischen den zwei Zeitpunkten 
$t$ und $t^\prime$ beschreiben. Aus der Bedingung für das detaillierte Gleichgewicht 
\eqref{equation::G_040} erhalten wir dann für die Verbundwahrscheinlichkeiten 
\begin{equation}
 P_F( x^\prime, x ) \ \exp\bigl( - \Delta\Sigma_F( x^\prime, x ) / k_B \bigr) \ = \  P_R( x, x^\prime )
 \label{equation::G_070}
\end{equation}
oder alternativ
\begin{equation}
 P_F( x^\prime, x ) \ = \  P_R( x, x^\prime ) \ \exp\bigl( - \Delta\Sigma_R( x, x^\prime ) / k_B \bigr)
 \label{equation::G_080}
\end{equation}
je nachdem ob die Exponentialfaktoren alle auf die linke oder auf die rechte Seite 
gebracht werden. Die Argumente der Exponentialfaktoren sind gegeben durch die Funktionen 
\begin{eqnarray}
 \Delta\Sigma_F( x^\prime, x ) &=& - \Delta\Sigma_R( x, x^\prime ) \nonumber\\
 &=& S(x^\prime) - S(x) \nonumber\\ 
 &&- k_B \bigl[ \ln P_{R,0}(x^\prime) - \ln P_{F,0}(x) \bigr] \ , \hspace{5mm}
 \label{equation::G_090}
\end{eqnarray}
welche sich im Vorzeichen für den Vorwärtsprozess und für den Rückwärtsprozess 
unterscheiden. 

Diese Funktionen ermöglichen die Definition einer neuen Variablen $\Delta\Sigma$, 
welche eine Art Entropie-Änderung des Zustands im Zeitintervall $\Delta t = t^\prime - t$ 
darstellt. Wir multiplizieren die Verbundwahrscheinlichkeiten \eqref{equation::G_050} 
und \eqref{equation::G_060} mit entsprechenden Delta-Funktionen und integrieren über 
die Variablen $x_i$ und $x^\prime_i$. Als Ergebnis erhalten wir dann 
Wahrscheinlichkeitsverteilungen für die Variable $\Delta\Sigma$, und zwar für den 
Vorwärtsprozess
\begin{equation}
 P_F(\Delta\Sigma) = \int Dx^\prime \int Dx \ \delta( \Delta\Sigma - \Delta\Sigma_F( x^\prime, x ) ) 
 \ P_F( x^\prime, x )
 \label{equation::G_100}
\end{equation}
und für den Rückwärtsprozess
\begin{equation}
 P_R(\Delta\Sigma) = \int Dx \int Dx^\prime \ \delta( \Delta\Sigma - \Delta\Sigma_R( x, x^\prime ) )
 \ P_R( x, x^\prime ) \ .
 \label{equation::G_110}
\end{equation}
Multiplizieren wir nun die Bedingungen \eqref{equation::G_070} oder \eqref{equation::G_080} 
auf beiden Seiten mit entsprechenden Delta-Funktionen und integrieren wir über die 
Variablen $x_i$ und $x^\prime_i$, so bekommen wir eine Bedingung für die 
Wahrscheinlichkeiten \eqref{equation::G_100} und \eqref{equation::G_110}, das wohl 
bekannte \emph{Fluktuations-Theorem von Crooks} \cite{Cr99}
\begin{equation}
 \frac{ P_F( + \Delta\Sigma ) }{ P_R( - \Delta\Sigma ) } \ = \ \exp( \Delta\Sigma / k_B ) \ .
 \label{equation::G_120}
\end{equation}
Die zwei unterschiedlichen Vorzeichen von $\Delta\Sigma$ in den Argumenten der 
Wahrscheinlichkeiten auf der linken Seite sind eine Folge der zwei Vorzeichen der 
Funktionen \eqref{equation::G_090} für den Vorwärtsprozess und für den Rückwärtsprozess.

Multiplizieren wir weiterhin in der Gleichung \eqref{equation::G_120} die Faktoren 
etwas um, und integrieren wir auch noch über die verbleibende Variable $\Delta\Sigma$, 
so erhalten wir das \emph{integrale Fluktuations-Theorem}
\begin{eqnarray}
 \langle \exp( - \Delta\Sigma / k_B ) \rangle 
 &=& \int d(\Delta\Sigma) \ P_F( \Delta\Sigma ) \ \exp( - \Delta\Sigma / k_B ) \nonumber\\ 
 &=& \int d(\Delta\Sigma) \ P_R( - \Delta\Sigma ) \ = \  1 \ .
 \label{equation::G_130}
\end{eqnarray}
Diese integrale Variante des Fluktuations-Theorems wurde in dieser Form erstmals 
allgemein von Crooks \cite{Cr99} und später speziell für kolloidale Teilchen in einem 
Lösungsmittel von Seifert \cite{Se05} hergeleitet. Sie entspricht einer 
Jarzynski-Gleichung \cite{Ja97A,Ja97B} für die Variable $\Delta\Sigma$, wie wir 
später sehen werden. 

Crooks unterteilt in seinen Darstellungen \cite{Cr98,Cr99,Cr00} das zeitliche Intervall 
$\Delta t = t^\prime - t$ in viele kleine infinitesimale Intervalle und betrachtet 
somit den gesamten stochastischen Prozess zwischen $t$ und $t^\prime$. Der Grund dafür 
ist, dass äußere Kräfte durch zeitabhängige Parameter beschrieben werden, welche korrekt 
berücksichtigt werden müssen. In unserem Fall liefert der GENERIC-Formalismus mit der 
Gleichung \eqref{equation::G_040} eine Bedingung für das detaillierte Gleichgewicht, 
in der die Exponentialfaktoren als Argumente bereits die Entropie haben. Somit reicht 
es aus, dass wir die Verbundwahrscheinlichkeiten \eqref{equation::G_050} und 
\eqref{equation::G_060} nur für zwei Zeiten, den Anfang $t$ und das Ende $t^\prime$ 
des Intervalls, definieren und betrachten. 

Die Fluktuations-Theoreme \eqref{equation::G_120} und \eqref{equation::G_130} haben 
zwar eine recht einfache Form. Die eigentliche Schwierigkeit liegt jedoch in der 
Bedeutung und der Interpretation der Variablen $\Delta\Sigma$, definiert in 
\eqref{equation::G_090}. In der Hydrodynamik ist $S(x)$ die Entropie. Diese lässt sich 
auf mikroskopischer Ebene durch die Maximierung \eqref{equation::B_210} unter den 
Nebenbedingen \eqref{equation::B_220} und \eqref{equation::B_230} definieren. 
Folglich ist der erste Term in der Formel \eqref{equation::G_090} die Entropie-Änderung 
$\Delta S = S(x^\prime) - S(x)$ des stochastischen Prozesses von der Anfangszeit $t$ 
zur Endzeit $t^\prime$. Andererseits besteht der zweite Term aus den Logarithmen der 
beiden Verteilungen der Anfangszustände für den Vorwärtsprozess $P_{F,0}(x)$ und für 
den Rückwärtsprozess $P_{R,0}(x^\prime)$. Diesen Beitrag kann man auch als eine Art 
Entropie interpretieren. Er ist jedoch zunächst völlig willkürlich, weil die beiden 
Verteilungen der Anfangszustände beliebig gewählt werden können. Die genaue 
physikalische Bedeutung der Variablen $\Delta\Sigma$ wird also erst dann festgelegt, 
wenn man für die Anfangsverteilungen $P_{F,0}(x)$ und $P_{R,0}(x^\prime)$ eine 
konkrete Wahl trifft.

Eine ausführliche und allgemeine Darstellung des Fluktuations-Theorems in seinen 
verschiedenen Varianten ist in dem Übersichtsartikel von Seifert \cite{Se12} zu finden. 
Die Grundlage dort ist eine Langevin-Gleichung für kolloidale Teilchen in einem 
Lösungsmittel. Es wird eine Entropievariable ähnlich wie $\Delta\Sigma$ in 
\eqref{equation::G_090} betrachtet, mit ähnlicher Struktur und mit ebenfalls zwei 
Beiträgen. Der erste Term ist die Entropie-Änderung im Medium oder Lösungsmittel, 
vergleichbar mit unserer Entropie-Änderung $\Delta S = S(x^\prime) - S(x)$ in der 
Flüssigkeit. Der zweite Term wird als Entropie-Änderung des kolloidalen 
Vielteilchen-Systems interpretiert. Wir stellen fest, dass unsere Betrachtungen des 
Fluktuations-Theorems für die Hydrodynamik einer normalen Flüssigkeit im Rahmen des 
GENERIC-Formalismus zu ähnlichen Ergebnissen führt wie die Darstellung von Seifert 
\cite{Se12} für kolloidale Systeme.

Im Folgenden verwenden wir als spezielle Wahl für die Anfangsverteilungen $P_{F,0}(x)$ 
und $P_{R,0}(x^\prime)$ die Boltzmann-Verteilungen für das thermische Gleichgewicht 
\eqref{equation::F_180} zusammen mit dem großkanonischen thermodynamischen Potential 
\eqref{equation::F_160}. Wir setzen also
\begin{eqnarray}
 P_{F,0}(x) &=& P_\mathrm{eq}(x) \ = \  Z^{-1} 
 \  \exp\bigl( - \Omega(x) / k_B T \bigr) \ , \hspace{10mm}
 \label{equation::G_140} \\
 P_{R,0}(x^\prime) &=& P_\mathrm{eq}(x^\prime) \ = \  Z^{\prime-1}
 \  \exp\bigl( - \Omega(x^\prime) / k_B T \bigr) \ . \hspace{10mm}
 \label{equation::G_150}
\end{eqnarray}
Wir nehmen dabei an, dass auf die Flüssigkeit über äußere Parameter gewisse orts- und 
zeitabhängige Kräfte ausgeübt werden. Dann sind die beiden Gleichgewichtsverteilungen 
\eqref{equation::G_140} und \eqref{equation::G_150} für die zwei Zeiten $t$ und 
$t^\prime$ unterschiedlich. Insbesondere sind die beiden Normierungsfaktoren $Z$ 
und $Z^\prime$ verschieden. Aus diesen können wir formal die Änderung einer 
\emph{freien Energie} $\Delta F $ definieren, gegeben durch
\begin{equation}
 \Delta F \ = \  - k_B T \, [ \ln Z^\prime - \ln Z ] \ .
 \label{equation::G_160}
\end{equation}
Setzen wir nun die Anfangsverteilungen \eqref{equation::G_140} und 
\eqref{equation::G_150} in die Gleichung \eqref{equation::G_090} ein so erhalten 
wir für die Entropievariable $\Delta\Sigma$ die zwei Funktionen
\begin{eqnarray}
 \Delta\Sigma_F( x^\prime, x ) &=& - \Delta\Sigma_R( x, x^\prime ) \nonumber\\
 &=& \bigl\{ [ E(x^\prime) - E(x) ] 
 - \mathbf{v} \cdot [ \mathbf{P}(x^\prime) - \mathbf{P}(x) ] \nonumber\\
 && - \mu [ N(x^\prime) - N(x) ] - \Delta F \bigr\} / T
 \label{equation::G_170}
\end{eqnarray}
mit unterschiedlichen Vorzeichen für den Vorwärtsprozess und für den Rückwärtsprozess. 
Offensichtlich ist die Änderung der Entopie $\Delta S = S(x^\prime) - S(x)$ 
heraus gefallen.

Wir nehmen als spezielles Beispiel an, die Flüssigkeit sei in einem Volumen 
eingesperrt, welches durch äußere Kräfte zeitlich verändert wird. Auf diese Weise 
wird die Flüssigkeit komprimiert und expandiert. Der erste Term in der Formel 
\eqref{equation::G_170}, die Änderung der Energie $\Delta E = E(x^\prime) - E(x)$, 
lässt sich dann als eine der Flüssigkeit zugefügte Arbeit $\Delta W$ interpretieren. 
Für den Vorwärtsprozess und für den Rückwärtsprozess finden wir also mit unterschiedlichen 
Vorzeichen die Funktionen für die Arbeit
\begin{eqnarray}
 \Delta W_F( x^\prime, x ) &=& - \Delta W_R( x, x^\prime ) \nonumber\\
 &=& E(x^\prime) - E(x) \ .
 \label{equation::G_180}
\end{eqnarray}
Wird die Flüssigkeit über ein äußeres Potential in einem Volumen eingesperrt, so 
gilt keine Translations-Invarianz und keine Impulserhaltung. In diesem Fall muss 
in den Boltzmann-Verteilungen \eqref{equation::G_140} und \eqref{equation::G_150}
der Lagrange-Parameter für die Geschwindigkeit null sein, also $\mathbf{v} = \mathbf{0}$. 
Folglich ist der zweite Term in \eqref{equation::G_170} null. Die Erhaltung der 
Teilchenzahl bewirkt weiterhin, dass wir den dritten Term in \eqref{equation::G_170} 
null setzen dürfen gemäß $N(x^\prime) - N(x) \to 0$. Dies ist erlaubt, weil die 
Verbundwahrscheinlichkeiten \eqref{equation::G_050} und \eqref{equation::G_060} 
entsprechende Delta-Funktionen als Faktoren enthalten. Der letzte Term in der Formel 
\eqref{equation::G_170} wird schließlich mit der Änderung der freien Energie 
\eqref{equation::G_160} identifiziert.

Als Ergebnis erhalten wir aus \eqref{equation::G_170} für den Vorwärtsprozess 
\begin{equation}
 \Delta\Sigma_F( x^\prime, x ) \ = \  [ \Delta W_F( x^\prime, x ) - \Delta F ] / T \ .
 \label{equation::G_190}
\end{equation}
Es gilt also ein linearer Zusammenhang zwischen der Entropievariable $\Delta\Sigma$ 
und der Arbeit $\Delta W$. Folglich erhalten wir aus \eqref{equation::G_100} und 
\eqref{equation::G_110} durch Variablentransformation entsprechende 
Verteilungs-Funktionen für die Arbeit $P_F( \Delta W )$ und $P_R( \Delta W )$. 
Für diese Verteilungs-Funktionen bekommen wir aus \eqref{equation::G_120} das 
\emph{Fluktuations-Theorem von Crooks} \cite{Cr98}
\begin{equation}
 \frac{ P_F( + \Delta W ) }{ P_R( - \Delta W ) } \ = \ \exp( [ \Delta W - \Delta F ] / T ) \ .
 \label{equation::G_200}
\end{equation}
Führen wir weiterhin eine Integration über die Arbeit $\Delta W$ durch, so erhalten 
wir entsprechend zu \eqref{equation::G_130} das integrale Fluktuations-Theorem
\begin{equation}
 \langle \exp( - \Delta W / k_B T ) \rangle \ = \  \exp( - \Delta F / k_B T ) \ ,
 \label{equation::G_210}
\end{equation}
welches erstmals von Jarzynski \cite{Ja97A,Ja97B} hergeleitet wurde und unter dem 
Namen \emph{Jarzynski-Gleichung} bekannt ist. Während die Jarzynski-Gleichung 
\eqref{equation::G_210} ursprünglich für die durch äußere Kräfte verrichtete Arbeit 
$\Delta W$ hergeleitet wurde, kann man \eqref{equation::G_130} als Jarzynski-Gleichung 
für die Entropievariable $\Delta\Sigma$ betrachten.

Zum Schluss betrachten wir das thermische Gleichgewicht. Hier sind die Energie, 
der Impuls und die Teilchenzahl erhalten. Weiterhin sind die Normierungsfaktoren 
$Z$ und $Z^\prime$ gleich, so dass die frei Energie \eqref{equation::G_160} null ist. 
Folglich sind alle Terme in der Definition der Entropievariable \eqref{equation::G_170} 
null, und die Verteilungsfunktionen \eqref{equation::G_100} und \eqref{equation::G_110} 
vereinfachen sich zu Delta-Funktionen. Im thermischen Gleichgewicht erhalten wir 
also die Verteilungsfunktionen für die Entropievariable
\begin{equation}
 P_{F,\mathrm{eq}}( \Delta\Sigma ) \ = \  P_{R,\mathrm{eq}}( \Delta\Sigma ) \ = \  \delta( \Delta\Sigma )
 \label{equation::G_220}
\end{equation}
und für die Arbeit
\begin{equation}
 P_{F,\mathrm{eq}}( \Delta W ) \ = \  P_{R,\mathrm{eq}}( \Delta W ) \ = \  \delta( \Delta W ) \ .
 \label{equation::G_230}
\end{equation}
Die Fluktuations-Theoreme \eqref{equation::G_120}, \eqref{equation::G_130} und 
\eqref{equation::G_200}, \eqref{equation::G_210} werden mit diesen Verteilungsfunktionen 
in trivialer Weise erfüllt. Im Ergebnis stellen wir fest: Die Fluktuations-Theoreme 
gelten also sowohl im thermischen Gleichgewicht als auch im Nichtgleichgewicht. 

Die Exponentialfunktion ist eine konvexe Funktion. Daher folgt aus dem integralen 
Fluktuations-Theorem \eqref{equation::G_130} die Ungleichung 
\begin{equation}
 \langle \Delta\Sigma\rangle \geq 0 \ .
 \label{equation::G_240}
\end{equation}
Das Gleichheitszeichen gilt genau dann, wenn die Verteilungsfunktionen 
\eqref{equation::G_100} und \eqref{equation::G_110} Delta-Funktionen sind. Dies ist 
mit \eqref{equation::G_220} im thermischen Gleichgewicht der Fall. Wir kommen somit 
zu dem Schluss: Die Entropie-Variable $\Delta\Sigma$ und die zugehörigen 
Fluktuations-Theoreme sind mit dem zweiten Hauptsatz der Thermodynamik vereinbar. 
Im Nichtgleichgewicht wächst die Entropie-Variable im Mittel an, im thermischen 
Gleichgewicht bleibt sie konstant.

Wir fassen zusammen und kommen zu folgendem Ergebnis. Die meisten Überlegungen zum 
und die verschiedenen Varianten des Fluktuations-Theorems, die in dem Übersichtsartikel 
von Seifert \cite{Se12} für kolloidale Teilchen in einem Lösungsmittel beschrieben 
werden, lassen sich auf die Hydrodynamik einer normalen Flüssigkeit und den 
GENERIC-Formalismus übertragen. Unbefriedigend ist jedoch die Interpretation der 
Entropie-Variable $\Delta\Sigma$, definiert durch die Funktionen \eqref{equation::G_090}. 
Für die Hydrodynamik nicht klar, was der zweite Term in \eqref{equation::G_090} 
physikalisch bedeuten soll.

\subsection{Ein modifiziertes Fluktuations-Theorem \break für die Entropie}
\label{section::7C}
In der Hydrodynamik benötigen wir ein Fluktuations-Theorem, in dem die Variable 
ausschließlich die Entropie-Änderung $\Delta\Sigma = \Delta S = S(x^\prime) - S(x)$ 
ohne irgend einen zusätzlichen zweiten Term ist. Ein solches Fluktuations-Theorem 
wollen wir hier nun herleiten. Ausgangspunkt ist wieder die bedingte Wahrscheinlichkeit 
für die relevanten hydrodynamischen Variablen $P( x^\prime | x )$. Wir berechnen daraus 
eine bedingte Wahrscheinlichkeit für die Entropie-Änderung $\Delta S$ mit der Formel 
\begin{equation}
 P( \Delta S | x ) = \int Dx^\prime \ \delta( \Delta S - [ S(x^\prime) - S(x) ] ) 
 \ P( x^\prime | x ) \ .
 \label{equation::G_250}
\end{equation}
Wir betrachten im Folgenden nur den \emph{Vorwärtsprozess}. Der Anfangszustand wird 
beschrieben durch die hydrodynamischen Variablen $x_i$ zur früheren Zeit $t$. Der 
Endzustand wird beschrieben durch die hydrodynamischen Variablen $x^\prime_i$ zur 
späteren Zeit $t^\prime$. Das Zeitintervall $\Delta t = t^\prime - t $ ist 
\emph{positiv}.

Wir berechnen zunächst die bedingte Wahrscheinlichkeit für die relevanten 
hydrodynamischen Variablen $P( x^\prime | x )$ als Lösung der Fokker-Planck-Gleichung 
\eqref{equation::F_010}. Für infinitesimal kleine Zeitintervalle $\Delta t$ erhalten 
wir eine Gauß-Verteilung in den hydrodynamischen Variablen $x^\prime_i$ des Endzustandes. 
Die Position und die Breite der Verteilung werden durch die zwei Kramers-Moyal-Koeffizienten 
$K^{(1)}_i(x)$ und $K^{(2)}_{ij}(x)$ für die hydrodynamischen Variablen des Anfangszustandes 
$x_i$ festgelegt. 

Im nächsten Schritt setzen wir diese Verteilung in die Formel \eqref{equation::G_250} 
ein und berechnen mittels Integration über $x^\prime_i$ die bedingte Wahrscheinlichkeit 
für die Entropie-Änderung $P( \Delta S | x )$. Im Ergebnis erhalten wir wiederum eine 
Gauß-Verteilung, welche dargestellt wird durch 
\begin{eqnarray}
 P( \Delta S | x ) &=& [ 4 \pi \, K^{(2)}_S(x) \ \Delta t ]^{-1/2} \nonumber \\ 
 && \times \exp \biggl\{ - \, \frac{ [ \Delta S - K^{(1)}_S(x) \, \Delta t ]^2 }{ 4 \, K^{(2)}_S(x) \, \Delta t } 
 \biggr\} \qquad
 \label{equation::G_260}
\end{eqnarray}
mit den zwei Kramers-Moyal-Koeffizienten $K^{(1)}_S(x)$ und $K^{(2)}_S(x)$ für 
die \emph{Entropie}. Um diese einfache Formel zu erhalten, haben wir im Argument der 
Exponentialfunktion eine Entwicklung nach Potenzen von kleinen $\Delta t$ durchgeführt. 
Daher gilt die Formel \eqref{equation::G_260} zunächst nur für infinitesimal kleine 
Zeitintervalle $\Delta t$. Die zwei Kramers-Moyal-Koeffizienten für die Entropie 
$K^{(1)}_S(x)$ und $K^{(2)}_S(x)$ lassen sich durch die zwei Kramers-Moyal-Koeffizienten 
$K^{(1)}_i(x)$ und $K^{(2)}_{ij}(x)$ für die hydrodynamischen Variablen darstellen. 
Wir finden die zwei Formeln 
\begin{eqnarray}
 K^{(1)}_S(x) &=& \frac{ \partial S(x) }{ \partial x_i } \, K^{(1)}_i(x) 
 \ + \ \frac{ \partial^2 S(x) }{ \partial x_i \partial x_k } \, K^{(2)}_{ik}(x) \ , \hspace{8mm}
 \label{equation::G_270} \\
 K^{(2)}_S(x) &=& \frac{ \partial S(x) }{ \partial x_i }
 \, K^{(2)}_{ik}(x) \, \frac{ \partial S(x) }{ \partial x_k } \ ,
 \label{equation::G_280} 
\end{eqnarray}
welche für eine beliebige Funktion $S = S(x)$ gelten. Der jeweils erste Term erklärt 
sich als Transformationsformel des Differentialkalküls. Der zweite Term von 
\eqref{equation::G_270} enthält die zweiten Ableitungen von $S(x)$ und ist eine 
Besonderheit der stochastischen Prozesse.

Wir haben jetzt eine explizite Wahrscheinlichkeitsverteilung, für die wir ein spezielles 
modifiziertes Fluktuations-Theorem herleiten können. Wir berechnen den Quotienten 
wie auf der linken Seite des ursprünglichen Fluktuations-Theorems \eqref{equation::G_120} 
und setzen die Formel \eqref{equation::G_260} explizit ein. Nach einigen Umformungen 
finden wir dann als Ergebnis 
\begin{equation}
 \frac{ P( + \Delta S | x ) }{ P( - \Delta S | x ) } \ = 
 \ \exp \bigl\{ \alpha(x) \, \Delta S / k_B \bigr\}
 \label{equation::G_290}
\end{equation}
mit dem dimensionslosen Faktor
\begin{equation}
 \alpha(x) = k_B \, \frac{ K^{(1)}_S(x) }{ K^{(2)}_S(x) } \ .
 \label{equation::G_300}
\end{equation}
Die Gleichung \eqref{equation::G_290} stellt offensichtlich eine Erweiterung und 
Modifizierung des Fluktuations-Theorems dar. Im Argument der Exponentialfunktion auf 
der rechten Seite kommt ein zusätzlicher Faktor $\alpha(x)$ in Form eines Verhältnisses 
der beiden Kramers-Moyal-Koeffizienten vor. In seiner ursprünglichen Form 
\eqref{equation::G_120} gilt das Fluktuations-Theorem nur dann, wenn der Faktor 
$\alpha(x) = 1$ ist und somit die Bedingung 
\begin{equation}
 K^{(1)}_S(x) \ = \ K^{(2)}_S(x) / k_B
 \label{equation::G_310}
\end{equation}
erfüllt wird. Ansonsten stellt unsere Gleichung \eqref{equation::G_290} eine 
Erweiterung und Modifizierung des Fluktuations-Theorems dar.

Wir weisen darauf hin, dass unsere Gleichung \eqref{equation::G_290} nur für 
infinitesimal kleine Zeitintervalle $\Delta t$ hergeleitet wurde. Eine Erweiterung 
auf größere endliche Zeitintervalle $\Delta t$ ist ohne Änderung möglich, wenn die 
zwei Kramers-Moyal-Koeffizienten $K^{(1)}_S(x(t))$ und $K^{(2)}_S(x(t))$ nahezu 
konstant sind und nicht über die Pfade $x_i(t)$ von der Zeit abhängen. Dies ist der 
Fall, wenn sich die normale Flüssigkeit in einem \emph{stationären} 
Nichtgleichgewichts-Zustand befindet und die Effekte der Fluktuationen klein sind. 
Beispiele sind eine stationäre laminare Strömung oder ein stationärer Wärmetransport. 
In diesen Fällen kann das Zeitintervall $\Delta t$ sehr groß sein. Andererseits wird 
für eine turbulente Strömung oder Wärmetransport mit chaotischer Konvektion das 
Zeitintervall $\Delta t$ entsprechend klein sein.

\subsection{Explizite Berechnung der Kramers-Moyal-Koeffizienten}
\label{section::7D}
Um den Zusammenhang zwischen dem Fluktuations-Theorem in der ursprünglichen Form 
\eqref{equation::G_120} und unserer allgemeineren Gleichung \eqref{equation::G_290} 
mit dem dimensionslosen Faktor \eqref{equation::G_300} zu verstehen, benötigen wir nun 
explizite Formeln für die zwei Kramers-Moyal-Koeffizienten. Den ersten Koeffizienten 
haben wir mit Gl.\ \eqref{equation::F_360} zuvor in Kapitel \ref{section::6C} 
berechnet. Lassen wir das Zeitargument weg, so haben wir
\begin{eqnarray}
 K^{(1)}_S(x) &=& \frac{ \partial S(x) }{ \partial x_i }
 \, M_{ik}(x) \, \frac{ \partial S(x) }{ \partial x_k } \nonumber \\
 && + \, \frac{ \partial }{ \partial x_k } \left( \frac{ \partial S(x) }{ \partial x_i } 
 \, k_B \, M_{ik}(x) \right) \ .
 \label{equation::G_320}
\end{eqnarray}
Den zweiten Koeffizienten können wir aus der Langevin-Gleichung für die Entropie 
\eqref{equation::F_330} entnehmen. Wir betrachten dazu den zweiten fluktuierenden 
Term auf der rechten Seite und setzen dort die fluktuierenden Kräfte 
\eqref{equation::F_020} ein. Unter Verwendung von \eqref{equation::F_040} und der 
Beziehung $K^{(2)}_{ik}(x(t)) = k_B M_{ik}(x(t))$ erhalten wir dann
\begin{equation}
 K^{(2)}_S(x) \ = \ \frac{ \partial S(x) }{ \partial x_i }
 \, k_B \, M_{ik}(x) \, \frac{ \partial S(x) }{ \partial x_k } \ .
 \label{equation::G_330}
\end{equation}
Die zwei Kramers-Moyal-Koeffizienten \eqref{equation::G_320} und \eqref{equation::G_330} 
erfüllen offensichtlich im allgemeinen nicht die notwendige Bedingung 
\eqref{equation::G_310} für das Fluktuations-Theorem.

Alternativ lassen sich die zwei Kramers-Moyal-Koeffizienten für die Entropie auch 
mit den Formeln \eqref{equation::G_270} und \eqref{equation::G_280} berechnen. 
Wir benötigen dazu die zwei Kramers-Moyal-Koeffizienten für die hydrodynamischen 
Variablen in expliziter Form. Letztere bekommen wir aus der Langevin-Gleichung des 
GENERIC-Formalismus \eqref{equation::F_130} durch Vergleich mit der allgemeinen Form 
der stochastischen Theorie \eqref{equation::F_050} oder \eqref{equation::F_120}. 
Diesen Vergleich müssen wir sorgfältig durchführen und dabei beachten, dass alle 
Langevin-Gleichungen im Stratonovich-Formalismus definiert sind. Wir verwenden weiterhin 
die Bedingung \eqref{equation::F_200} für die Matrix $B_{im}(x(t))$, so dass der dritte 
Term in der allgemeinen Gleichung \eqref{equation::F_120} null ist. Anschließend setzen 
wir die so erhaltenen Kramers-Moyal-Koeffizienten für die hydrodynamischen Variablen 
in die Formeln \eqref{equation::G_270} und \eqref{equation::G_280} ein. Wegen der 
Nebenbedingung \eqref{equation::F_140} fällt ein reversibler Term weg. Als Ergebnis 
erhalten wir somit wiederum \eqref{equation::G_320} und \eqref{equation::G_330}.

Der erste Kramers-Moyal-Koeffizient \eqref{equation::G_320} enthält zwei Beiträge, 
einen dissipativen und einen fluktuierenden. Entsprechend zerlegen wir 
\begin{equation}
 K^{(1)}_S(x) \ = \ K^{(1)}_{S,\mathrm{diss}}(x) \ + \ K^{(1)}_{S,\mathrm{fluc}}(x)
 \label{equation::G_340} \ .
\end{equation}
Vergleichen wir den ersten Term von \eqref{equation::G_320} mit dem zweiten Koeffizienten 
\eqref{equation::G_330}, so finden wir 
\begin{equation}
 K^{(1)}_{S,\mathrm{diss}}(x) \ = \ K^{(2)}_S(x) / k_B \ .
 \label{equation::G_350}
\end{equation}
Diese Gleichung stimmt offensichtlich mit der Bedingung \eqref{equation::G_310} überein. 
Der dissipative Anteil des ersten Koeffizienten würde also die Bedingung für das 
Fluktuations-Theorem erfüllen. Die Abweichungen und Modifizierungen liefert 
folglich der fluktuierende Anteil des ersten Koeffizienten. Für diesen entnehmen 
wir aus dem zweiten Term von \eqref{equation::G_320} die Formel 
\begin{equation}
 K^{(1)}_{S,\mathrm{fluc}}(x) \ = \ k_B \, \frac{ \partial }{ \partial x_i } \left( M_{ik}(x) 
 \, \frac{ \partial S(x) }{ \partial x_k } \right) \ ,
 \label{equation::G_360}
\end{equation}
wobei die Reihenfolge der Faktoren etwas umgeordnet und die Indizes umbenannt wurden.

Wenn wir die Abweichungen des modifizierten Fluktuations-Theorem \eqref{equation::G_290} 
vom ursprünglichen \eqref{equation::G_120} verstehen wollen, dann müssen wir den 
dimensionslosen Faktor \eqref{equation::G_300} genauer untersuchen. Unter Verwendung 
der Gleichungen \eqref{equation::G_350} und \eqref{equation::G_340} formen wir diesen 
dimensionslosen Faktor um und erhalten 
\begin{equation}
 \alpha(x) = \frac{ K^{(1)}_S(x) }{ K^{(1)}_{S,\mathrm{diss}}(x) } 
 = \frac{ K^{(1)}_S(x) }{ K^{(1)}_S(x) - K^{(1)}_{S,\mathrm{fluc}}(x) } \ .
 \label{equation::G_370}
\end{equation}
Die Abweichungen vom Fluktuations-Theorem in der ursprünglichen Form werden klein 
sein, wenn der fluktuierende Anteil $K^{(1)}_{S,\mathrm{fluc}}(x)$ deutlich kleiner 
als der gesamte erste Kramers-Moyal-Koeffizient $K^{(1)}_S(x)$ ist und somit in guter 
Näherung $\alpha(x)\approx 1$ gilt. Umgekehrt werden wesentliche Abweichungen in 
Erscheinung treten, wenn $K^{(1)}_{S,\mathrm{fluc}}(x)$ in die Größenordnung von 
$K^{(1)}_S(x)$ kommt oder deutlich größer wird. Um die Abschätzungen durchzuführen, 
müssen wir also sowohl den fluktuierenden Anteil $K^{(1)}_{S,\mathrm{fluc}}(x)$ als 
auch den gesamten ersten Kramers-Moyal-Koeffizienten $K^{(1)}_S(x)$ explizit berechnen.

Wir betrachten im Folgenden wieder die normale Flüssigkeit und fügen den hydrodynamischen 
Variablen $x_i(\mathbf{r})$ die Ortsvariable $\mathbf{r}$ im Argument hinzu. Wir verwenden  
den lokale Ansatz für die Onsager-Matrix \eqref{equation::D_590} und setzen diesen in 
die Formel \eqref{equation::G_360} ein. Dann erhalten wir für den fluktuierenden Anteil 
des ersten Kramers-Moyal-Koeffizienten
\begin{widetext}
\begin{equation}
 K^{(1)}_{S,\mathrm{fluc}}[x] = - \, k_B \, \int d^d r \ \frac{ \delta }{ \delta x_i(\mathbf{r}) } 
 \left( \partial_m \, N_{im,kn}(x(\mathbf{r})) \, \partial_n 
 \ \frac{ \delta S[x] }{ \delta x_k(\mathbf{r}) } \right) \ .
 \label{equation::G_380}
\end{equation}
Dieser ist ein Funktional in den hydrodynamischen Variablen $x_i(\mathbf{r})$, und 
die partiellen Ableitungen wurden durch Funktional-Ableitungen ersetzt. Für die weitere 
Berechnung fügen wir zwei räumliche Delta-Funktionen ein, so dass wir drei Integrale 
über räumliche Koordinaten haben. Wir führen dann zuerst die Funktional-Ableitungen 
aus und berechnen anschließend die Integrale über die Delta-Funktionen. Nach sorgfältiger 
Bildung eines Grenzwertes erhalten wir dann
\begin{equation}
 K^{(1)}_{S,\mathrm{fluc}}[x] = + \, k_B \, \int d^d r \ N_{im,kn}(x(\mathbf{r})) 
 \, \lim_{ \mathbf{r}^\prime \to \mathbf{r} } \left( \partial^{\ }_m \partial^\prime_n 
 \ \frac{ \delta^2 S[x] }{ \delta x_i(\mathbf{r}) \, \delta x_k(\mathbf{r}^\prime) } \right) \ .
 \label{equation::G_390}
\end{equation}
Wegen einer partiellen Integration über den Raum hat sich das Vorzeichen geändert, und 
einer der beiden Differentialoperatoren hat einen Strich bekommen. Das bedeutet, 
$\partial^{\ }_m$ wirkt auf die eine Ortsvariable $\mathbf{r}$, und $\partial^\prime_n$ 
wirkt auf die andere Ortsvariable $\mathbf{r}^\prime$. Wir nehmen nun an, dass wie in 
der Hydrodynamik üblich ein \emph{lokales} thermisches Gleichgewicht gilt. Das bedeutet, 
die gesamte Entropie $S[x]$ lässt sich als Integral über die lokale Entropiedichte 
$\sigma(\mathbf{r})$ darstellen gemäß $S[x] = \int d^d r \, \sigma(\mathbf{r})$. 
Dann berechnen wir die zweite Funktional-Ableitung
\begin{equation}
 \frac{ \delta^2 S[x] }{ \delta x_i(\mathbf{r}) \, \delta x_k(\mathbf{r}^\prime) } = 
 \frac{ \partial^2 \sigma(\mathbf{r}) }{ \partial x_i(\mathbf{r}) \, \partial x_k(\mathbf{r}) }
 \, \delta( \mathbf{r} - \mathbf{r}^\prime )
 \label{equation::G_400}
\end{equation}
und erhalten den fluktuierenden Anteil des Koeffizienten
\begin{equation}
 K^{(1)}_{S,\mathrm{fluc}}[x] = - \, k_B \, \int d^d r \ N_{im,kn}(x(\mathbf{r})) 
 \, \frac{ \partial^2 \sigma(\mathbf{r}) }{ \partial x_i(\mathbf{r}) \, \partial x_k(\mathbf{r}) }
 \, \partial_m \partial_n \delta( \mathbf{0} ) \ .
 \label{equation::G_410}
\end{equation}
Die räumlichen Differentialoperatoren $\partial_m$ und $\partial_n$ wirken jetzt nur 
noch auf die räumliche Delta-Funktion. Wir führen hier den Grenzwertprozess explizit 
aus mit
\begin{equation}
 \partial_m \partial_n \delta( \mathbf{0} ) = - \lim_{ \mathbf{r}^\prime \to \mathbf{r} } 
 \partial^{\ }_m \partial^\prime_n \delta( \mathbf{r} - \mathbf{r}^\prime ) \ .
 \label{equation::G_420}
\end{equation}
Weil der Strich am Differentialoperator wieder weggefallen ist, hat sich das Vorzeichen 
wieder geändert. 

Gegenwärtig identifizieren wir mit $x_i(\mathbf{r})$ die hydrodynamischen Variablen 
einer normalen Flüssigkeit, wobei eine der Variablen die Energiedichte 
$\varepsilon(\mathbf{r})$ ist. Alternativ können wir einen anderen Satz von 
hydrodynamischen Variablen verwenden, wobei eine der Variablen die Entropiedichte 
$\sigma(\mathbf{r})$ ist. Ein solche Transformation wurde in Kapitel \ref{section::4D} 
beschrieben und angewendet. Diese können wir hier auch auf die Formel 
\eqref{equation::G_410} anwenden. Dazu tauschen wir die Onsager-Matrix 
$N_{im,kn}(x(\mathbf{r}))$ aus durch die neue Matrix $\Lambda_{im,kn}(x(\mathbf{r}))$ 
mit einer etwas einfacheren Struktur. Weiterhin ersetzen wir oben in der zweiten 
Ableitung die Entropiedichte $\sigma(\mathbf{r})$ durch die Energiedichte 
$\varepsilon(\mathbf{r})$ und drehen das Vorzeichen um. Als Ergebnis erhalten wir dann 
\begin{equation}
 K^{(1)}_{S,\mathrm{fluc}}[x] = + \, k_B \, \int d^d r \ \Lambda_{im,kn}(x(\mathbf{r})) 
 \, \frac{ \partial^2 \varepsilon(\mathbf{r}) }{ \partial x_i(\mathbf{r}) \, \partial x_k(\mathbf{r}) }
 \, \partial_m \partial_n \delta( \mathbf{0} ) \ .
 \label{equation::G_430}
\end{equation}
Wir bemerken, dass die Umformungen von \eqref{equation::G_410} auf \eqref{equation::G_430} 
exakt sind ohne irgendwelche Näherungen. Im nächsten Schritt setzen wir für die 
hydrodynamischen Variablen $x_i(\mathbf{r})$ explizit die Massendichte $\rho(\mathbf{r})$, 
die Impulsdichte $\mathbf{j}(\mathbf{r})$ und die Entropiedichte $\sigma(\mathbf{r})$ 
ein. Aus Symmetriegründen sind viele Nichtdiagonalelemente der Onsager-Matrix 
$\Lambda_{im,kn}(x(\mathbf{r}))$ null. Da es für die Massendichte $\rho$ keine Dissipation 
gibt, fallen weiterhin die Terme mit den Massendichten komplett weg. Als Ergebnis erhalten 
wir dann
\begin{equation}
 K^{(1)}_{S,\mathrm{fluc}}[x] = + \, k_B \, \int d^d r \ \left[ \Lambda^{jj}_{im,kn}(x) 
 \, \frac{ \partial^2 \varepsilon }{ \partial j_i \, \partial j_k } 
 + \Lambda^{\sigma\sigma}_{m,n}(x) \, \frac{ \partial^2 \varepsilon }{ \partial \sigma^2 } 
 \right] \, \partial_m \partial_n \delta( \mathbf{0} ) \ ,
 \label{equation::G_440}
\end{equation}
wobei die Onsager-Matrizen $\Lambda^{jj}_{im,kn}(x)$ und $\Lambda^{\sigma\sigma}_{m,n}(x)$ 
explizit durch die Formeln \eqref{equation::D_770} und \eqref{equation::D_780} gegeben 
sind, mit nur drei Parametern, der Scherviskosität $\eta$, der Volumenviskosität $\zeta$ 
und der Wärmeleitfähigkeit $\varkappa$. Die zweiten Ableitungen der Energiedichte 
berechnen wir mit den thermodynamischen Relationen
\begin{equation}
 \frac{ \partial^2 \varepsilon }{ \partial j_i \, \partial j_k } 
 = \left( \frac{ \partial v_k }{ \partial j_i } \right)_{\sigma,\rho} 
 = \delta_{ik} \, \frac{ 1 }{ \rho } \ , \qquad
 \frac{ \partial^2 \varepsilon }{ \partial \sigma^2 } 
 = \left( \frac{ \partial T }{ \partial \sigma } \right)_{\mathbf{j},\rho}
 = \frac{ T }{ \rho \, c_V } \ .
 \label{equation::G_450}
\end{equation}
Wegen der Invarianz unter Galilei-Transformationen hat die Energiedichte die Struktur 
\eqref{equation::D_370}. Das bedingt das einfache Ergebnis für die zweite Ableitung 
nach den Impulsdichten $\mathbf{j}(\mathbf{r})$. Die zweite Ableitung nach der 
Entropiedichte $\sigma(\mathbf{r})$ ist komplizierter und führt auf die spezifische 
Wärme pro Masseneinheit bei konstantem Volumen $c_V$ als zusätzlichen Parameter. Die 
Formel \eqref{equation::G_440} lässt sich damit nun explizit auswerten.

Wir stellen fest, dass das Ergebnis \eqref{equation::G_440} als konstanten Faktor die 
zweiten räumlichen Ableitungen einer Delta-Funktion bei Argument Null enthält. Dieser 
Faktor ist unendlich groß und stellt eine Ultraviolett-Divergenz dar. Eine normale 
Flüssigkeit besteht bekanntlich aus Atomen oder Molekülen. Somit wird es eine minimale 
Längenskala $\ell$ geben, mit der wir eine Regularisierung durchführen müssen. Wir 
ersetzen daher die Delta-Funktion durch eine Gauß-Funktion gemäß
\begin{equation}
 \delta( \mathbf{r} ) = [ 2\pi \ell^2 ]^{-d/2} \ \exp( - \mathbf{r}^2 / 2 \ell^2 )
 \label{equation::G_460}
\end{equation}
und berechnen damit die zweiten räumlichen Ableitungen der Delta-Funktion bei Argument Null
\begin{equation}
 \partial_m \partial_n \delta( \mathbf{0} ) = - \, \delta_{mn} \, (2\pi)^{-d/2} \, \ell^{-(d+2)} \ .
 \label{equation::G_470}
\end{equation}
Setzen wir nun die Onsager-Matrizen \eqref{equation::D_770} und \eqref{equation::D_780}, 
die zweiten Ableitungen der Energiedichte \eqref{equation::G_450} und die zweite 
räumliche Ableitung der Delta-Funktion \eqref{equation::G_470} ein, so erhalten wir 
aus der Formel \eqref{equation::G_440} das explizite Ergebnis
\begin{equation}
 K^{(1)}_{S,\mathrm{fluc}}[x] = - \, k_B \, \int d^d r \ \left[ 
 ( d + 2 ) ( d - 1 ) \, \frac{ \eta }{ \rho } + d \, \frac{ \zeta }{ \rho } 
 + d \, \frac{ \varkappa }{ \rho \, c_V } \right] 
 \, \frac{ 1 }{ (2\pi)^{d/2} \, \ell^{d+2} } \ .
 \label{equation::G_480}
\end{equation}
\end{widetext}
Die minimale Länge $\ell$ der Regularisierung steht im Nenner mit einer hohen Potenz. 
Wenn diese sehr klein ist und dem mittleren Abstand zwischen den Atomen oder 
Molekülen entspricht, dann wird das Ergebnis für den fluktuierende Anteil des ersten 
Kramers-Moyal-Koeffizienten $K^{(1)}_{S,\mathrm{fluc}}[x]$ extrem groß. Wenn 
andererseits die minimale Länge $\ell$ hinreichend groß ist, dann wird das Ergebnis 
sehr klein.

Wir benötigen weiterhin den gesamten ersten Kramers-Moyal-Koeffizienten $K^{(1)}_S(x)$. 
Um die Größenordnung abzuschätzen, reicht hier eine Mittlere-Feld-Näherung aus, welche 
die Fluktuationen weglässt. Wir bestimmen dazu für eine gegebene physikalische Situation 
eines Nichtgleichgewichts die mittleren Felder für Temperatur $\langle T(\mathbf{r},t) \rangle$, 
Geschwindigkeit $\langle \mathbf{v}(\mathbf{r},t) \rangle$ und chemisches Potential 
$\langle \mu(\mathbf{r},t) \rangle$. Diese setzten wir dann in die Formel 
\begin{equation}
 \langle R \rangle = \langle \partial_k v_i \rangle \, \langle \Lambda^{jj}_{ik,mn} \rangle 
 \, \langle \partial_n v_m \rangle 
 + \langle \partial_k T \rangle \, \langle \Lambda^{\sigma\sigma}_{kn} \rangle \, 
 \langle \partial_n T \rangle
 \label{equation::G_490}
\end{equation}
ein, um die mittlere erzeugte Wärme pro Volumen und Zeit $\langle R(\mathbf{r},t) \rangle$ 
zu berechnen. Die Formel \eqref{equation::G_490} ist die Mittlere-Feld-Näherung 
von \eqref{equation::D_1090}. Betrachten wir die hydrodynamische Gleichung 
\eqref{equation::D_550}, so finden wir für den ersten Kramers-Moyal-Koeffizienten 
$K^{(1)}_S(x)$ schließlich die Formel in Mittlerer-Feld-Näherung
\begin{equation}
 K^{(1)}_S[x] \approx \int d^d r \ \frac{ \langle R \rangle }{ \langle T \rangle } \ .
 \label{equation::G_500}
\end{equation}

\subsection{Abhängigkeit von der minimalen Länge}
\label{section::7E}
Um die Abhängigkeit von der minimalen Länge $\ell$ besser zu verstehen, 
gehen wir zurück zu der Definition der Entropie $S(t)$ durch die Maximierung 
\eqref{equation::B_210} unter den Nebenbedingungen \eqref{equation::B_220} und 
\eqref{equation::B_230}. Die Entropie $S(t)=S[x(t)]$ ist ein Funktional, das sehr stark 
davon abhängen wird, welche der relevanten hydrodynamischen Variablen $x_i(\mathbf{r},t)$ 
und folglich wieviele physikalische Freiheitsgrade in die Nebenbedingungen 
\eqref{equation::B_220} eingehen. Wir treffen die Auswahl dieser Freiheitsgrade so,
dass nur räumliche Variationen auf Längenskalen betrachtet werden, die größer als 
die minimale Länge $\ell$ sind. Auf diese Weise wird eine Regularisierung der 
Ultraviolett-Divergenzen erreicht. Wir stellen dazu die hydrodynamischen Variablen 
durch ein Fourier Integral dar gemäß 
\begin{equation}
 x_i(\mathbf{r},t) = \int \frac{ d^d k }{ (2\pi)^d }  \ \theta( k_\mathrm{max} - |\mathbf{k}| ) 
 \, e^{ i \mathbf{k} \cdot \mathbf{r} } \, x_i(\mathbf{k},t) \ ,
 \label{equation::G_510}
\end{equation}
wobei die Theta-Funktion die Wellenvektoren $\mathbf{k}$ auf eine Kugel mit 
Radius $k_\mathrm{max}=2\pi/\ell$ einschränkt. Die Nebenbedingungen im Ortsraum 
\eqref{equation::B_220} lassen sich nun durch entsprechende Nebenbedingungen 
im Fourier-Raum
\begin{equation}
 \mathrm{Sp}\{ \tilde{\varrho}(t) \, a_i(\mathbf{k}) \} = x_i(\mathbf{k},t)
 \label{equation::G_520}
\end{equation}
ersetzen. Die Anzahl dieser Nebenbedingen wird durch die Anzahl der $\mathbf{k}$-Vektoren 
bestimmt, also durch das Volumen der Kugel mit Radius $k_\mathrm{max}=2\pi/\ell$ im 
Fourier-Raum. Für einen $d$-dimensionalen Raum mit Volumen $V$ finden wir die Anzahl 
der physikalischen Freiheitsgrade 
\begin{eqnarray}
 N_F &=& V \int \frac{ d^d k }{ (2\pi)^d } \ \theta( k_\mathrm{max} - |\mathbf{k}| ) \nonumber\\
 &=& \frac{ V }{ (2\pi)^d } \frac{ \Omega_d }{ d } \, ( k_\mathrm{max} )^d 
 = \frac{ \Omega_d }{ d } \frac{ V }{ \ell^d } \ ,
 \label{equation::G_530}
\end{eqnarray}
wobei $\Omega_d = 2 \, \pi^{d/2} / \Gamma(d/2)$ die Oberfläche der $d$-dimensionalen 
Einheitskugel definiert.

Ist die minimale Länge $\ell$ klein, so wird die Anzahl der relevanten physikalischen 
Freiheitsgrade und folglich die Anzahl der Nebenbedingen $N_F$ groß. Man kann in diesem 
Fall erwarten, dass die Fluktuationen der Entropie $S(t)$ auch entsprechend groß werden. 
Dies sieht man explizit an dem fluktuierenden Anteil des ersten Kramers-Moyal-Koeffizienten 
\eqref{equation::G_480}, welcher in diesem Fall sehr groß wird. Ist umgekehrt die 
minimale Länge $\ell$ groß, so wird die Anzahl der Nebenbedingen $N_F$ klein, und 
die Fluktuationen der Entropie $S(t)$ werden klein. Entsprechend wird der fluktuierenden 
Anteil des ersten Kramers-Moyal-Koeffizienten \eqref{equation::G_480} sehr klein.

Die minimale Länge $\ell$ kann man als Flußparameter einer Renormierungsgruppe 
interpretieren. Vergrößert man $\ell$ so wird die Längenskala für die minimale Auflösung 
des Modells, also der hydrodynamischen Gleichungen, vergröbert. Es ändern sich die 
Parameter des Modells wie z.B.\ die Kramers-Moyal-Koeffizienten. Die Eigenschaften des 
physikalischen Systems, also der normalen Flüssigkeit, bleiben jedoch unverändert. 
Im Limes $\ell \to \infty$ kann man einen Fixpunkt im Raum der Parameter erwarten. 
Für den fluktuierenden Anteil des ersten Kramers-Moyal-Koeffizienten finden wir 
\begin{equation}
 K^{(1)}_{S,\mathrm{fluc}}[x] \to 0 \quad \mbox{für} \quad \ell \to \infty \ .
 \label{equation::G_540}
\end{equation}
Das bedeutet, im Infrarot-Limes $\ell \to \infty$ wird der fluktuierenden Anteil des 
ersten Kramers-Moyal-Koeffizienten, welcher den Korrekturterm zum Fluktuations-Theorem 
darstellt, \emph{irrelevant}. Wir kommen somit zu dem Ergebnis: Auf großen Längenskalen 
$\ell$ wird das Fluktuations-Theorem in seiner ursprünglichen Form gültig bleiben.
Für kleine Längenskalen $\ell$ wird der Korrekturterm \eqref{equation::G_480} wichtig 
werden. 

Es findet also ein kontinuierlicher Übergang statt von kleinen Längenskalen $\ell$, 
für welche der Korrekturterm dominiert, zu großen Längenskalen, für welche der 
Korrekturterm irrelevant ist. Wir wollen nun herausfinden, bei welcher kritischen 
Längenskala $\ell_c$ dieser Übergang stattfindet. Betrachten wir den dimensionslosen 
Faktor \eqref{equation::G_370} im modifizierten Fluktuations-Theorem 
\eqref{equation::G_290}, welcher die Abweichungen vom ursprünglichen Fluktuations-Theorem 
\eqref{equation::G_120} beschreibt, so finden wir eine sinnvolle Definition für die 
kritische minimale Länge $\ell_c$ mit der Bedingung
\begin{equation}
 \left| K^{(1)}_{S,\mathrm{fluc}}[x] \right| = \left| K^{(1)}_S[x] \right| 
 \quad \Longleftrightarrow \quad \ell = \ell_c \ .
 \label{equation::G_550}
\end{equation}
Die beiden erforderlichen Kramers-Moyal-Koeffizienten werden in den Formeln 
\eqref{equation::G_480} und \eqref{equation::G_500} zusammen mit 
\eqref{equation::G_490} definiert. Für $K^{(1)}_{S,\mathrm{fluc}}[x]$ haben wir 
eine einfache explizite Formel. Für die Berechnung von $K^{(1)}_S[x]$ muss man 
eine konkrete physikalische Situation betrachten, wie z.B.\ eine Scherströmung 
mit einem Geschwindigkeits-Gradienten oder einen Wärmetransport mit einem 
Temperaturgradienten. Als Ergebnis erhält man dann einen eindeutigen konkreten 
Wert für die kritische minimale Länge $\ell_c$.

Eine grobe und universelle Abschätzung der kritischen minimalen Länge $\ell_c$ ist 
jedoch einfacher möglich. In der Experimentalphysik werden üblicherweise die SI-Einheiten 
als Einheiten für Messgrößen und Parameter verwendet. Diese Einheiten haben die 
Eigenschaft, dass die Zahlenwerte der Größen in der Nähe der Eins liegen mit einer 
Toleranz von einigen wenigen Größenordnungen. Wir wollen hier mal $2.5$ Größenordnungen 
nach oben und nach unten annehmen. Das entspricht Zahlenwerten zwischen $0.003$ und 
$300$. In den obigen Formeln \eqref{equation::G_480} und \eqref{equation::G_500} 
zusammen mit \eqref{equation::G_490} weichen nur zwei Größen deutlich davon ab, und 
zwar die Boltzmann-Konstante $k_B=1.38 \times 10 ^{-23} \, \mathrm{J/K}$ und die 
minimale Länge $\ell \ll 1 \, \mathrm{m}$, welche beide sehr kleine Zahlenwerte haben. 
Diese kommen nur in dem Verhältnis $k_B / \ell^{d+2}$ vor. Die kritische minimale Länge 
$\ell_c$ bekommen wir daher aus der Bedingung, dass der Zahlenwert dieses Verhältnisses 
in SI-Einheiten bei Eins liegt, also
\begin{equation}
 k_B / \ell^{d+2} = 1 \, \mathrm{ J / K \, m^{d+2} } \ .
 \label{equation::G_560}
\end{equation}
Nehmen wir an, die Dimension des Raumes sei $d=3$, so finden wir die kritische 
minimale Länge 
\begin{equation}
 \ell_c = 2.7 \times 10^{-5} \, \mathrm{m} = 27 \, \mathrm{\mu m} \ .
 \label{equation::G_570}
\end{equation}
Wegen dem Exponenten $d+2=5$ in \eqref{equation::G_560} reduziert sich die Toleranz 
auf eine halbe Größenordnung nach oben und nach unten. Somit liegt also die kritische 
minimale Länge im Intervall $10^{-5} \, \mathrm{m} \lesssim \ell_c \lesssim 10^{-4} \, \mathrm{m}$. 
Den kleineren Wert erwarten wir für Flüssigkeiten mit größeren Dichten wie Wasser, 
und den größeren Wert erwarten wir für Gase mit kleineren Dichten wie Luft.

\subsection{Erwartungswert und Varianz der Entropie-Änderung}
\label{section::7F}
Mit der Verteilungsfunktion \eqref{equation::G_260} lassen sich der Erwartungswert und die 
Varianz der Entropie-Änderung $\Delta S$ berechnen und durch die zwei Kramers-Moyal-Koeffizienten 
darstellen. Wir finden für den Erwartungswert
\begin{equation}
 \langle \Delta S \rangle = K^{(1)}_S(x) \, \Delta t
 \label{equation::G_580}
\end{equation}
und für die Varianz
\begin{equation}
 \langle [ \Delta S - \langle \Delta S \rangle ]^2 \rangle = 2 \, K^{(2)}_S(x) \, \Delta t \ .
 \label{equation::G_590}
\end{equation}
Den zweiten Kramers-Moyal-Koeffizienten können wir über \eqref{equation::G_350} durch 
den dissipativen Anteil des ersten Koeffizienten ausdrücken. Verwenden wir weiterhin 
\eqref{equation::G_340} und \eqref{equation::G_580}, so erhalten wir 
\begin{equation}
 \langle [ \Delta S - \langle \Delta S \rangle ]^2 \rangle / ( k_B )^2 = 2 \, \Bigl[ \langle \Delta S \rangle 
 - K^{(1)}_{S,\mathrm{fluc}}(x) \, \Delta t \Bigr] \Big/ k_B \ .
 \label{equation::G_600}
\end{equation}
Setzen wir hier den fluktuierenden Anteil des ersten Koeffizienten \eqref{equation::G_480} 
ein, so erhalten wir für die Hydrodynamik einer normalen Flüssigkeit das explizite Ergebnis
\begin{widetext}
\begin{equation}
 \langle [ \Delta S - \langle \Delta S \rangle ]^2 \rangle / ( k_B )^2 \ = \ 2 \, \langle \Delta S \rangle / k_B 
 \ + \ 2 \int d^d r \ \left[ ( d + 2 ) ( d - 1 ) \, \frac{ \eta }{ \rho } 
 + d \, \frac{ \zeta }{ \rho } + d \, \frac{ \varkappa }{ \rho \, c_V } \right] 
 \, \frac{ \Delta t }{ (2\pi)^{d/2} \, \ell^{d+2} } \ .
 \label{equation::G_610}
\end{equation}
\end{widetext}
Die Entropie-Änderung $\Delta S$ ist eine extensive Größe. Daher ist sie proportional 
zum Volumen $V$ des Systems. Ebenso ist sie proportional zum Zeitintervall $\Delta t$. 
Daher enthalten in den Formeln \eqref{equation::G_600} und \eqref{equation::G_610} 
beide Terme auf den rechten Seiten jeweils einen Faktor $V \Delta t$, der ausgeklammert 
werden kann. Folglich wird die relative Größe der beiden Terme zueinander nicht durch 
das Volumen des Systems $V$ und nicht durch das betrachtete Zeitintervall $\Delta t$ 
beeinflusst.

Wir interpretieren die zwei Terme in der folgenden Weise. Der erste Term in 
\eqref{equation::G_610} is proportional zu dem Mittelwert der Entropie-Änderung. 
Er ist null im thermischen Gleichgewicht und positiv in einem Nichtgleichgewichts-Zustand. 
Folglich ist der erste Term der \emph{Nichtgleichgewichts-Beitrag} zu der Varianz 
der Entropie-Änderung. Andererseits ist der zweite Term in \eqref{equation::G_610} 
ein Integral über lokale Größen, die in lokalen thermischen Gleichgewichten 
berechnet werden. Aus diesem Grunde interpretieren wir den zweiten Term als den 
Beitrag der \emph{Gleichgewichts-Fluktuationen} zu der Varianz der Entropie-Änderung.

Das Fluktuations-Theorem in seiner ursprünglichen Form erlaubt nur den ersten Term 
in den Formeln \eqref{equation::G_600} und \eqref{equation::G_610}. Das bedeutet, 
in der Varianz der Entropie-Änderung wird nur der Nichtgleichgewichts-Beitrag 
berücksichtigt. Daher interpretieren wir den Korrekturterm in der Modifizierung 
des Fluktuations-Theorems als Beitrag der Gleichgewichts-Fluktuationen. In den letzten 
zwei Abschnitten haben wir herausgefunden, dass der Korrekturterm stark von der 
minimalen Länge $\ell$ abhängt, welche eine Regularisierung der Effekte auf kurzen 
Längenskalen bewirkt. Ob in den Formeln \eqref{equation::G_600} und \eqref{equation::G_610} 
nun der erste oder der zweite Term dominiert, hängt davon ab, ob die minimale Länge 
$\ell$ größer oder kleiner als die kritische minimale Länge $\ell_c$ ist, welche 
in den Gleichungen \eqref{equation::G_550}-\eqref{equation::G_570} bestimmt wurde. 

Zusammenfassend stellen wir also fest. Wir finden eine Erweiterung und Modifizierung 
des Fluktuations-Theorems und berechnen einen Korrekturterm. Ob dieser Korrekturterm 
nun dominiert oder irrelevant ist, hängt von der minimalen Längenskala $\ell$ 
ab, auf der die relevanten hydrodynamischen Variablen $x_i(\mathbf{r},t)$ variieren 
dürfen. Für $\ell < \ell_c$ dominiert der Korrekturterm. Für $\ell > \ell_c$ wird 
er klein und irrelevant.

\subsection{Modifizierte Jarzynski-Gleichung \break für die Entropie}
\label{section::7G}
Zum Schluss leiten wir eine modifizierte Version der Jarzynski-Gleichung 
\eqref{equation::G_130} her, in welcher die Entropie-Änderung 
$\Delta S = S(x^\prime) - S(x)$ die Variable ist. Wir berechnen zunächst die 
linke Seite des Fluktuations-Theorems \eqref{equation::G_120} und setzen hier die 
bedingten Wahrscheinlichkeiten \eqref{equation::G_260} ein. Verschieben wir die 
Argumente der bedingten Wahrscheinlichkeiten noch mit einem Zusatzterm 
$K^{(1)}_{S,\mathrm{fluc}}(x) \Delta t$, so erhalten wir
\begin{equation}
 \frac{ P( + \Delta S + K^{(1)}_{S,\mathrm{fluc}}(x) \Delta t | x ) }{ P( - \Delta S 
 + K^{(1)}_{S,\mathrm{fluc}}(x) \Delta t | x ) } \ = \ \exp( \Delta S / k_B ) \ .
 \label{equation::G_620}
\end{equation}
Die rechte Seite stimmt dann offensichtlich mit dem Fluktuations-Theorem in der ursprünglichen 
Form überein. Wir multiplizieren die Faktoren etwas um und verschieben die Entropie-Änderung 
nochmals gemäß $\Delta S \to \Delta S - K^{(1)}_{S,\mathrm{fluc}}(x) \Delta t$. Dann finden 
wir die äquivalente Gleichung
\begin{eqnarray}
 \exp( - [ \Delta S - K^{(1)}_{S,\mathrm{fluc}}(x) \Delta t ] / k_B ) \ P( \Delta S | x ) \ = \nonumber\\
 = \ P( - \Delta S + 2 \, K^{(1)}_{S,\mathrm{fluc}}(x) \Delta t | x ) \ .
 \label{equation::G_630}
\end{eqnarray}
Diese Gleichung integrieren wir nun über $\Delta S$. Wegen der Normierung der bedingten 
Wahrscheinlichkeitsverteilung ergibt die rechte Seite eins, und wir finden die 
modifizierte Jarzynski-Gleichung
\begin{equation}
 \bigl\langle \exp\bigl( - \bigl[ \Delta S - K^{(1)}_{S,\mathrm{fluc}}(x) \Delta t 
 \bigr] \big/ k_B \bigr) \bigr\rangle \ = \ 1 \ .
 \label{equation::G_640}
\end{equation}
Setzen wir hier die Entropie-Änderung $\Delta S = S(x^\prime) - S(x)$ ein, so können wir 
den Erwartungswert alternativ mit der allgemeineren bedingten Wahrscheinlichkeit 
$P( x^\prime | x )$ berechnen, wobei über die hydrodynamischen Variablen des 
Endzustandes $x^\prime_i$ integriert wird.

Die Gleichungen \eqref{equation::G_620}-\eqref{equation::G_640} gelten zunächst nur 
für infinitesimale Zeitintervalle $\Delta t$. Die modifizierte Jarzynski-Gleichung 
\eqref{equation::G_640} kann jedoch exakt und ohne Näherungen auf ein beliebig großes 
endliches Zeitintervall $\Delta t$ erweitert werden. Der Grund dafür ist, dass die 
rechte Seite von \eqref{equation::G_640} eins ist und nicht von den hydrodynamischen 
Variablen des Anfangszustandes $x_i$ abhängt. Wir zerlegen nun das endliche 
Zeitintervall $\Delta t$ in eine unendliche Anzahl infinitesimaler Zeitintervalle. 
Für jedes dieser infinitesimalen Zeitintervalle gibt es eine modifizierte 
Jarzynski-Gleichung. Wir multiplizieren alle diese modifizierten Jarzynski-Gleichungen 
zusammen, klammern die Integrationen über die hydrodynamischen Variablen alle aus, 
und erhalten somit das Ergebnis 
\begin{equation}
 \Bigl\langle \exp\Bigl( - \Bigl[ \Delta S - \int K^{(1)}_{S,\mathrm{fluc}}(x(t)) 
 \, dt \Bigr] \Big/ k_B \Bigr) \Bigr\rangle \ = \ 1 \ .
 \label{equation::G_650}
\end{equation}
Das Produkt der unendlich vielen bedingten Wahrscheinlichkeiten $P( x^\prime | x )$ 
für jedes infinitesimale Zeitintervall ergibt eine Verbundwahrscheinlichkeit für 
einen Pfad von den hydrodynamischen Variablen $x_i(t)$. Somit wird der Erwartungswert 
in \eqref{equation::G_650} zu einem Pfadintegral. Lediglich die hydrodynamischen 
Variablen des Anfangszustandes $x_i$ sind noch unbestimmte Variablen, weil wir 
von bedingten Wahrscheinlichkeiten ausgegangen sind. Wenn wir zusätzlich noch mit 
einer Wahrscheinlichkeitsverteilung für den Anfangszustand $P_0(x)$ multiplizieren 
und über $x_i$ integrieren, gibt es keine freien Variablen mehr. Das Pfadintegral 
integriert dann wie üblich über alle Variablen $x_i(t)$ des gesamten endlichen 
Zeitintervalls $\Delta t$.

Die Gl.\ \eqref{equation::G_650} ist die modifizierte Jarzynski-Gleichung 
für beliebig große endliche Zeitintervalle $\Delta t$. Sie ist exakt gültig 
und unterscheidet sich von der ursprünglichen Jarzynski-Gleichung \eqref{equation::G_130} 
\cite{Ja97A,Ja97B} durch den Zusatzterm mit dem fluktuierenden Anteil des ersten 
Kramers-Moyal-Koeffizienten $K^{(1)}_{S,\mathrm{fluc}}(x(t))$. Die ursprüngliche 
Jarzynski-Gleichung \eqref{equation::G_130} erhalten wir im Spezialfall 
$K^{(1)}_{S,\mathrm{fluc}}(x(t))=0$ zurück. Wir haben den fluktuierenden Anteil des 
ersten Koeffizienten für eine normale Flüssigkeit explizit berechnet. Das Ergebnis 
\eqref{equation::G_480} hängt stark von der minimalen Länge $\ell$ ab, mit der 
eine Regularisierung durchgeführt wurde so dass nur Variationen der hydrodynamischen 
Variablen auf Längenskalen oberhalb dieser minimalen Länge berücksichtigt werden. 
Ob der Zusatzterm nun klein und irrelevant oder groß und dominierend ist hängt 
davon ab, ob die minimale Länge $\ell$ größer oder kleiner als die kritische 
minimale Länge $\ell_c$ ist, welche in \eqref{equation::G_550}-\eqref{equation::G_570} 
definiert wurde. Folglich darf der Zusatzterm in der modifizierten 
Jarzynski-Gleichung \eqref{equation::G_650} für eine normale Flüssigkeit im 
allgemeinen nicht vernachlässigt werden. 

Zum Schluss haben wir noch eine Bemerkung für das thermische Gleichgewicht, in 
dem $\langle \Delta S \rangle_\mathrm{eq}=0$ gilt. Während in diesem Fall die 
ursprüngliche Jarzynski-Gleichung eine scharfe bedingte Wahrscheinlichkeit 
$P( \Delta S | x ) = \delta( \Delta S )$ mit Breite null erzwingt, bewirkt der 
Zusatzterm, dass die modifizierte Jarzynski-Gleichung weniger einschränkend ist 
und erlaubt, dass auch im thermischen Gleichgewicht die bedingte Wahrscheinlichkeit 
$P( \Delta S | x )$ eine endliche Breite haben darf.

\subsection{Vergleich der Fluktuations-Theoreme}
\label{section::7H}
Wir haben das Fluktuations-Theorem im Rahmen des GENERIC-Formalismus in zwei 
verschiedenen Varianten hergeleitet und untersucht, zum einen in der ursprünglichen 
Form (Abschnitte \ref{section::7A} und \ref{section::7B}) für eine entropieartige 
Variable $\Delta\Sigma$ und zum anderen in einer modifizierten Form (Abschnitte 
\ref{section::7C} bis \ref{section::7G}) für die Entropie-Änderung $\Delta S$. 

Im modifizierten Fluktuations-Theorem haben wir einen Zusatz-Term gefunden, 
welcher mit den Fluktuationen der Entropie im thermischen Gleichgewicht 
zusammenhängt. Dieser Zusatzterm ist wichtig für die Konsistenz der Theorie. 
In dem ursprünglichen Fluktuations-Theorem gibt es diesen Zusatz-Term nicht. 
Dafür hat die Variable $\Delta\Sigma$ neben der Entropie-Änderung $\Delta S$ 
einen zweiten Term. Es gibt also entweder \emph{explizit} im Fluktuations-Theorem 
oder \emph{implizit} in der Variablen einen Zusatz-Term.

Das ursprüngliche Fluktuations-Theorem ist eine Folge der Mikroreversibilität und 
der Existenz eines detaillierten Gleichgewichts. Andererseits wurde das modifizierte 
Fluktuations-Theorem mit einer expliziten Verteilungsfunktion für die bedingte 
Wahrscheinlichkeit hergeleitet und so auf die Kramers-Moyal-Koeffizienten der Entropie 
zurückgeführt. Die explizite Struktur dieser Kramers-Moyal-Koeffizienten liefert 
dann das Fluktuations-Theorem und den Zusatz-Term. Diese Struktur wird durch den 
GENERIC-Formalismus bestimmt und hängt wiederum mit der Mikroreversibilität und 
dem detaillierten Gleichgewicht zusammen. In sofern ist auch das modifizierte 
Fluktuations-Theorem letztendlich eine Folge der Zeitumkehrinvarianz des zugrunde 
mikroskopischen physikalischen Systems und der daraus folgenden Existenz eines 
detaillierten Gleichgewichts.

\section{Abschließende Bemerkungen}
\label{section::8}
Aus der mikroskopischen Theorie für Quantenvielteilchensysteme wurden mit Methoden der 
Quantenstatistik und Projektionsoperatoren die verallgemeinerten hydrodynamischen Gleichungen 
hergeleitet. Zunächst sind die Gleichungen exakt. Sie sind räumlich und zeitlich nicht lokal 
und enthalten Gedächtniseffekte und Fluktuationen. Die Gleichungen haben eine besondere 
grundlegende Struktur. Auf der rechten Seite stehen drei Arten von Termen: reversible, 
dissipative und fluktuierende. Bereits die exakten Gleichungen lassen sich auf eine 
verallgemeinerte Form des GENERIC-Formalismus von Grmela und Öttinger \cite{GO97A,GO97B,Ot05} 
bringen.

Die Näherungen ändern an dieser grundlegenden Struktur nichts. Vernachlässigt man im ersten 
Schritt die Gedächtniseffekte (Markov-Näherung), so bekommen die hydrodynamischen Gleichung 
eine Form, welche der ursprünglichen Version des GENERIC-Formalismus entspricht. Betrachtet 
man eine normale Flüssigkeit und vernachlässigt nicht lokale Effekte, so folgen die 
hydrodynamischen Gleichungen mit Fluktuationen, wie sie aus den Lehrbüchern \cite{LL06,LL09} 
bekannt sind. Durch Symmetrieargumente findet man, dass die Effekte von Dissipation und 
Fluktuation in einer normalen Flüssigkeit durch drei Parameter beschrieben werden, die 
Scherviskosität $\eta$, die Volumenviskosität $\zeta$ und die Wärmeleitfähigkeit $\varkappa$.

Die exakten verallgemeinerten hydrodynamischen Gleichungen sind invariant unter der Zeitumkehr, 
weil wir annehmen, dass die zugrunde liegende mikroskopische Theorie diese Eigenschaft hat. 
Wir untersuchen, wie sich die Zeitumkehrinvarianz auf die drei Terme der rechten Seite verteilt. 
Die reversiblen Terme sind natürlich invariant. In Folge ist auch die Summe der dissipativen 
und der fluktuierenden Terme invariant in unter der Zeitumkehr. Die dissipativen und 
fluktuierenden Terme im einzelnen brechen jedoch die Zeitumkehrinvarianz. Man sieht dies 
insbesondere an der Bewegungsgleichung für die Entropie. Der dissipative Term ist hier 
quadratisch und positiv definit. Er führt zu einem Anwachsen der Entropie mit der Zeit und 
bildet die Grundlage für den zweiten Hauptsatz der Thermodynamik. Dem kann nur der fluktuierende 
Term entgegenwirken. Nur dieser kann die Entropie wieder absenken. Solange man also die 
fluktuierenden Terme berücksichtigt, bleibt die Zeitumkehrinvarianz bestehen. Daran ändern 
auch die Näherungen nichts, welche zu den hydrodynamischen Gleichungen der normalen Flüssigkeit 
führen.

Werden die Gedächtniseffekte vernachlässigt, so bekommen die hydrodynamischen Gleichungen die 
Form von Langevin-Gleichungen, wobei die fluktuierenden Terme durch gaußische stochastische 
Kräfte modelliert werden. Wir vergleichen mit der allgemeinen Theorie für stochastischer 
Prozesse und stellen fest, dass die Hydrodynamik mit Fluktuationen zu einer stochastischen 
Theorie mit gaußischen Fluktuationen äquivalent ist. Eine zugehörige Fokker-Planck-Gleichung 
lässt sich finden, deren Lösung im thermischen Gleichgewicht eine großkanonische 
Boltzmann-Verteilung ist. Weiterhin zeigen wir, dass im thermischen Gleichgewicht die 
Entropie im Mittel konstant bleibt, wie es erwartet wird. Zwar bewirkt der positiv definite 
dissipative Term durch Fluktuationen auch im thermischen Gleichgewicht ein stetiges Anwachsen 
der Entropie. Dieser Effekt wird jedoch im Mittel exakt kompensiert durch den fluktuierenden 
Term. 

Es lässt sich also eine stochastische Theorie mit gaußischen stochastischen Kräften 
formulieren, die in sich konsistent und frei von Widersprüchen ist. Wir führen diese 
Überlegungen für die allgemeinen nichtlokalen hydrodynamischen Gleichungen in der 
GENERIC-Form durch. Sie sind jedoch nicht darauf beschränkt, sondern gelten auch für 
die speziellen lokalen hydrodynamischen Gleichungen einer normalen Flüssigkeit.
Als konkretes Beispiel haben wir immer eine normale Flüssigkeit betrachtet. Unsere 
Überlegungen zusammen mit dem GENERIC-Formalismus gelten jedoch allgemeiner und sind 
auch auf komplexere Flüssigkeiten anwendbar. Solche wären z.B.\ Mischungen aus 
unterschiedlichen Komponenten ohne und mit chemischen Reaktionen \cite{Ot09,Ba14,Ba15}.

Zum Schluss haben wir gezeigt, wie man die Herleitung des Fluktuations-Theorems nach 
Crooks \cite{Cr98,Cr99,Cr00} auf den GENERIC-Formalismus und die Hydrodynamik einer 
normalen Flüssigkeit übertragen kann. Es stellt sich jedoch heraus, dass die Variable 
des Fluktuations-Theorems nicht die Entropie-Änderung ist, sondern einen Zusatzterm 
enthält. Aus diesem Grunde haben wir alternativ für die Entropie-Änderung als Variable 
eine \emph{modifizierte Version} des Fluktuations-Theorems und der Jarzynski-Gleichung 
hergeleitet. Wir finden hier im Fluktuations-Theorem selbst einen Zusatzterm, der von dem 
fluktuierenden Anteil des ersten Kramers-Moyal-Koeffizienten für die Entropie erzeugt wird. 

Wir berechnen diesen Zusatzterm explizit für eine normale Flüssigkeit und finden zunächst eine 
Ultraviolett-Divergenz. Um ein endliches und physikalisch sinnvolles Ergebnis zu erhalten, 
müssen wir eine Regularisierung durchführen mit einer minimalen Längenskala, bis zu welcher 
die räumlichen Variationen und Fluktuationen der hydrodynamischen Variablen berücksichtigt 
werden. Der Zusatzterm hängt stark von dieser minimalen Länge ab. Je nachdem ob die minimale 
Länge größer oder kleiner als eine bestimmte kritische Länge ist, wird der Zusatzterm klein 
und irrelevant oder groß und dominierend sein. Wir schließen daraus, dass für eine normale 
Flüssigkeit das Fluktuations-Theorem und die Jarzynski-Gleichung im allgemeinen durch 
einen Zusatzterm modifiziert werden müssen, wenn die Variable die Entropie-Änderung 
in der Flüssigkeit sein soll.

\acknowledgments
Der Autor dankt Prof.\ Dr.\ M.\ Fuchs für inspirierende Diskussionen und 
Prof.\ Dr.\ H.\ C.\ Öttinger und Prof.\ Dr.\ U.\ Seifert für hilfreiche Kommentare zum 
Manuskript.


\end{document}